\documentclass[12pt,a4paper,oneside,reqno]{article}
\usepackage{adjustbox}
\usepackage{amsmath}
\usepackage[skip=2pt]{caption}  
\usepackage{amssymb}
\usepackage{float}
\usepackage[T1]{fontenc}
\usepackage{hhline}
\usepackage[utf8]{inputenc}
\usepackage{mathtools}
\usepackage{multicol}
\usepackage{multirow}
\usepackage{wasysym}
\usepackage{graphicx}
\usepackage{caption}
\usepackage{subcaption}
\usepackage{amsmath,amssymb} 
\usepackage{siunitx}
\usepackage[margin=1in]{geometry} 
\usepackage{enumerate}
\usepackage{hyperref}
\usepackage{graphicx}

\newcommand\blfootnote[1]{%
  \begingroup
  \renewcommand\thefootnote{}\footnote{#1}%
  \addtocounter{footnote}{-1}%
  \endgroup
}
\usepackage{amsmath,cases}

\usepackage{cite}
\usepackage{hyperref}
\hypersetup{
colorlinks=true,
linkcolor=blue,
filecolor=magneta,
urlcolor=cyan,
}
\usepackage{lineno}
\begin{document}
 \nocite{*} 

\title{Bifurcations and synchronization of coupled translational-rotational stick-slip oscillators}

\maketitle
\author
\begin{center}
{Grzegorz Kudra}$^{1,*}$, {Ali Fasihi$^{1}$}, {Mohammad Parsa Rezaei$^{1}$},  {Muhammad Junaid-U-Rehman$^{1}$}, {Jan Awrejcewicz}$^{1}$
\blfootnote{
{E-mail address:} (G. Kudra${}^*$) grzegorz.kudra@p.lodz.pl, (M. J. Rehman) muhammad-junaid.u-rehman@dokt.p.lodz.pl, (A. Fasihi) Ali.fasihi@dokt.p.lodz.pl, (M.P. Rezaei) mohammad-parsa.rezaei@p.lodz.pl, (J. Awrejcewicz) jan.awrejcewicz@dokt.p.lodz.pl}
\end{center}
\begin{center}
\emph{$^1$Department of Automation, Biomechanics, and Mechatronics, Lodz University of Technology, $1/15$ Stefanowski st., $90-537$ \L\'od\'z, Poland}\\
\end{center}

\begin{abstract}
\noindent This paper presents mathematical modeling and numerical analysis of bifurcation and synchronization phenomena in a system of coupled oscillators driven by a finite-power energy source and generating two-dimensional stick-slip translational-rotational vibrations. The mechanical system consists of rigid disks placed on moving belts and asymmetrically connected to a support via springs, with each disk simultaneously performing translational and rotational motion. The belts are driven by a common DC motor. The disk contacts the belt over a finite contact area, resulting in mutually coupled frictional force and torque through their dependence on the linear and angular slip velocity. The paper utilizes special approximations of the resultant frictional forces and torques based on generalizations of Padé's developments, in which special smoothing elements are introduced to avoid singularities when the relative motion between the disk and belt disappears. Furthermore, the model also provides a smooth approximation of the friction model, in which static friction is greater than kinetic friction. Numerical analysis was conducted based on direct numerical simulations, Poincaré maps, and bifurcation diagrams. It was shown that the system can exhibit a two-dimensional stick-slip phenomenon, and the system exhibits predominantly periodic dynamics, sometimes with a very long period. In this work, a system of two oscillators driven by a common, limited-power motor was studied, which can synchronize and oscillate in phase or in counterphase.
\end{abstract}

\section{Introduction}
Friction resulting from direct contact between two solid bodies is a common phenomenon in mechanical systems and engineering structures. Typical components where this phenomenon occurs include bearings, gears, hydraulic and pneumatic cylinders, valves, brakes, and wheels. While it is unavoidable and necessary for the operation of mechanical devices, it is often undesirable and requires modification or control. Therefore, it has long been the subject of intensive study and research aimed at understanding the physical basis of this phenomenon \cite{op10a,op10b,op11,op15} and methods of modeling it \cite{op13,op14,op16}. Friction research has been ongoing for a very long time and remains relevant today, stemming not only from the need to understand the phenomenon but also from important engineering needs, such as in automation, the design of drive systems, precise positioning mechanisms, and anti-lock braking systems in cars. Because friction is a highly nonlinear phenomenon, it can lead to steady-state errors, limit cycles, or even chaotic behavior. One method for reducing the adverse impact of friction on positioning is dithering, which involves introducing vibrations into the system to reduce stick-slip, dead zone, and hysteresis. These vibrations are typically implemented electronically in the control signal or mechanically in early autopilots \cite{op12}. In addition to sliding friction, during contact between solid bodies, a phenomenon called rolling friction, more commonly referred to as rolling resistance, may also occur, which results from asymmetric contact pressure in the contact area and is the result of elastic hysteresis \cite{op70,op71,op72,op1,op3,op5}.

The classically understood friction model represents the relationship between a single component of the friction force and the one-dimensional relative displacement of the contacting bodies. This relationship can have varying degrees of complexity, ranging from the classical Coulomb friction model to advanced relations that also incorporate additional internal state variables. Such friction models are directly applicable to the mathematical description and study of dynamic systems with frictional contact, where the same relative motion of the contacting surfaces occurs at each contact point. Because the classical Coulomb model is discontinuous, its continuous \cite{op19, op17} or smooth versions \cite{op18,op20} are often used. 
In the paper \cite{op22}, a smooth model of the friction force dependent on velocity was presented. To represent the friction force near zero relative velocity, an additional relation was used. Other commonly used friction models include the Karnopp model \cite{op23,op24}, the Dahl model \cite{op25,op26}, and the LuGre model \cite{op26,op27,op28}, which is particularly used in friction-based control systems. A certain extension of the LuGre model was introduced by Gonthier \cite{op32}. Even these classically understood, one-dimensional friction models lead to complex dynamic phenomena, the most famous of which is the stick–slip phenomenon \cite{op29,op30,op31,op2}.

It turns out, however, that real engineering problems very often involve frictional contacts, where the assumption of point contact or a uniform slip velocity distribution over the entire contact surface does not provide a correct solution. When bodies come into contact over a finite area, the pressure distribution is non-uniform, local stick–slip transitions and microslip appear. Full and accurate modeling then requires spatial discretization and the use of appropriate methods, such as the finite element method \cite{op37,op38, op39, op40}. Although there also exist specialized models and software, such as Kalker’s CONTACT \cite{op64,op65, op66, op69}, this leads to a drastic increase in simulation time, high computational cost, and is not suitable in situations where a fast yet realistic numerical simulation is required. Therefore, simple models are sought that nevertheless reproduce the resultant interaction forces between bodies well. Friction over a finite contact area not only generates a resultant force opposing relative motion, but also a moment arising from the distribution of elementary friction forces over that area. To correctly model this phenomenon, both the resultant force and the friction-induced moment must be considered together.  

The first solution to this problem was presented by Contensou \cite{op_Contensou} by means of his integral models, which were later developed by Zhuravlev, Kireenkov, and Kudra using Padé approximations and their generalizations.  
Kireenkov and Zhuravlev \cite{op33,op60, op35}, based on Contensou’s integral model, proposed a two-dimensional model in the form of Padé approximations of the friction force and moment as functions of velocity components. In \cite{op3,op1}, Kudra and Awrejcewicz proposed smooth approximations of the friction force and moment on an elliptic contact with a pressure distribution deformed due to rolling resistance, which constitute an extension and generalization of Kireenkov’s solutions. In \cite{op6}, the possibility of smoothing the model of friction forces and moments accounting for different coefficients of static and kinetic friction was additionally presented, allowing the stick–slip phenomenon to be reproduced. These models were then applied to the study of the dynamics of several dynamical systems, such as the Celtic stone \cite{op36,op4, op5}, including studies with a particular focus on rolling resistance \cite{op9}, the billiard ball \cite{op42}, an ellipsodal body operating as a tippe-top \cite{op_Ellipsoid}, a clutch \cite{op8}, and the contact of a car tire with the ground \cite{op34}.  

Friction force and moment models for finite contact areas can be classified as follows:  
\begin{enumerate}[(i)]
    \item Models involving spatial discretization with the use of appropriate numerical solution methods such as the finite element method \cite{op37,op38, op39, op40, op64,op65, op66, op69},
    \item Integral models, in which the pressure distribution is assumed to be known (e.g., Hertzian, deformed Hertzian, or uniform), and the friction force and moment are computed through integration \cite{op_Contensou},
    \item Approximation models (e.g., Padé approximant) of the integral models \cite{op33,op60, op35, op3,op1},
\end{enumerate}

On the other hand, models can be divided into:  
\begin{enumerate}[(i)]
    \item Smooth models, convenient for direct numerical simulations \cite{op3,op1, op6, op43},
    \item Nonsmooth models, in which differential inclusions appear, with two sub-approaches used for their solution: the model treated as hybrid is simulated using event-driven techniques, or time-stepping methods are employed to explicitly handle discontinuities~\cite{op62,op61,op63,op41}.
\end{enumerate}
The friction force and moment model adopted in the present study belongs to the class of smooth approximation models.

On the other hand, in engineering practice, systems with non-ideal energy sources are common \cite{op44}, where the drive is coupled to the dynamical system and exhibits phenomena absent in systems with ideal sources, such as the Sommerfeld effect. Studies show that limited-power actuators can alter the character of resonance passages, leading to nonlinear phenomena such as bifurcations and sudden amplitude jumps \cite{op45}. In \cite{op46}, the authors investigate the influence of source power on dynamic effects such as bifurcations and stability changes. Warmiński \cite{op47} studies mechanical systems with parametric and self-excitation, demonstrating the impact of limited drive power on system behavior during resonance and stability transitions. Shvets \cite{op48} analyzes the routes to chaos in non-ideal systems.

Nature and engineering systems abound with examples of synchronization phenomena. Clocks, insects, heart pacemakers, neurons, and applauding crowds all tend to act synchronously. Nonlinear dynamics provides an explanation of these phenomena, which turn out to be universal \cite{op50,op51}. There are also studies where the notion of synchronization appears, but it is enforced through a control system rather than spontaneous, while still being of strong practical significance \cite{op56,op52,op53,op54,op55}.

To the authors’ knowledge, there are no, or only very few, studies analyzing dry-friction systems exhibiting self-excited vibrations (such as a mass-on-belt setup) that simultaneously incorporate a non-ideal belt drive, e.g., a DC motor, in the mathematical model. In particular, there is a lack of studies on coupled oscillators of this type driven by a common motor, as well as on systems where the body on the belt can perform two-dimensional stick–slip vibrations combining rotational and translational motion. In this sense, the present work attempts to fill this gap and take first steps in the analysis of such systems, which may be of both practical and theoretical significance. The paper continues the works \cite{op2, op6}, where a similar setup was already analyzed, but limited to a single oscillator with constant belt velocity. In \cite{op2}, a hybrid event-driven model (handling transitions between stick and slip) was applied, whereas in \cite{op6} a continuous version of the model was employed.

The paper is organized as follows. Section \ref{sec:Mechanical system of single oscillator with constant belt velocity} develops the mathematical model of a single oscillator operating under constant belt velocity, structured into three subsections: Subsection \ref{subsec:System description} provides a detailed description of the physical model, while Subsections \ref{subsec:Equations of motion} present the formulation of the governing equations in both dimensional and nondimensional forms. Section \ref{sec:chain of oscillators driven by a DC motor} extends the analysis to a chain of oscillators actuated by a single DC motor. Numerical investigations, including bifurcation analysis and the examination of time responses, phase portraits, and Poincaré sections, are reported in Section \ref{sec:Numerical results}. Finally, Section \ref{sec:Concluding remarks} concludes the study with a summary of the key results and insights.

\section{Mechanical system of single oscillator with constant belt velocity}\label{sec:Mechanical system of single oscillator with constant belt velocity}
\subsection{System description}\label{subsec:System description}
\begin{figure}[H]
  \centering
  \includegraphics[scale=0.6]{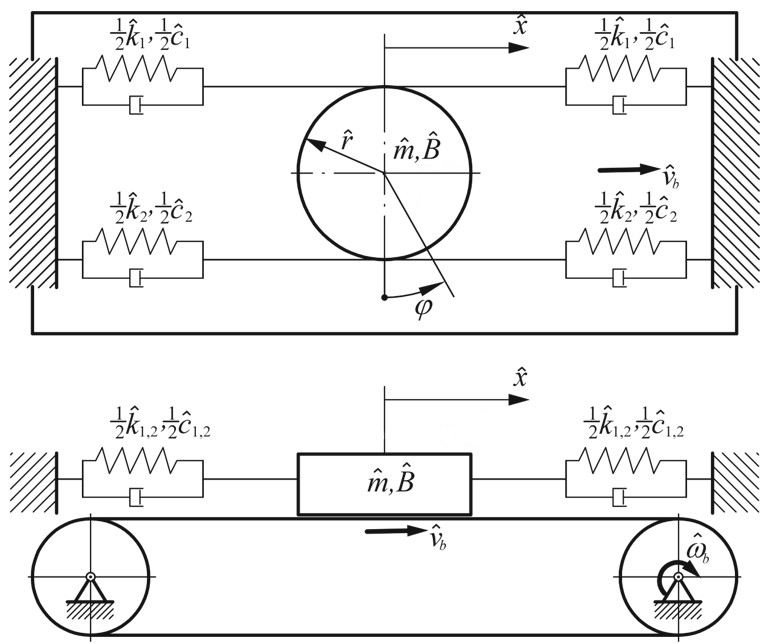} 
  \caption{2DOF system consisting of a single plane disk situated elastically on a belt moving with constant velocity \cite{op2}.}
  \label{fig:base_system}
\end{figure}

In Fig. \ref{fig:base_system}, the physical system under investigation is a single oscillator in the form of a rigid disk placed on a belt moving with constant velocity. The disk---characterized by its mass $\hat{m}$\footnote{
Here and throughout the manuscript, parameters with hats ( $\hat{\ }$ ) denote dimensional (physical) parameters. Due to convenience, we reserve parameters without hats for nondimensional forms as they are used in the analysis in the next sections.
}, radius $\hat{r}$, and moment of inertia $\hat{B}$--- is elastically supported through two linear springs of stiffness $\tfrac{1}{2}\hat{k}_1$ and $\tfrac{1}{2}\hat{k}_2$, and viscously damped by elements with coefficients $\tfrac{1}{2}\hat{c}_1$ and $\tfrac{1}{2}\hat{c}_2$. The springs are prestressed so that the cords remain taut during motion. The system is described by two generalized coordinates: the horizontal displacement of the disk’s center $\hat{x}$ along the belt axis and its angular rotation $\varphi$ about the center. The moving belt enforces a constant velocity $\hat{v}_b$, introducing tangential friction forces $\hat{T}$ and resisting torques $\hat{M}$ at the contact interface. These frictional effects are responsible for stick–slip transitions, which strongly influence the oscillator’s dynamics. This configuration forms a 2DOF translational–rotational oscillator in which the combined effects of elasticity, damping, and dry friction give rise to complex nonlinear behaviors, making it a fundamental model for studying bifurcations, energy dissipation, and stability in friction-induced vibrations.

\subsection{Equations of motion}\label{subsec:Equations of motion}

Using the parameters defined above, the governing equations of motion are obtained in dimensional form:

\begin{subequations}\label{eq:base_equations}
\begin{align}
&\hat{m}\,\hat{x}'' 
+ \hat{k}_{1}\left(\hat{x} - \varphi\, \hat{r}\right) 
+ \hat{k}_{2}\left(\hat{x} + \varphi\, \hat{r}\right) 
+ \hat{c}_{1}\left(\hat{x}' - \varphi' \hat{r}\right) 
+ \hat{c}_{2}\left(\hat{x}' + \varphi' \hat{r}\right) \nonumber \\
&\quad+ \mu\, \hat{N}\, T_{s}\!\left(\frac{\hat{x}' - \hat{v}_{b}}{\hat{r}},\, \varphi',\, \hat{\varepsilon}\right)
= 0,\\
&\hat{B}\,\varphi'' 
- \hat{k}_{1}\left(\hat{x} - \varphi\, \hat{r}\right)\hat{r} 
+ \hat{k}_{2}\left(\hat{x} + \varphi\, \hat{r}\right)\hat{r} 
- \hat{c}_{1}\left(\hat{x}' - \varphi'\hat{r}\right)\hat{r} 
+ \hat{c}_{2}\left(\hat{x}' + \varphi'\hat{r}\right)\hat{r} \nonumber \\
&\quad+ \hat{r}\mu\, \hat{N}\, M_{s}\!\left(\frac{\hat{x}' - \hat{v}_{b}}{\hat{r}},\, \varphi',\, \hat{\varepsilon}\right)
= 0,
\end{align}
\end{subequations}

\noindent where $(\;^{\prime}\;)$ denotes the derivative with respect to real time $\hat{t}$, and $\hat{N}=\hat{m} g$, $T_{s}(v_{s}, \omega_{s}, \varepsilon)$, and $M_{s}(v_{s}, \omega_{s}, \varepsilon)$ are special functions representing smooth approximations of the resultant friction force and torque on finite plane contact areas (see \cite{op6}). The smooth approximations eliminate singularities inherent in discontinuous friction models and enable the use of standard numerical integration methods for the differential equations. They are be defined as:
\begin{equation}
\begin{cases}
T_{s}(v_{s}, \omega_{s}, \varepsilon) 
= v_{s} \left( 
\sqrt{\frac{1}{v_{s}^{2} + \omega_{s}^{2} + \varepsilon^{2}}} 
+ \eta' \frac{\varepsilon^{3}}{(v_{s}^{2} + \omega_{s}^{2} + \varepsilon^{2})^{2}}
\right), \\
\\
M_{s}(v_{s}, \omega_{s}, \varepsilon) 
= \tfrac{2}{3}\,\omega_{s} \left( 
\sqrt{\frac{1}{v_{s}^{2} + \omega_{s}^{2} + \varepsilon^{2}}} 
+ \eta' \frac{\varepsilon^{3}}{(v_{s}^{2} + \omega_{s}^{2} + \varepsilon^{2})^{2}}
\right),
\end{cases}
\end{equation}
where $\eta'$ is a model parameter given by the piecewise polynomial approximations:
\begin{equation}\nonumber
\eta' =
\begin{cases}
-13.607 + 30.893\eta - 22.01\eta^{2} + 5.878\eta^{3}, & \text{for} ~~ \eta \in \langle 1, 1.3\rangle, \\[6pt]
-2.41 + 3.985\eta - 0.3581\eta^{2} + 0.0493\eta^{3}, & \text{for} ~~ \eta \in \langle 1.3, 2.7\rangle, \\[6pt]
-1.684 + 3.121\eta - 0.00483\eta^{2} + 0.000196\eta^{3}, & \text{for} ~~ \eta \in \langle 2.7, 10\rangle,
\end{cases}
\end{equation}
where $\eta = \mu_0/\mu$ is the ratio of static to kinetic friction coefficients. The polynomial approximations ensure that the difference between the actual static friction coefficient (maximum friction) and the desired static friction coefficient set by the approximation remains below $|\Delta \eta| < 0.001$ across all specified parameter ranges, providing both computational efficiency and high accuracy.

 To convert the dimensional equations of motion into the dimensionless form, non-dimensional variables are introduced as below
\begin{equation}\label{dimensionless}
    x = \frac{\hat{x}}{\hat{r}}, \quad 
t = \alpha \hat{t}, \quad 
\end{equation}
\noindent where \(\alpha = \sqrt{\tfrac{\hat{k}_{1} + \hat{k}_{2}}{\hat{m}}}\).
Using Eqs. (\ref{dimensionless}) and (\ref{eq:base_equations}), the governing equations can be expressed in the following dimensionless matrix form.

\begin{equation}\label{eq:dimensionless_single_oscillator}
\begin{split}
&\begin{bmatrix}
1 & 0 \\
0 & m
\end{bmatrix}
\begin{Bmatrix}
\ddot{x} \\ \ddot{\varphi}
\end{Bmatrix}
+
\begin{bmatrix}
c & c_{12} \\
c_{12} & c
\end{bmatrix}
\begin{Bmatrix}
\dot{x} \\ \dot{\varphi}
\end{Bmatrix}
+
\begin{bmatrix}
1 & k_{12} \\
k_{12} & 1
\end{bmatrix}
\begin{Bmatrix}
x \\ \varphi
\end{Bmatrix} +
\bar{\mu}
\begin{Bmatrix}
T_{s}(\dot{x} - v_{b},~\dot{\varphi},~\bar{\varepsilon}) \\
M_{s}(\dot{x} - v_{b},~\dot{\varphi},~\bar{\varepsilon})
\end{Bmatrix}
=
\begin{Bmatrix}
0 \\ 0
\end{Bmatrix}
\end{split}
\end{equation}

\noindent where $(\;^{.}\;)$ denotes differentiation with respect to the dimensionless time $t$.\\
Based on these nondimentional variables, the nondimensional parameters are obtained as follows:
\begin{equation}
\begin{split} \label{3}
&m = \frac{\hat{B}}{\hat{m}\hat{r}^{2}}, \quad k_{12} = \frac{\hat{k}_{2} - \hat{k}_{1}}{\hat{k}_{1} + \hat{k}_{2}}, \quad c = \frac{\hat{c}_{1} + \hat{c}_{2}}{\sqrt{\hat{m}(\hat{k}_{1} + \hat{k}_{2})}}, \\ 
&c_{12} = \frac{\hat{c}_{2} - \hat{c}_{1}}{\sqrt{\hat{m}(\hat{k}_{1} + \hat{k}_{2})}}, \quad \bar{\mu} = \frac{\mu \hat{m} g}{\hat{r}(\hat{k}_{1} + \hat{k}_{2})},  \quad
v_{b} = \frac{\hat{v}_{b}}{\alpha \hat{r}}, \quad \bar{\varepsilon} = \frac{\varepsilon}{\alpha}.
\end{split}
\end{equation}

\subsection{Numerical examples}\label{subsec:Numerical examples}

For initial validation, we first adopt parameter values from our previous work~\cite{op6}:
\begin{align}\label{eq:Literature parameter values}
m &= 90, \quad k_{12} = 0.85, \quad c = 0.0001, \quad c_{12} = 0, \quad \bar{\mu} = 5, \quad v_b = 0.15, \quad \bar{\varepsilon} = 10^{-5}
\end{align}

\begin{figure}
    \centering
    \begin{subfigure}{0.49\textwidth}
        \centering
        \includegraphics[width=\textwidth]{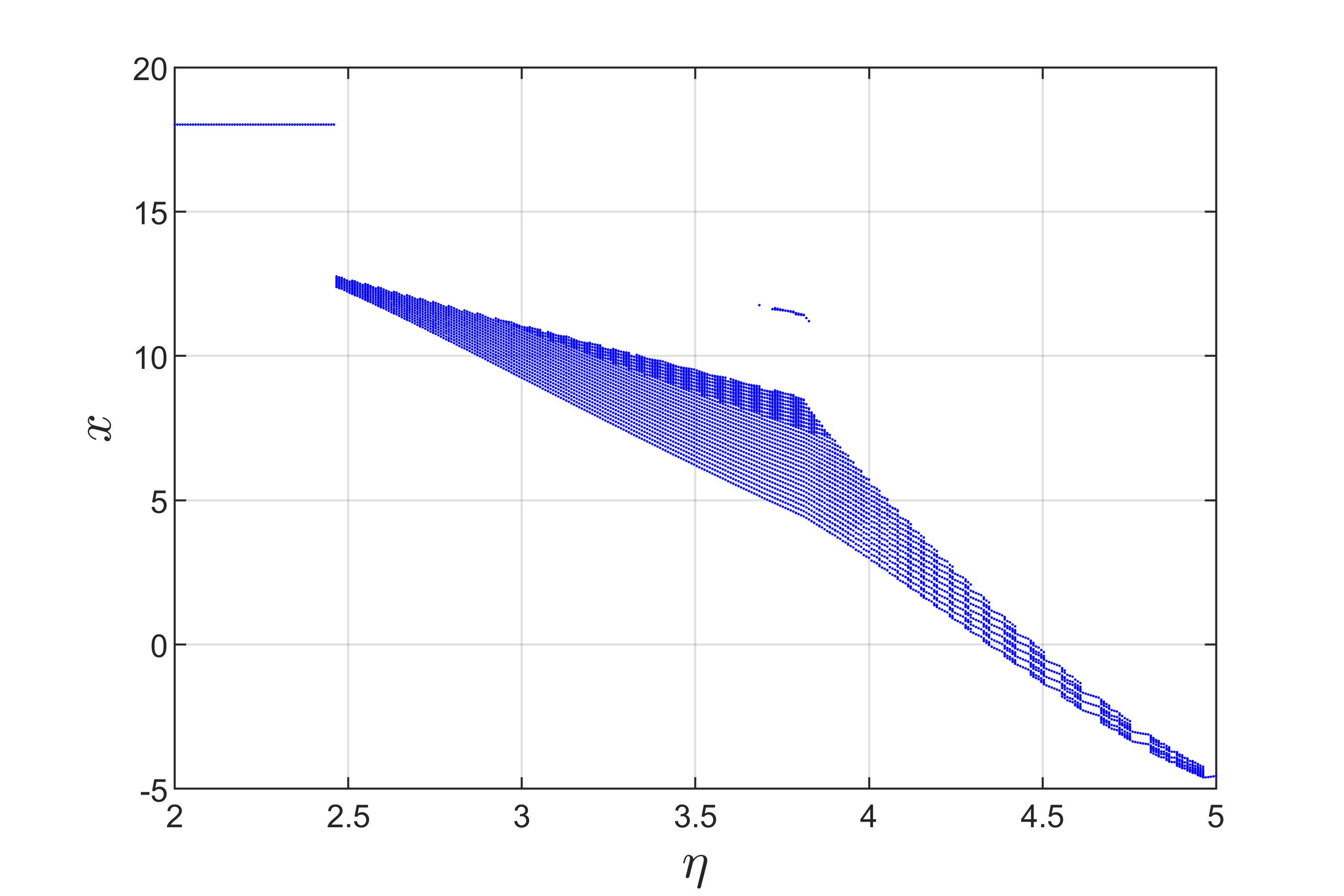}
        \caption{}
        \label{fig2a}
    \end{subfigure}
    \vspace{0.3cm}
    \begin{subfigure}{0.49\textwidth}
        \centering
        \includegraphics[width=\textwidth]{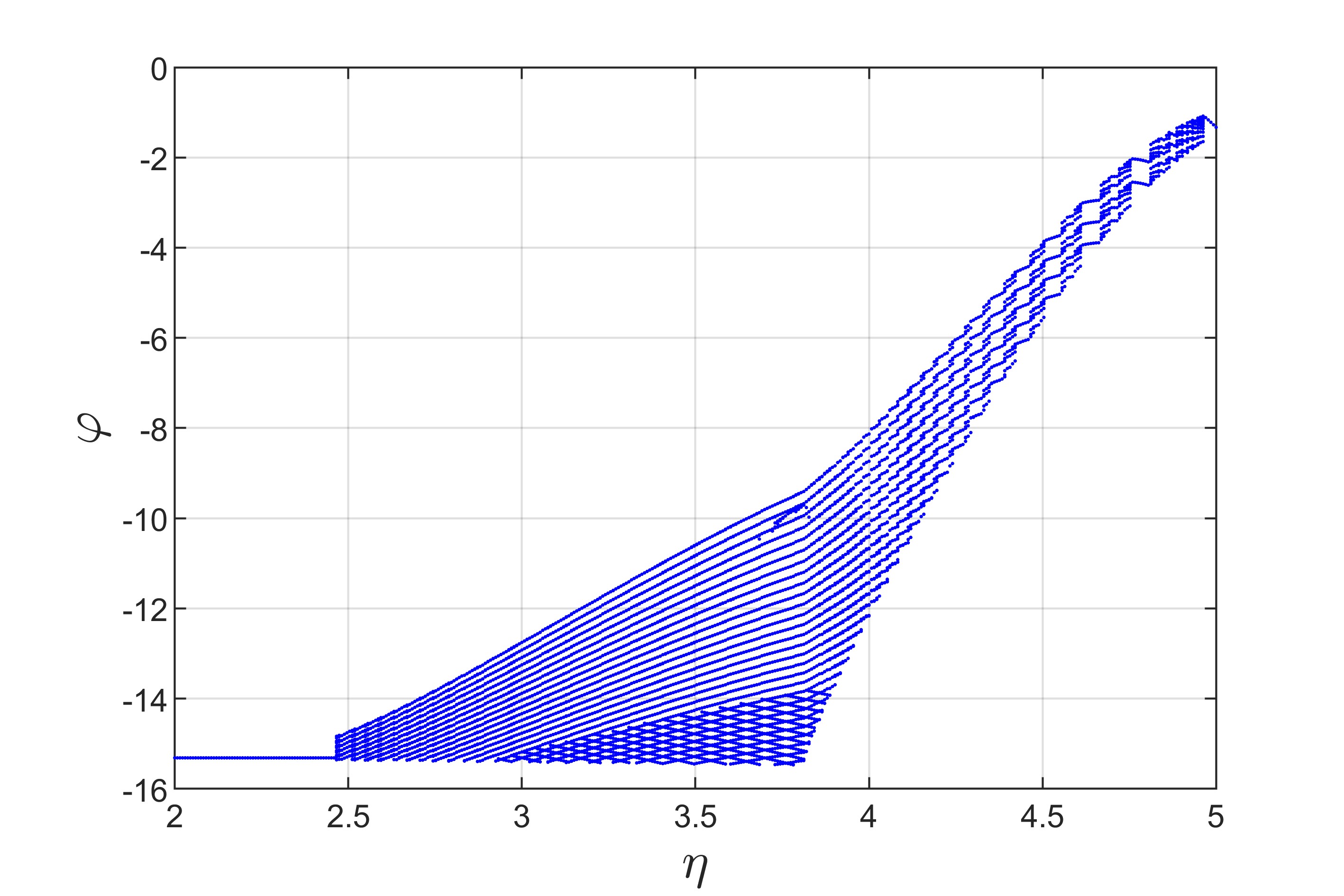}
        \caption{}
        \label{fig2b}
    \end{subfigure}
    \caption{
     Bifurcation diagrams of the Poincaré map defined by the local minima of $x$, as the friction  coefficient ratio ($\eta$) is swept as the control parameter: 
     (a) displacement $x$; (b) rotational coordinate $\varphi$; For each $\eta$, integration is restarted from $(x_0,\,\dot{x}_0,\,\varphi_0,\,\dot{\varphi}_0) = (0,\,0,\,0,\,0)$; Parameters follow Eq.~\ref{eq:Literature parameter values}.
     }

     \label{fig2}
\end{figure}

\begin{figure}[H]
    \centering
    \begin{subfigure}{0.48\textwidth}
        \includegraphics[width=\linewidth]{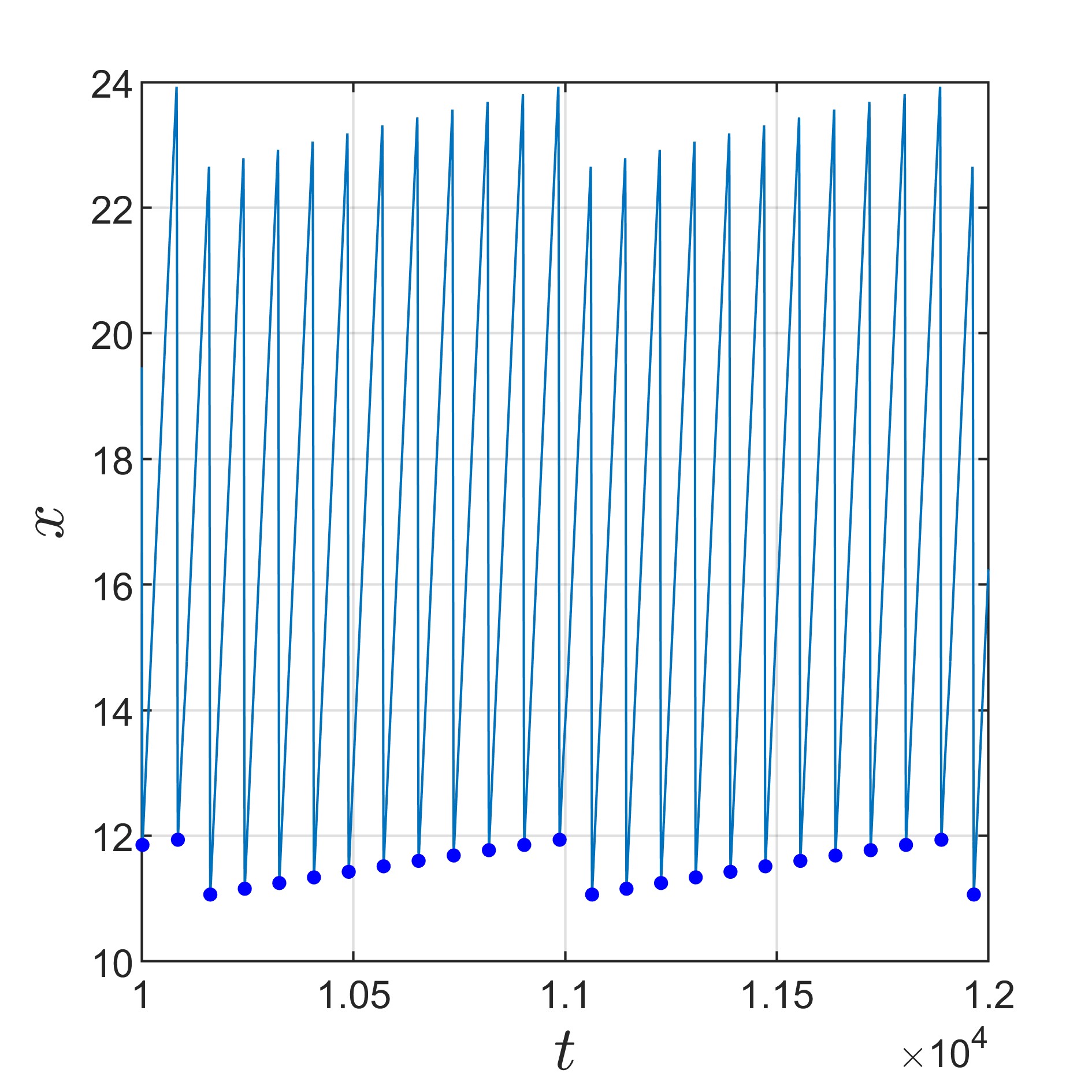}
        \caption{ }
        \label{fig3a}
    \end{subfigure}
    \begin{subfigure}{0.48\textwidth}
        \includegraphics[width=\linewidth]{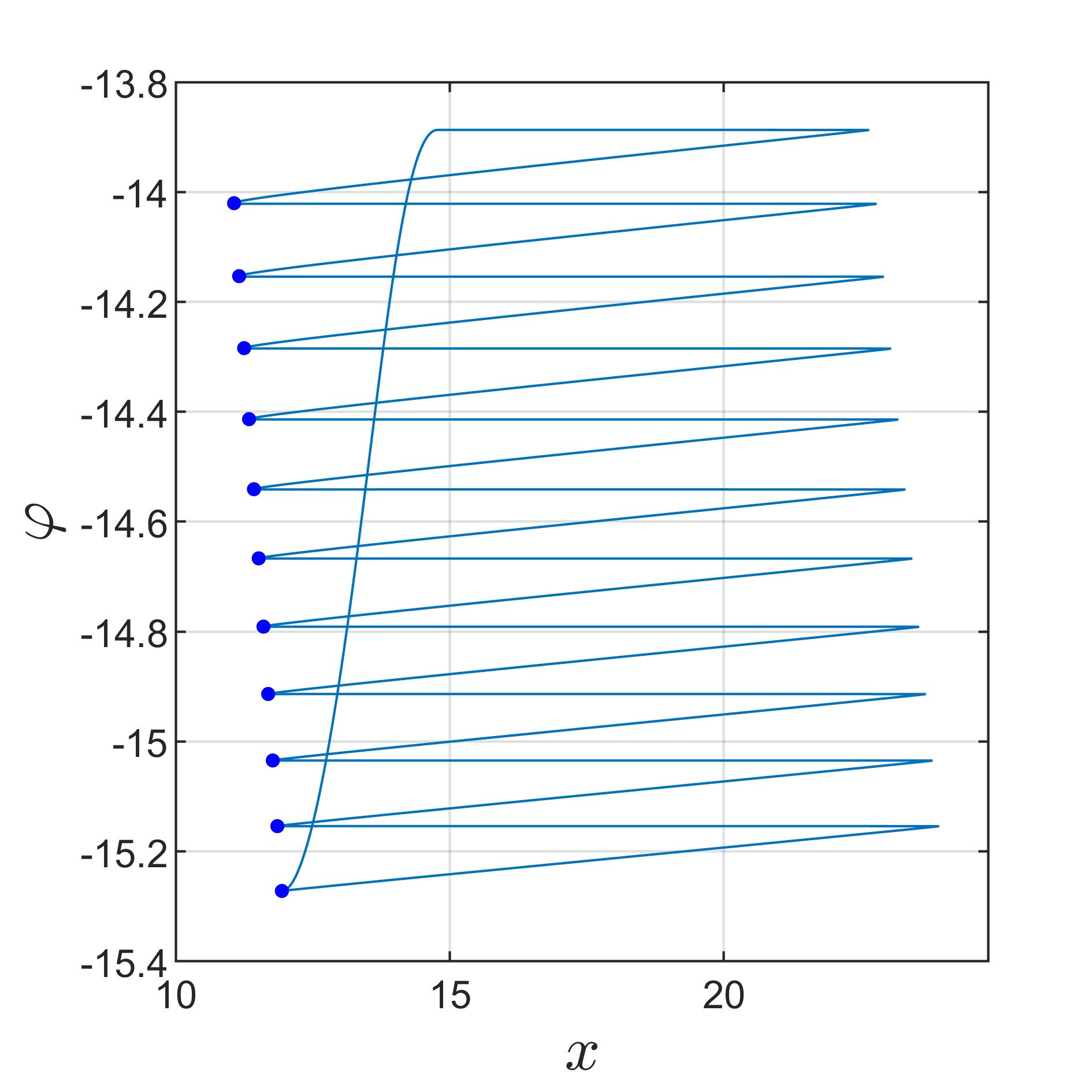}
        \caption{ }
        \label{fig3b}
    \end{subfigure}
        \begin{subfigure}{0.48\textwidth}
        \includegraphics[width=\linewidth]{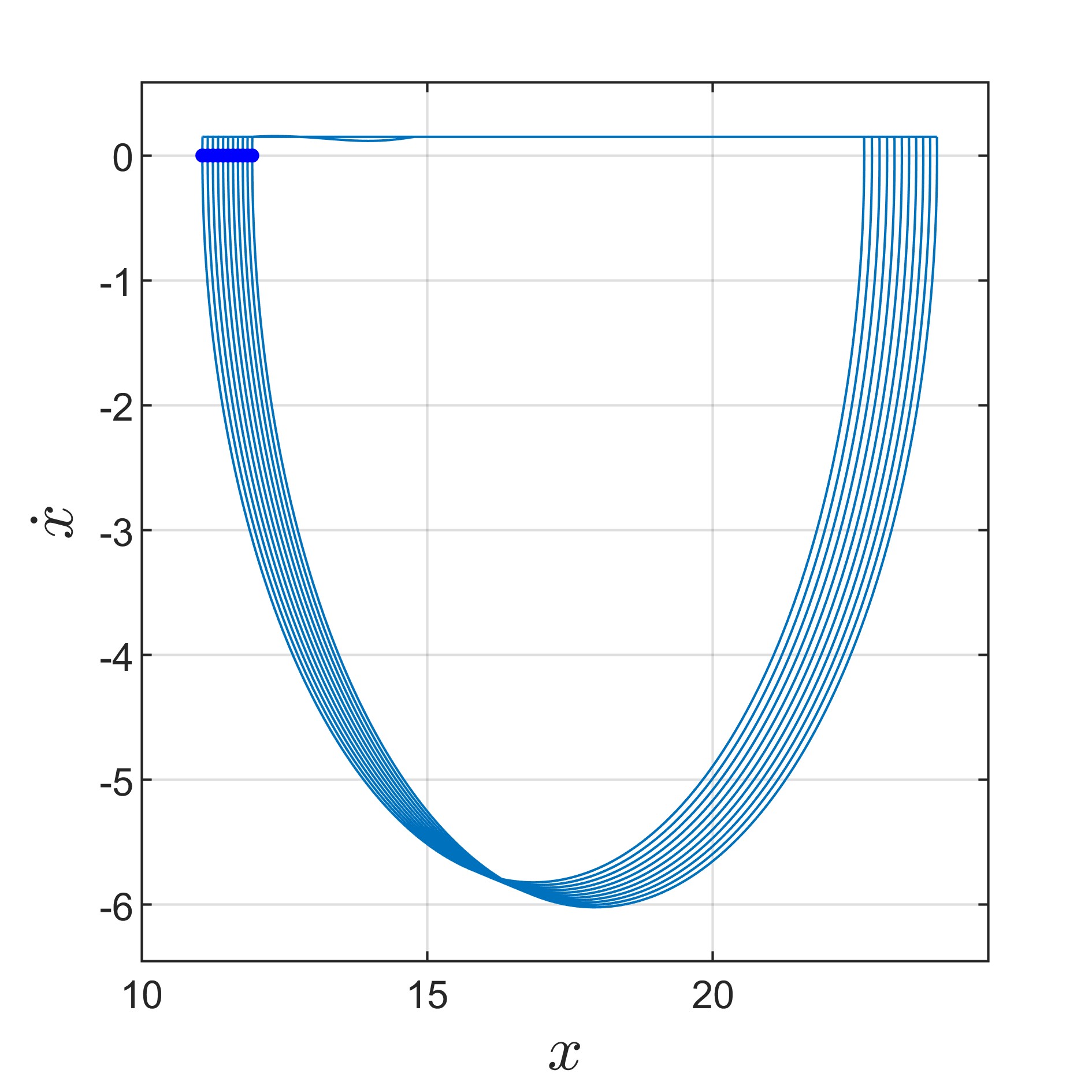}
        \caption{ }
        \label{fig3c}
    \end{subfigure}
           \begin{subfigure}{0.48\textwidth}
        \includegraphics[width=\linewidth]{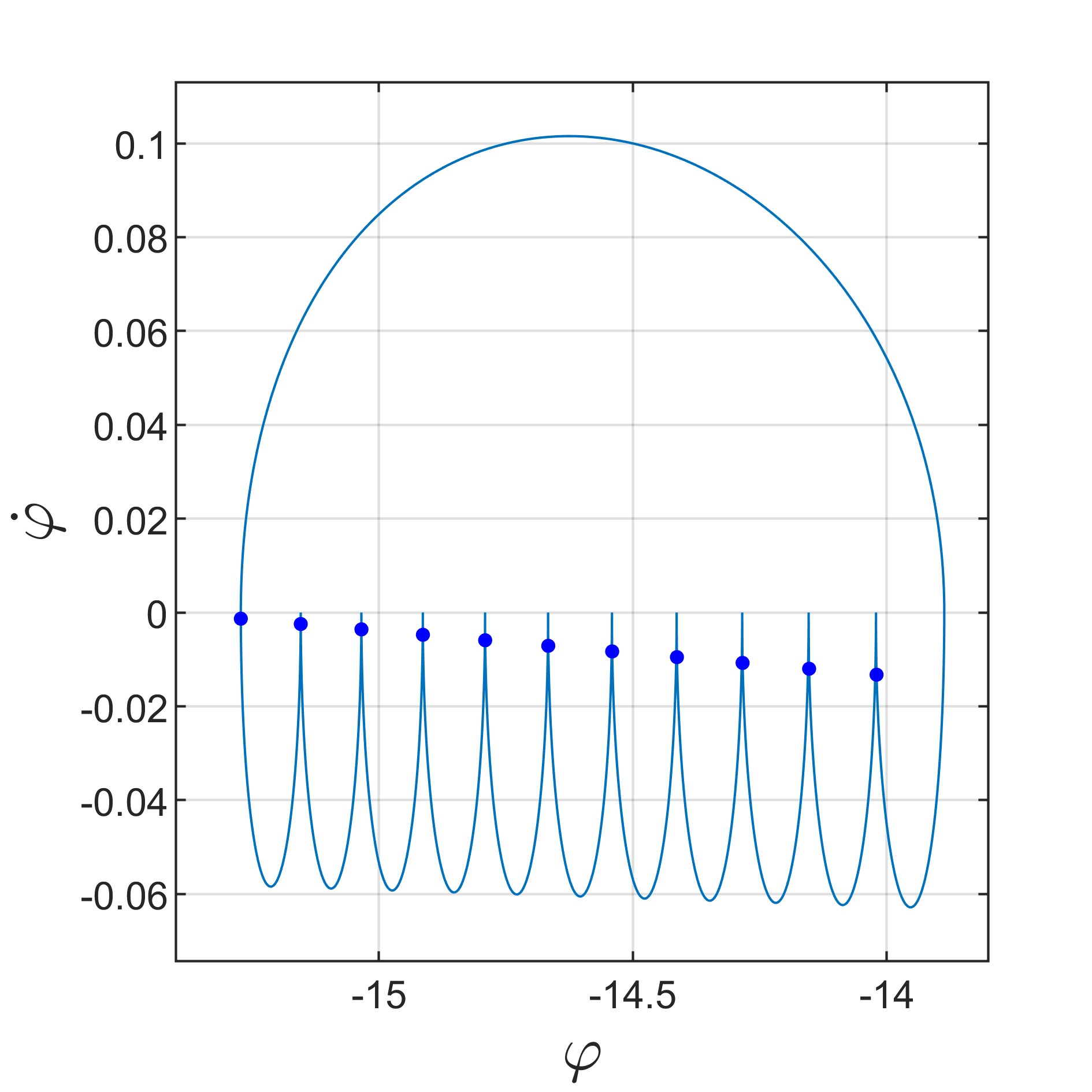}
        \caption{ }
        \label{fig3d}
    \end{subfigure}
\caption{
Time history (a), phase plots (b--d), and corresponding Poincaré points at $\eta = 2.7$;  
Integration is performed from the initial condition $(x_{0}, \dot{x}_{0}, \varphi_{0}, \dot{\varphi}_{0}) = (0,0,0,0)$; Other parameters follow Eq.~\ref{eq:Literature parameter values}.
}
    \label{fig3}
\end{figure}

\noindent
Figure~\ref{fig2} illustrates the bifurcation diagram of the single translational–rotational oscillator using these parameter values. Using Poincaré sampling at local minima of displacement $x$, the system response is mapped as the control parameter ($\eta$) varies.
When $\eta$ increases, the oscillator first displays simple periodic motion. As $\eta$ increases more, the oscillator moves through different stages of multi-periodic behavior. At first, there is simple periodic motion; then, after a sudden jump in amplitude, the system shifts to higher period regimes—starting with period-6, then period-7, and period-8 orbits, with the period continuing to rise up to period-28. As $\eta$ increases further, the period decreases sequentially, producing period-5, period-4, period-3, and period-2 solutions. Ultimately, for large $\eta$, the oscillator returns to single-periodic motion. 
Fig. \ref{fig3} presents representative time histories, phase portraits, and Poincaré maps for the system at $\eta = 2.7$, corresponding to one of the multi-periodic regimes identified in the previous bifurcation plot. The time response exhibits period-11 oscillations, characterized by a repeating pattern over 11 cycles. The phase portraits display intricate trajectories which do not close after a single loop, but instead require 11 cycles to return to their initial state. The Poincaré map further confirms this behavior by revealing exactly 11 distinct points, demonstrating the presence of a stable period-11 orbit. This example highlights the rich dynamical behavior that arises as variations in the friction ratio $\eta$ destabilize simple periodic solutions and induce higher-order periodic regimes.

\begin{figure}[H]
  \centering
  \includegraphics[scale=0.13]{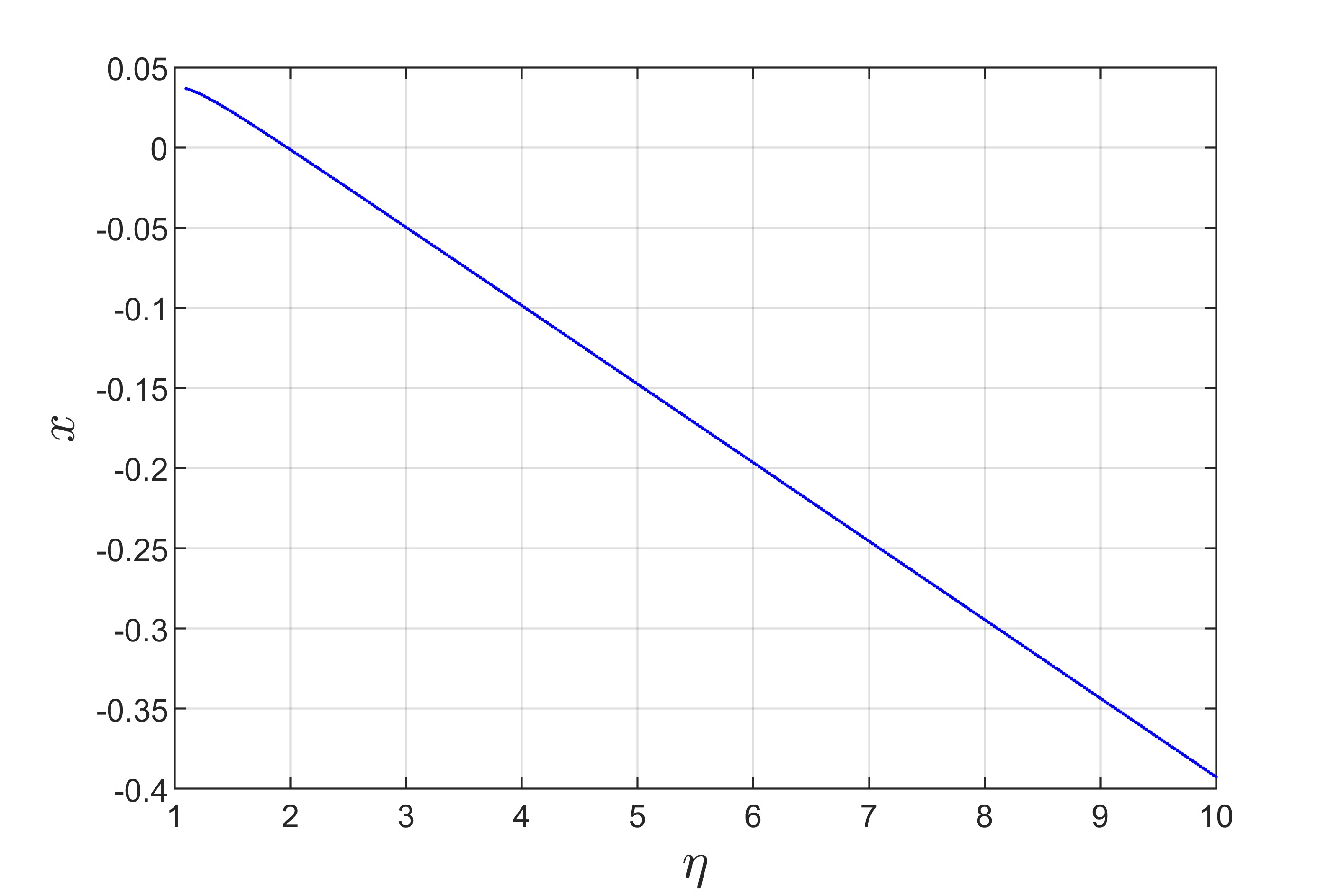} 
  \caption{
Bifurcation diagram of the Poincaré map defined by the local minima of $x$, as the friction coefficient ratio ($\eta$) is swept as the control parameter for the symmetric spring configuration ($\hat{k}_1 = \hat{k}_2 = 5000$ and thus $k_{12} = 0$); 
For each $\eta$, integration is restarted from $(x_0,\,\dot{x}_0,\,\varphi_0,\,\dot{\varphi}_0) = (0,\,0,\,0,\,0)$; Other parameters follow Eq.~\ref{eq:nondimensional parameter values}.
}
  \label{fig4}
\end{figure}

Parameter values for the following analysis are selected as presented below. Unless stated otherwise, the nondimensional parameters in Eq.~\ref{eq:nondimensional parameter values} --derived from the physical values in Eq.~\ref{eq:dimensional parameter values}-- are used for all calculations in this work. Any changes to these parameters for specific figures are stated in the corresponding captions.
\begin{align}\label{eq:dimensional parameter values}
\hat{m} &= 5 \ \text{(kg)}, \quad 
\hat{r} = 0.1 \ \text{(m)}, \quad 
\hat{B} = 0.1 \ \text{(kg} \cdot \text{m}^2), \quad 
\hat{k}_1 = 9000 \ \text{(N/m)},  \notag \\
\hat{k}_2 &= 1000 \ \text{(N/m)}, \quad 
\hat{c}_1 = \hat{c}_2 = 0 \ \text{(N}\cdot\text{s/m)}, \quad 
\hat{v}_{b0} = 0.05 \ \text{(m/s)}, \quad 
g = 9.81 \ \text{(m/s}^2)
\end{align}

\begin{align}\label{eq:nondimensional parameter values}
 m &= 2, \qquad k_{12} = -0.8, \qquad c = 0, \qquad c_{12} = 0, \qquad \bar{\mu} = 0.0491, \notag \\ 
\eta &= 1.5, \qquad T_b = 0.5, \qquad v_{b0} = 0.0112, \qquad \beta = 0.0014, \quad \bar{\varepsilon} = 10^{-5}
\end{align}
Fig.~\ref{fig4} shows the bifurcation diagram for a symmetric spring configuration, where both springs have equal stiffness. In this case, the system reduces to the classical model of a mass on a moving belt, and the rotational angle $\varphi$ remains zero in steady state. The bifurcation diagram displays a single line corresponding to simple periodic motion throughout the entire range of the control parameter, with no bifurcations present. As $\eta$ increases, only the minimum of $x$ decreases, while the system consistently maintains single-periodic behavior. The Poincaré map always reveals a single point, confirming the lack of complex dynamical transitions and reflecting the simplified nature of the symmetric system.

\begin{figure}[H]
    \centering
    \begin{subfigure}{0.48\textwidth}
        \includegraphics[width=\linewidth]{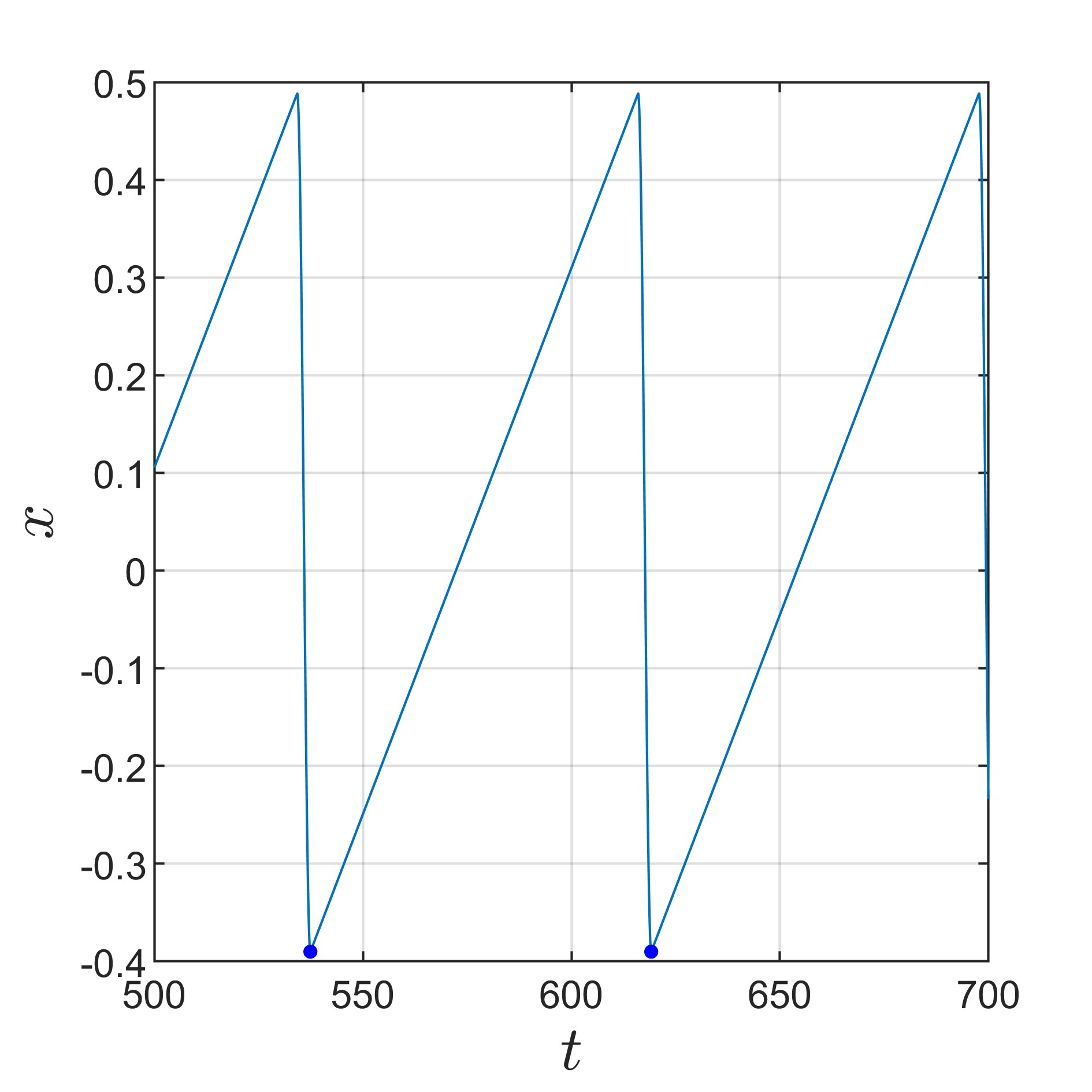}
        \caption{ }
        \label{fig5a}
    \end{subfigure}
    \begin{subfigure}{0.48\textwidth}
        \includegraphics[width=\linewidth]{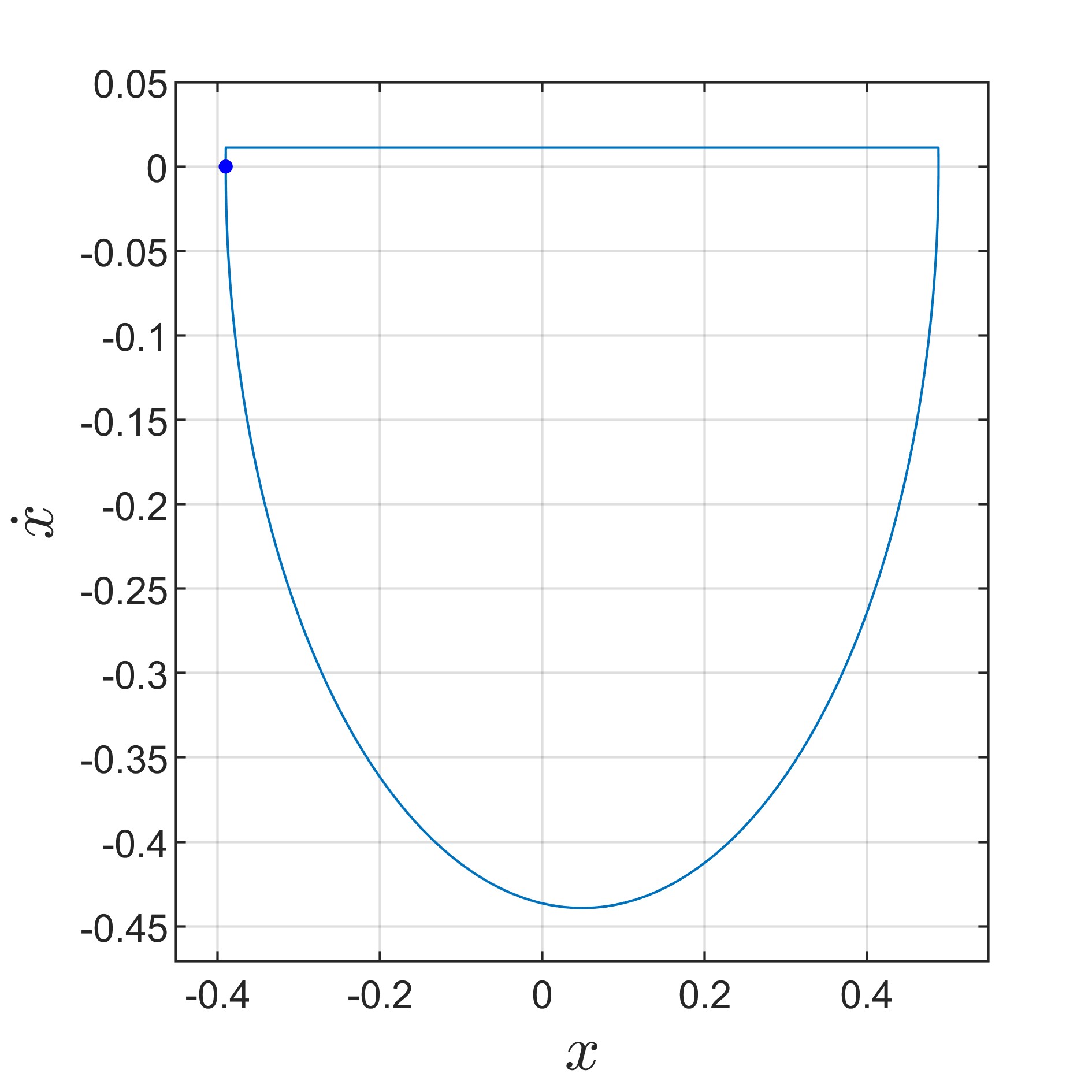}
        \caption{ }
        \label{fig5b}
    \end{subfigure}
\caption{
Time history (a) and phase plot (b) with the corresponding Poincaré point at $\eta = 9.95$, for the symmetric spring configuration ($\hat{k}_1 = \hat{k}_2 = 5000$ N/m, so $k_{12} = 0$); Integration is started from $(x_{0},\,\dot{x}_{0},\,\varphi_{0},\,\dot{\varphi}_{0}) = (0,\,0,\,0,\,0)$; Other parameters follow Eq.~\ref{eq:nondimensional parameter values}.
}
    \label{fig5}
\end{figure}

Fig.~\ref{fig5} presents the time history (Fig.~\ref{fig5a}), phase plot (Fig.~\ref{fig5b}), and the corresponding Poincaré point for the symmetric spring case at $\eta = 9.95$. The displacement response is regular and purely periodic, with the phase trajectory forming a closed loop and the Poincaré section consisting of a single point. These results confirm that in the absence of stiffness asymmetry, and for high friction ratios, the oscillator reduces to a simple two-dimensional stick–slip system, exhibiting no rotational motion and no irregular behavior.

\begin{figure}[H]
  \centering
  \includegraphics[scale=0.13]{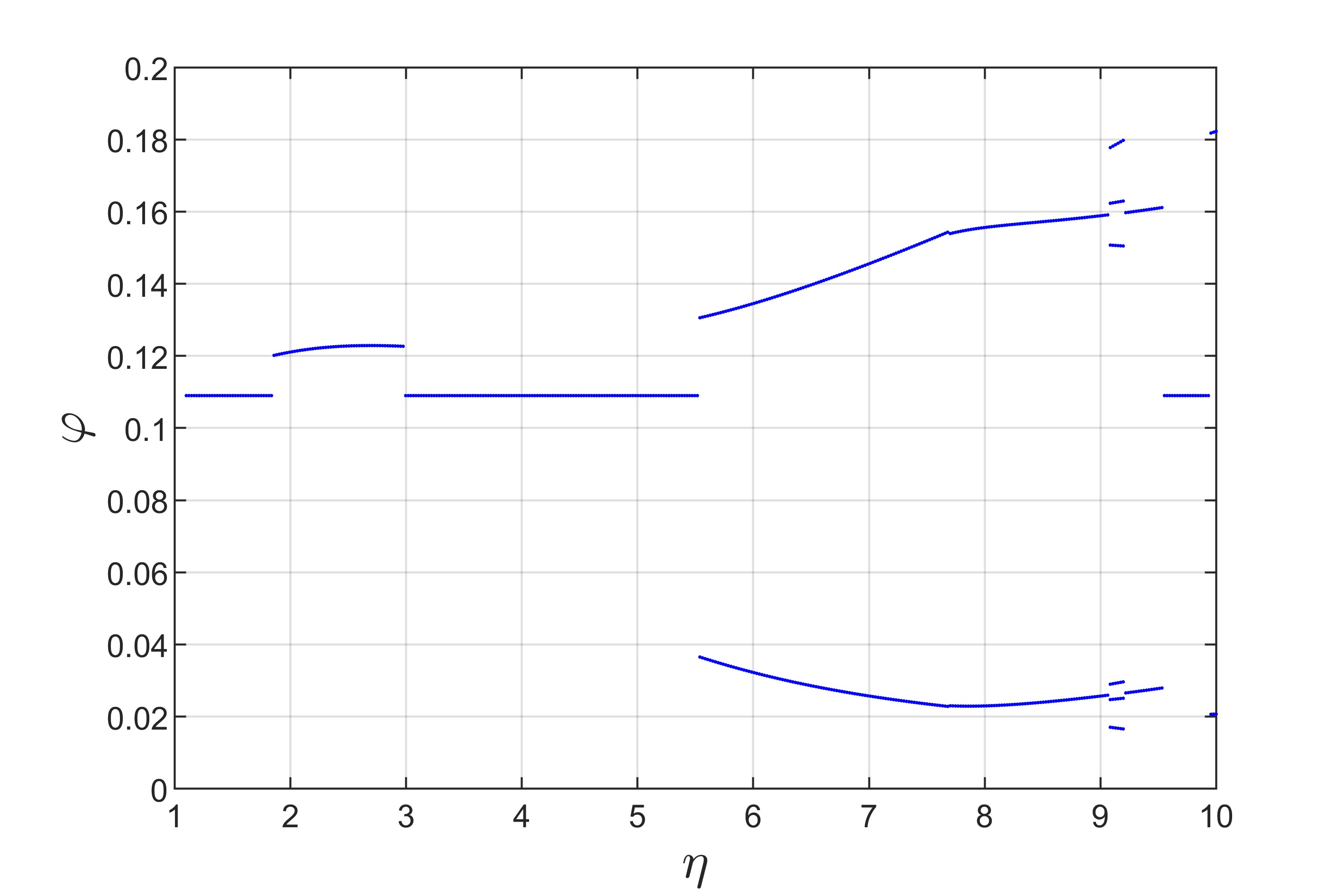} 
\caption{
Bifurcation diagram of the Poincaré map defined by the local minima of (a) $x_1$ and (b) $\varphi_1$, as the friction coefficient ratio ($\eta$) is swept as the control parameter; For each $\eta$, integration is restarted from $(x_{10},\,\dot{x}_{10},\,\varphi_{10},\,\dot{\varphi}_{10}) = (0,\,0,\,0,\,0)$; Parameters follow Eq.~\ref{eq:nondim_parameters}.
}
  \label{fig6}
\end{figure}

Fig.~\ref{fig6} shows the bifurcation diagram for the asymmetric spring configuration, where $k_{12} = -0.8$. With unequal stiffnesses, the rotational coordinate $\varphi$ becomes active and influences the dynamics. As the friction coefficient ratio $\eta$ increases, the system initially displays simple periodic motion. At $\eta \approx 1.85$, the amplitude of the periodic response abruptly increases, followed by a rapid return to its previous value with a second jump. For $\eta \approx 5.5$, a bifurcation occurs and the system enters a period-2 regime, where the steady-state response alternates between two values. After a brief interval with period-6 motion, the system jumps back to simple periodic oscillations near $\eta = 9.5$. For a short range, the simple periodic regime persists before another transition around $\eta = 9.99$, where period-2 reappears. This sequence highlights how introducing stiffness asymmetry activates rotational-translational coupling and leads to richer dynamical phenomena, including abrupt amplitude jumps and period-2 transitions.

\begin{figure}[H]
    \centering
    \begin{subfigure}{0.48\textwidth}
        \includegraphics[width=\linewidth]{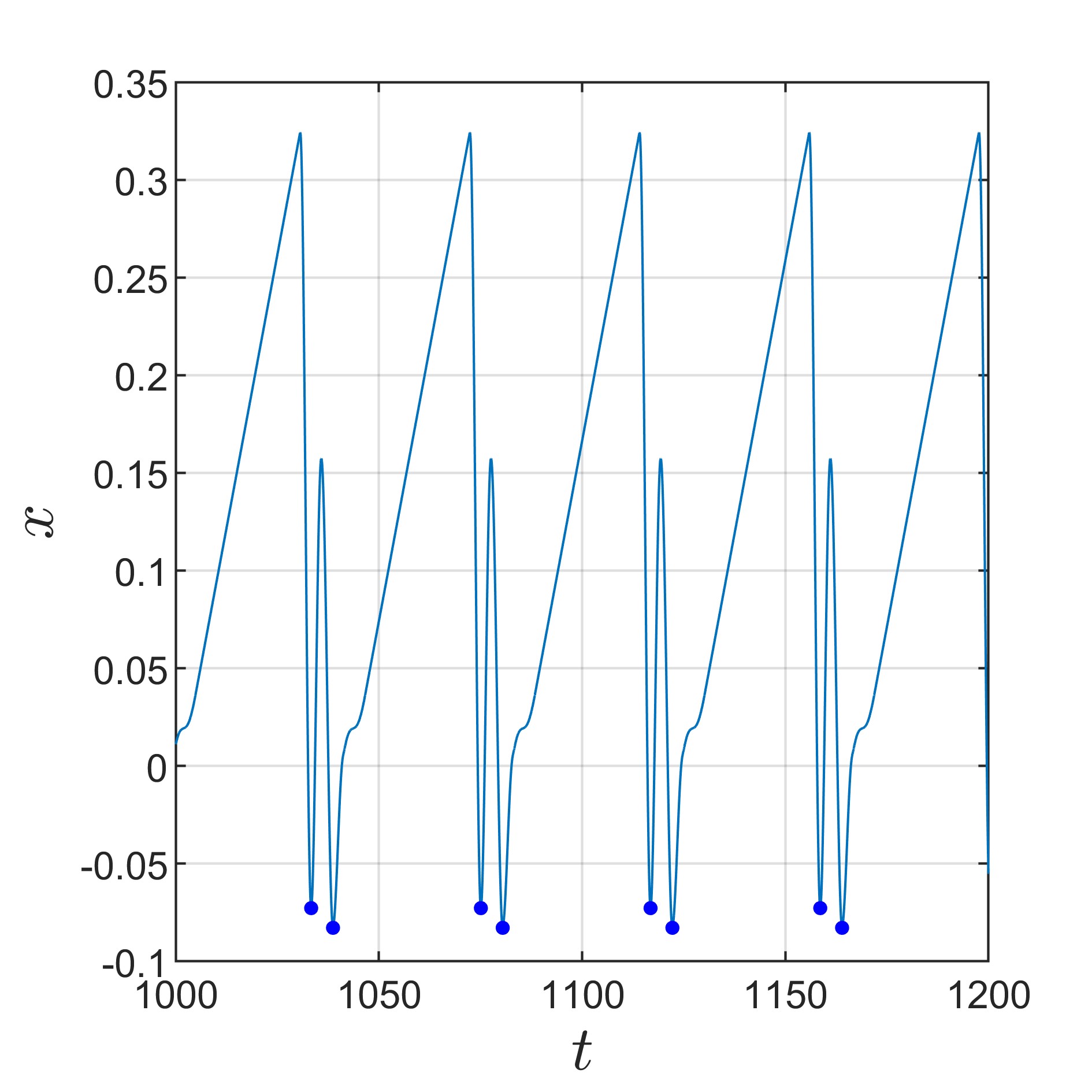}
        \caption{ }
        \label{fig7a}
    \end{subfigure}
    \begin{subfigure}{0.48\textwidth}
        \includegraphics[width=\linewidth]{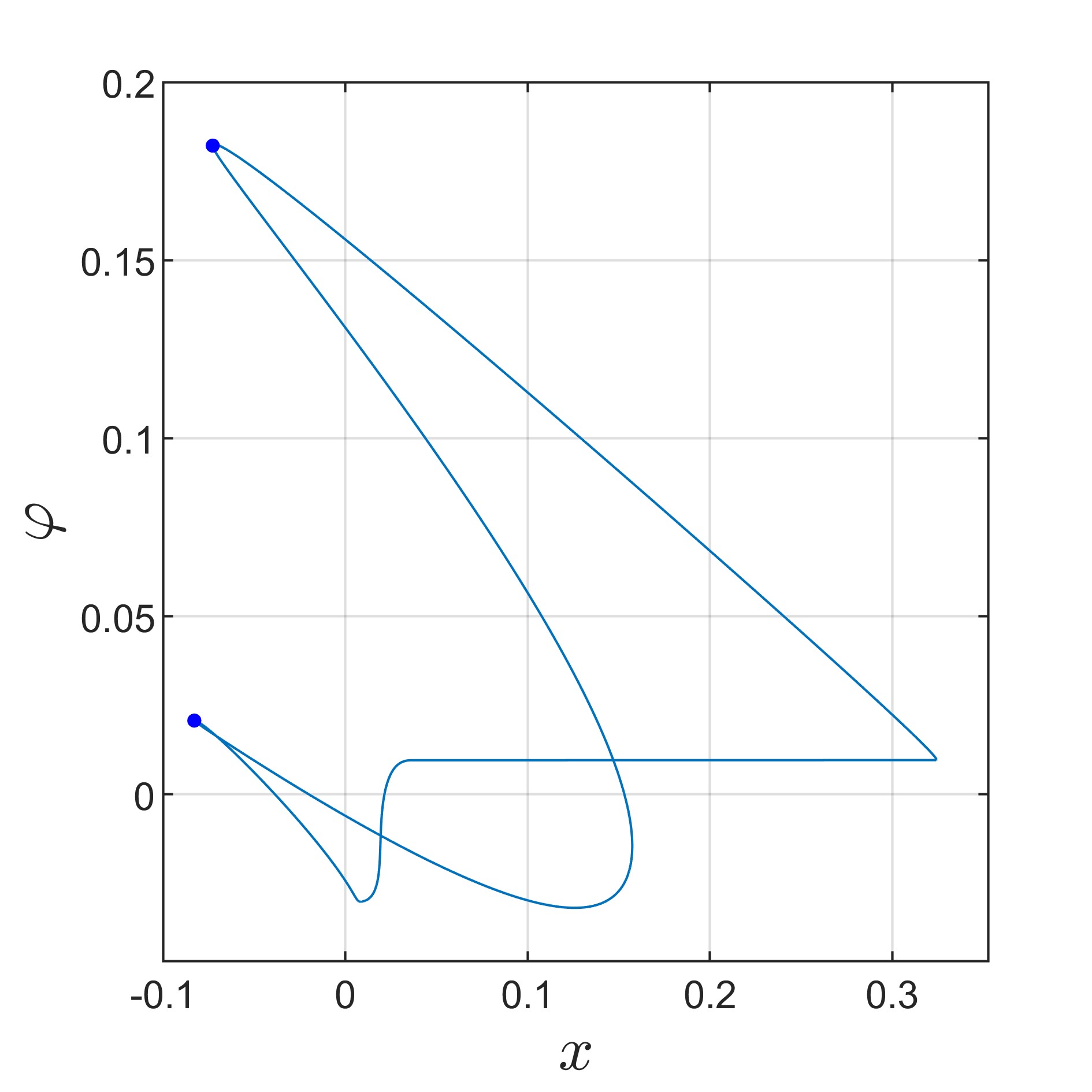}
        \caption{ }
        \label{fig7b}
    \end{subfigure}
        \begin{subfigure}{0.48\textwidth}
        \includegraphics[width=\linewidth]{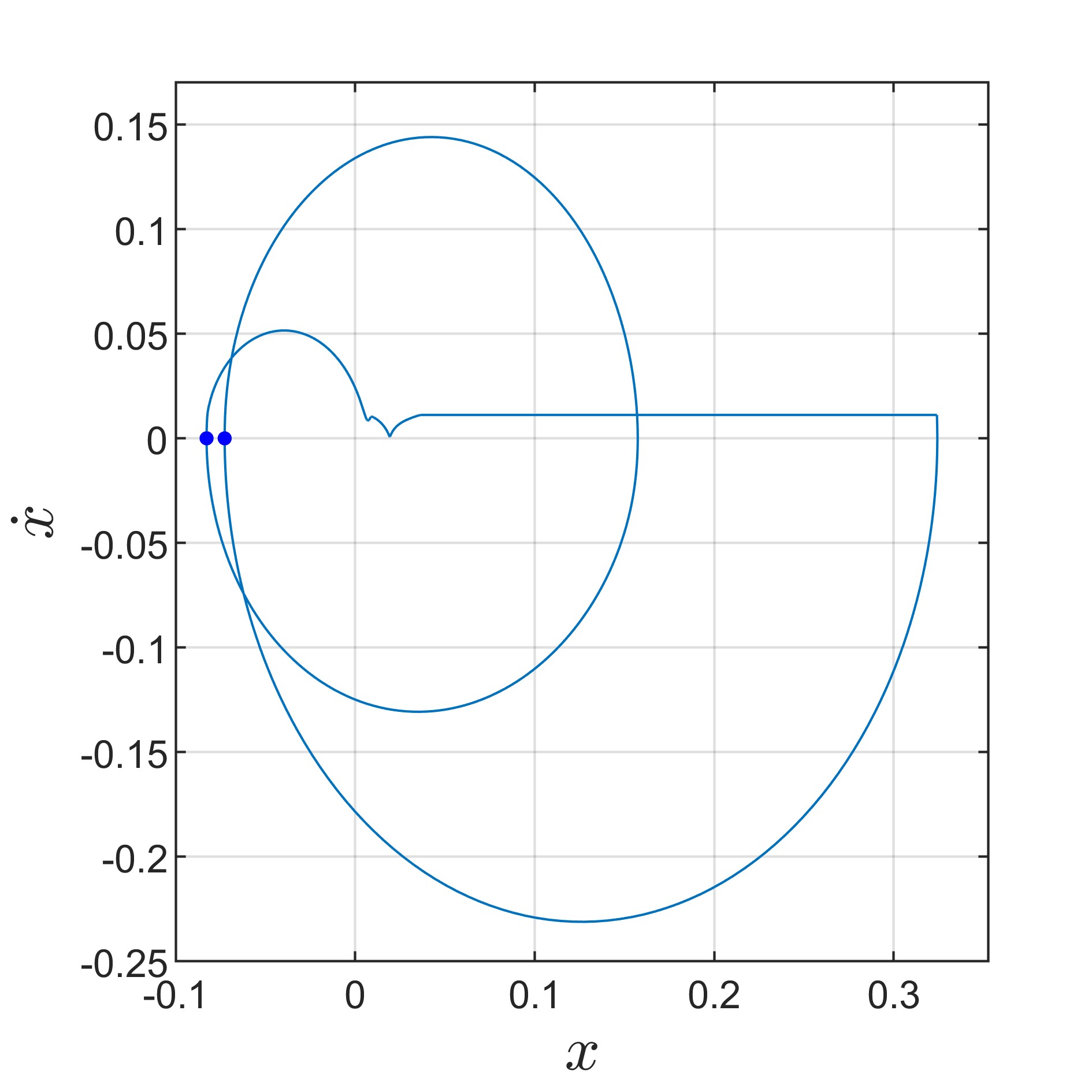}
        \caption{ }
        \label{fig7c}
    \end{subfigure}
           \begin{subfigure}{0.48\textwidth}
        \includegraphics[width=\linewidth]{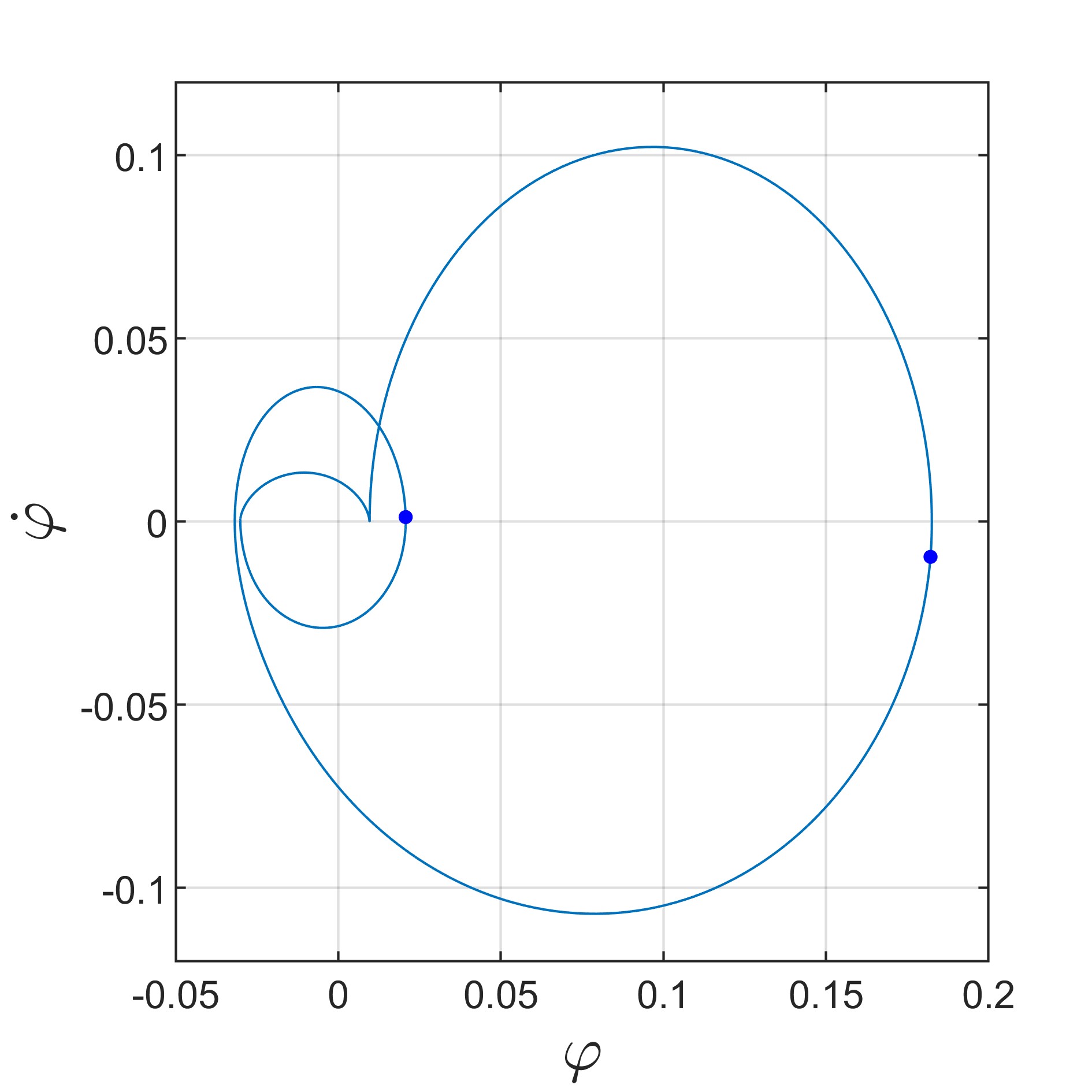}
        \caption{ }
        \label{fig7d}
    \end{subfigure}
\caption{
Time history (a) and phase plots (b--d), along with the corresponding Poincaré points (defined as local minima of $x$), at $\eta = 9.99$ for the asymmetric spring configuration; Integration is performed from the initial condition $(x_{0},\,\dot{x}_{0},\,\varphi_{0},\,\dot{\varphi}_{0}) = (0,\,0,\,0,\,0)$; Other parameters follow Eqs.~\ref{eq:nondimensional parameter values}.
}
\label{fig7}
\end{figure}

\noindent Accurately connects the bifurcation sequence found in Fig.~\ref{fig6} to the dynamical structure of Fig.~\ref{fig7} shows the time history and phase space for $\eta = 9.99$ in the asymmetric spring configuration. The displacement and rotational responses demonstrate period-doubled oscillations, with the time history showing alternating amplitudes and the phase trajectory forming a closed loop after two cycles. The Poincaré map presents two distinct points, confirming period doubling as the prevailing steady-state. This illustrates how, for high friction ratios, the interaction of stiffness asymmetry and dry friction leads to period-doubled oscillations rather than simple periodic motion.

\section{Chain of oscillators driven by a DC motor}\label{sec:chain of oscillators driven by a DC motor}
This section extends the model to a chain of N oscillators powered by a single DC motor, where the belt and oscillator dynamics affect each other directly. The formulation starts with the DC motor model, then develops the equations that link the motor, belt, and oscillator chain.

\subsection{DC motor model}\label{subsec:DC motor model}
The dynamics of the armature-controlled DC motor are governed by the following circuit equation:
\begin{equation}\label{eq:dcmotor_voltage}
u = L\dot{i} + Ri + K_{E}\hat{\omega}_{0},
\end{equation}
where $u$ is the applied voltage, $i$ is the armature current, and $\hat{\omega}_{0}$ denotes the angular velocity of the motor shaft prior to gear transmission. The parameters $L$ and $R$ represent the motor's inductance and resistance, respectively, while $K_E$ is the back-electromotive force (back-EMF) constant, characterizing the proportionality between the motor shaft speed and the internally induced voltage that opposes the input voltage.

The torque produced by the DC motor (before accounting for the gear transmission) is given by
\begin{equation}\label{eq:dcmotor_torque}
\hat{M}_0 = K_T i,
\end{equation}
where $K_T$ is the torque constant.

The relationship between the motor and the mechanical load is defined by the gear transmission ratio:
\begin{equation}\label{eq:gear_ratio}
i_{g} = \frac{\hat{\omega}_{0}}{\hat{\omega}_{b}} = \frac{\hat{M}}{\hat{M}_{0}},
\end{equation}
where $\hat{\omega}_{b}$ is the angular velocity at the output shaft of the transmission, and $\hat{M}$ is the gear output torque.

Combining the above relationships---Eq.~\eqref{eq:dcmotor_voltage} for the electrical subsystem and Eq.~\eqref{eq:gear_ratio} for the transmission---yields the following form of the circuit equation expressed in terms of the mechanical output:
\begin{equation}\label{eq:dcmotor_output}
u = L \dot{i} + R\frac{\hat{M}}{i_{g} K_{T}} + K_{E} i_{g} \hat{\omega}_{b}.
\end{equation}

\noindent By substituting for the output torque $\hat{M}$ from Eq.~\eqref{eq:dcmotor_output}, we obtain:
\begin{equation}\label{eq:torque_output_general}
\hat{M} = \frac{i_{g} K_{T}}{R} \left(u - K_{E} i_{g} \hat{\omega}_{b} - L\dot{i}\right).
\end{equation}

\noindent When inductance is neglected, i.e. $L=0$, Eq.~\eqref{eq:torque_output_general} simplifies to:
\begin{equation}\label{eq:torque_output_simple}
\hat{M} = \frac{K_{T}}{R} i_{g} u - \frac{K_{E} K_{T}}{R} i_{g}^{2} \hat{\omega}_{b}.
\end{equation}
This compact torque expression enables a direct formulation of the chain dynamics driven by the DC motor.
\subsection{Model of chain oscillator driven by DC motor}

\begin{figure}[H]
  \centering
  \includegraphics[scale=0.48]{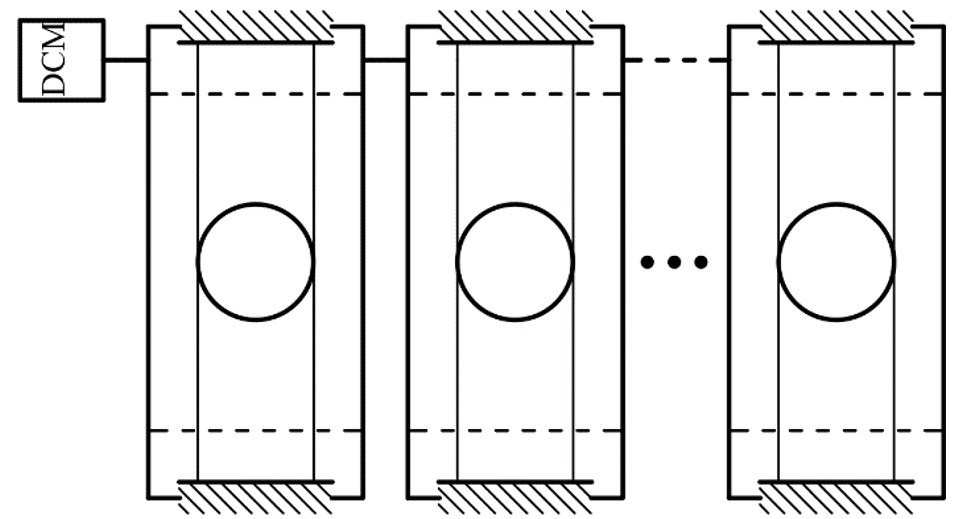} 
  \caption{The system of coupled oscillators driven by a single DC motor (DCM).}
  \label{fig:dc_motor_system}
\end{figure}

A schematic of the coupled oscillator chain driven by a DC motor is shown in Fig.~\ref{fig:dc_motor_system}.

The dimensional equation of motion for the masses coupled to the DC motor, reduced to the The dimensional equation of motion describing the dynamics of all masses connected to the DC motor—represented in terms of the angular velocity of the gearbox output shaft—can be written as:

\begin{equation}\label{eq:rotor_eom_dimensional}
\hat{I}\,{\hat{\omega}}^\prime_{b} + \hat{c}\,\hat{\omega}_{b} 
= \frac{K_{T}}{R} i_{g} u - \frac{K_{E}K_{T}}{R} i_{g}^{2} \hat{\omega}_{b} 
+ \hat{r}_{b}\mu \hat{N} \sum_{i=1}^{N} T_{s}\!\left(\frac{{\hat{x}}^\prime_{i} - \hat{v}_{b}}{\hat{r}}, {\varphi}^\prime_{i}, \hat{\varepsilon}\right),
\end{equation}
where $\hat{I}$ is the equivalent mass moment of inertia, $\hat{c}$ the equivalent damping coefficient, and $N$ the number of oscillators. Taking into account that $\hat{\omega}_{b}\hat{r}_{b} = \hat{v}_{b}$, where $\hat{r}_{b}$ is the radius of the belt pulley, the equation can be rewritten as:

To nondimensionalize the equation of motion, introduce the following dimensionless quantities:
\begin{equation}\label{eq:dimensional_scaling}
    \dot{v}_{b} = \frac{{\hat{v}^\prime}_{b}}{\alpha^{2}\hat{r}},\quad 
    v_{b} = \frac{\hat{v}_{b}}{\alpha \hat{r}},
\end{equation}
which yield the following nondimensionalized equation:
\begin{equation}\label{eq:eom_vb_nondim}
T_{b}\,\dot{v}_{b} + v_{b} 
= v_{b0} + \frac{\beta}{N} \sum_{i=1}^{N} 
T_{s}\!\left(\dot{x}_{i} - v_{b}, \dot{\varphi}_{i}, \bar{\varepsilon}\right),
\end{equation}
where $T_b$ and $v_{b0}$ are the motor time constant and applied belt velocity, respectively. The nondimensional parameters $T_b$, $v_{b0}$, and $\beta$ are defined as:
\begin{equation}\label{eq:nondim_parameters}
\begin{cases}
T_{b} = \dfrac{\alpha R \hat{I}}{R \hat{c} + K_{E}K_{T} i_{g}^{2}}, \\[1.5ex]
v_{b0} = \dfrac{\hat{r}_{b} K_{T} i_{g}}{\alpha \hat{r} \left(R \hat{c} + K_{E} K_{T} i_{g}^{2}\right)} u,\footnote{Here, $u$ is assumed to be a constant voltage.} \\[1.5ex]
\beta = \dfrac{N \hat{r}_{b} K_{T} i_{g}}{\alpha \hat{r} \left(R \hat{c} + K_{E} K_{T} i_{g}^{2}\right)}.
\end{cases}
\end{equation}

\noindent Building upon the single-oscillator formulation in Eq.~\eqref{eq:dimensionless_single_oscillator}, the governing equations can be extended to the following dimensionless form for a system of $N$ coupled oscillators driven by a DC motor:
\begin{equation}\label{eq:dimensionless_oscillators}
\begin{split}
&\begin{bmatrix}
1 & 0 \\
0 & m
\end{bmatrix}
\begin{Bmatrix}
\ddot{x}_{i} \\ \ddot{\varphi}_{i}
\end{Bmatrix}
+
\begin{bmatrix}
c & c_{12} \\
c_{12} & c
\end{bmatrix}
\begin{Bmatrix}
\dot{x}_{i} \\ \dot{\varphi}_{i}
\end{Bmatrix}
+
\begin{bmatrix}
1 & k_{12} \\
k_{12} & 1
\end{bmatrix}
\begin{Bmatrix}
x_{i} \\ \varphi_{i}
\end{Bmatrix} \\
& \qquad +
\bar{\mu}
\begin{Bmatrix}
T_{s}( \dot{x}_{i} - v_{b}, \dot{\varphi}_{i}, \bar{\varepsilon}) \\
M_{s}( \dot{x}_{i} - v_{b}, \dot{\varphi}_{i}, \bar{\varepsilon})
\end{Bmatrix}
=
\begin{Bmatrix}
0 \\ 0
\end{Bmatrix}, 
\qquad i = 1,2,\ldots,N.
\end{split}
\end{equation}

\noindent where the velocity, governed by the collective friction forces, evolves according to
\begin{equation}\label{eq:dimensionless_belt}
T_{b}\,\dot{v}_{b} + v_{b} 
= v_{b0} + \frac{\beta}{N} \sum_{i=1}^{N} 
T_{s}( \dot{x}_{i} - v_{b}, \dot{\varphi}_{i}, \bar{\varepsilon} ).
\end{equation}

\section{Numerical results}\label{sec:Numerical results}

\begin{figure}[H]
    \centering
    \begin{subfigure}{0.49\textwidth}
        \includegraphics[width=\linewidth]{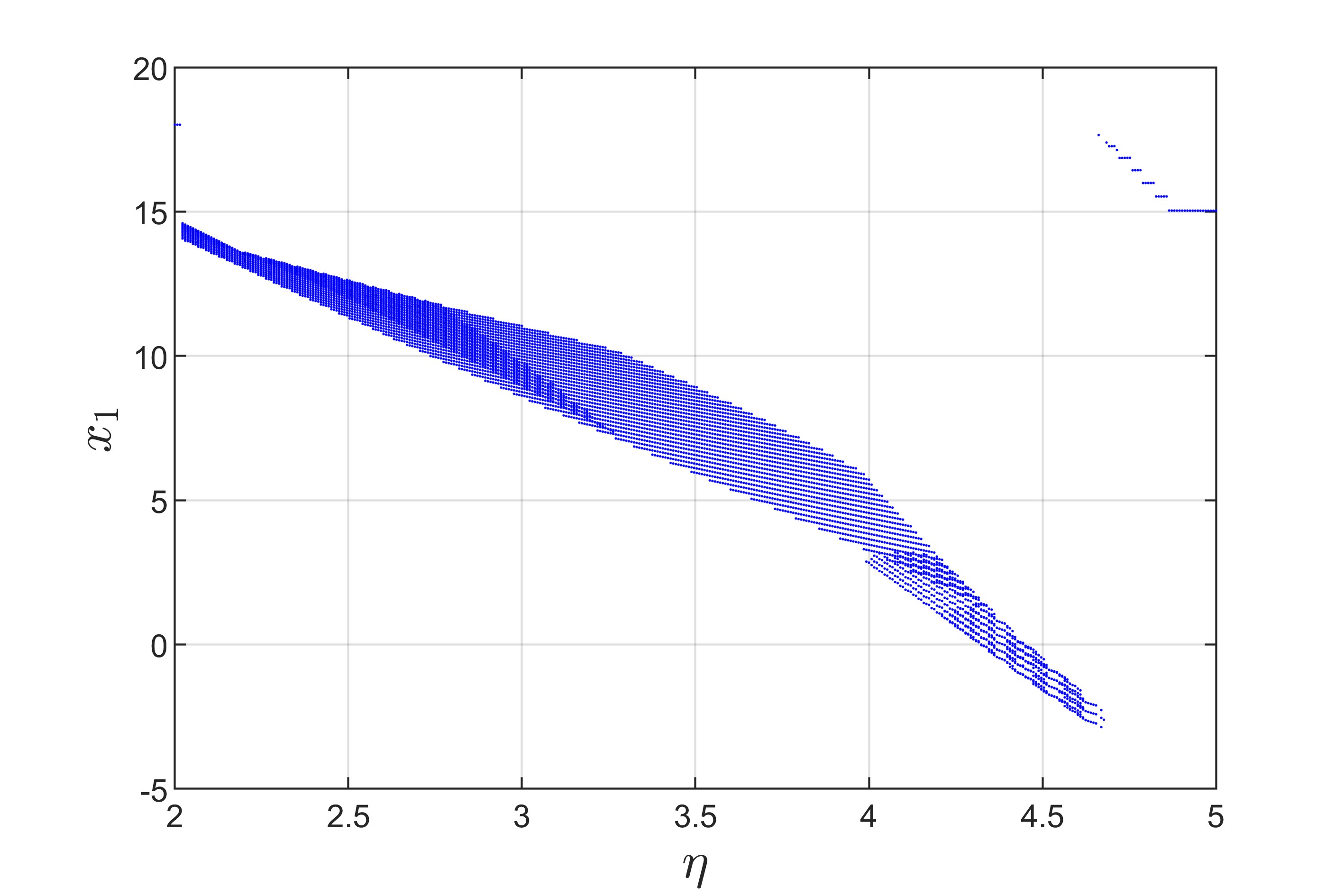}
        \caption{}
        \label{fig9a}
    \end{subfigure}
    \begin{subfigure}{0.49\textwidth}
        \includegraphics[width=\linewidth]{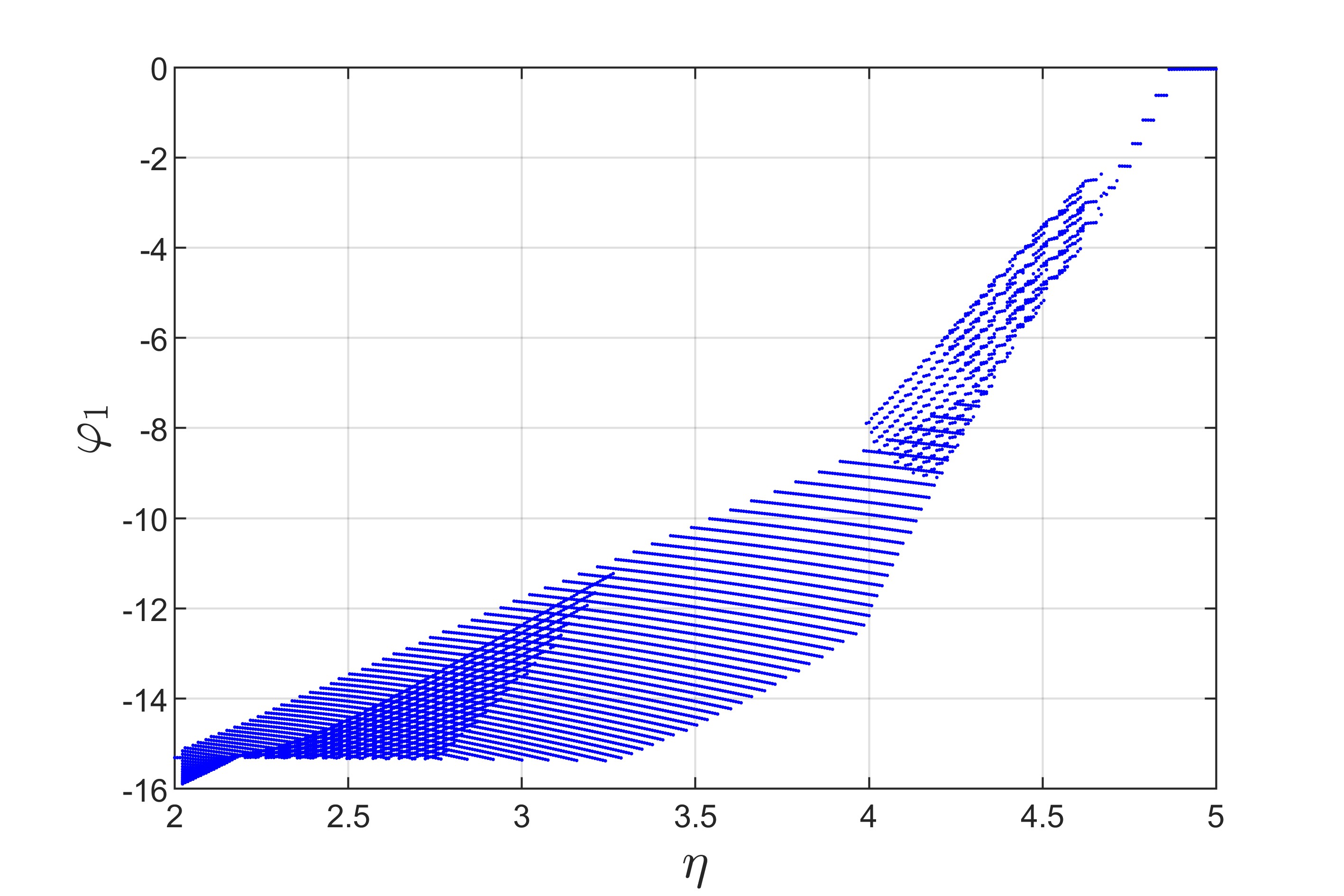}
        \caption{}
        \label{fig9b}
    \end{subfigure}
\caption{
Bifurcation diagrams of the Poincaré map defined by the local minima of (a) $x_1$ and (b) $\varphi_1$, as the friction coefficient ratio ($\eta$) is swept as the control parameter; For each $\eta$, integration is restarted from $(x_{10},\,\dot{x}_{10},\,\varphi_{10},\,\dot{\varphi}_{10}) = (0,\,0,\,0,\,0)$; Parameters follow Eq.~\ref{eq:Literature parameter values}.
}
    \label{fig9}
\end{figure}

As we have established the equations for coupled oscillators driven by a DC motor, this section investigates the three cases outlined in Section~\ref{subsec:Equations of motion} under the numerical examples.\footnote{In all numerical simulations, the initial value of the belt velocity is taken as $v_{b}(0) = 0$.} We begin by considering the effect of introducing the DC motor into a system with a single oscillator (see Fig. \ref{fig2}). Fig.~\ref{fig9} shows the bifurcation diagram of this system. The main pattern—marked by multi-periodic behavior for different values of $\eta$—looks similar to what is seen in the constant belt velocity case. However, adding the DC motor leads to clear differences in the details. First, the threshold value of $\eta$ where the amplitude jump and multi-periodicity start is much lower than in the constant velocity setup. Second, the sequence where the system adds more periods begins sooner, reaches a maximum at period-25, and then periods are removed at lower $\eta$ values than for the ideal motor case. Third, the system undergoes a sharp jump to a high-amplitude, single-period pattern, which stays stable for higher $\eta$, although there are still small amplitude changes. These distinctions highlight how the limited power from the DC motor affects the bifurcation structure and stability of the oscillator.

\begin{figure}[H]
    \centering
    \begin{subfigure}{0.48\textwidth}
        \includegraphics[width=\linewidth]{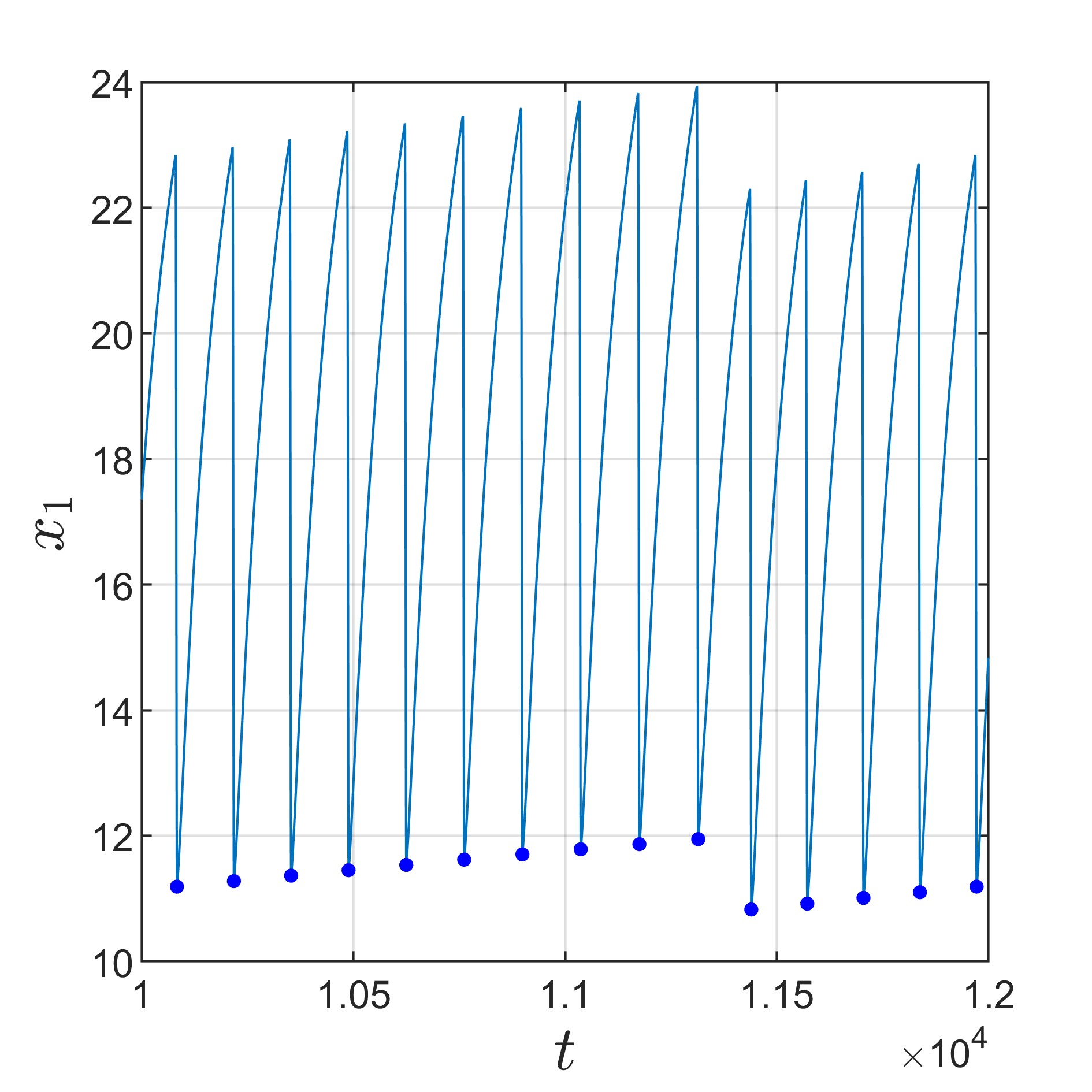}
        \caption{ }
        \label{fig10a}
    \end{subfigure}
    \begin{subfigure}{0.48\textwidth}
        \includegraphics[width=\linewidth]{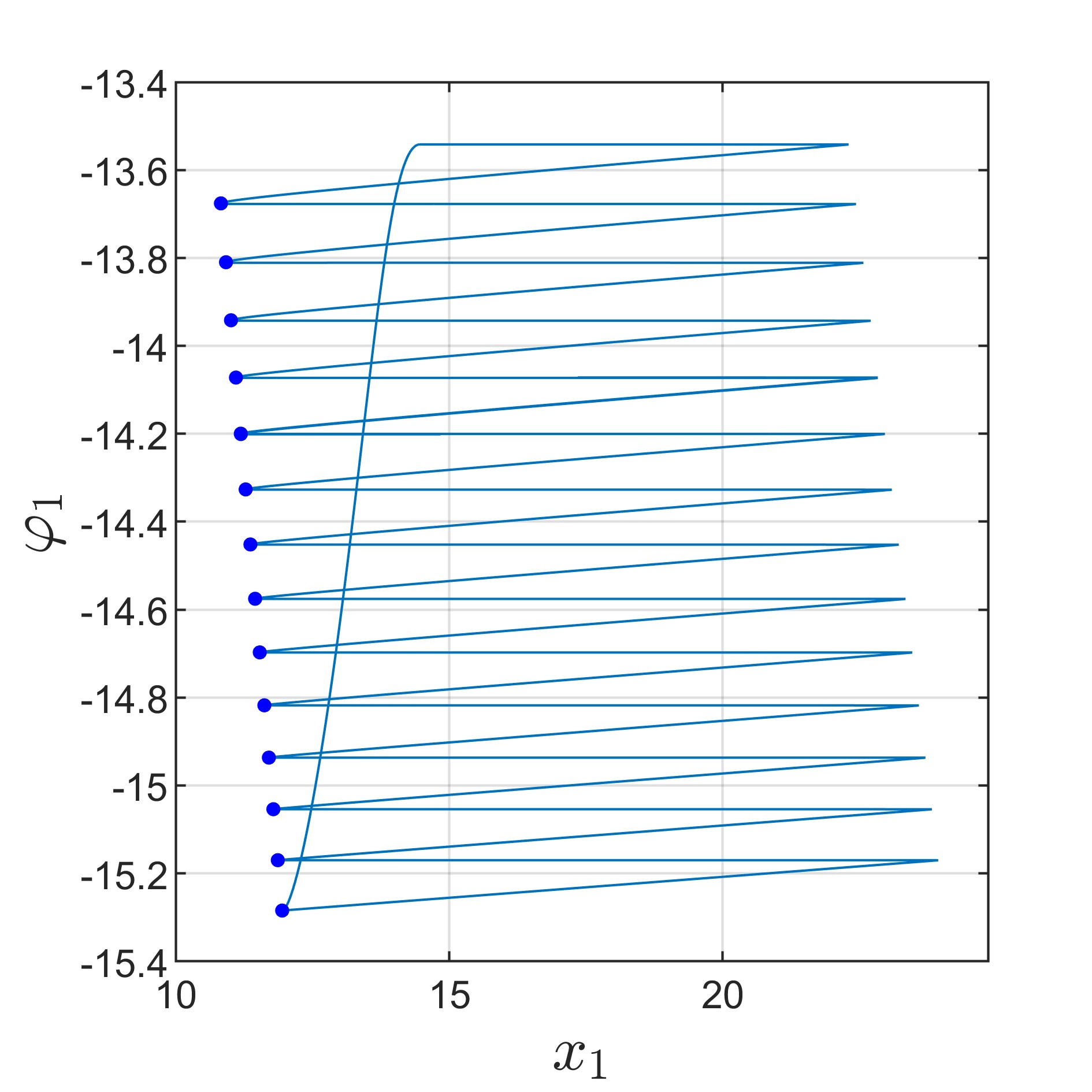}
        \caption{ }
        \label{fig10b}
    \end{subfigure}
        \begin{subfigure}{0.48\textwidth}
        \includegraphics[width=\linewidth]{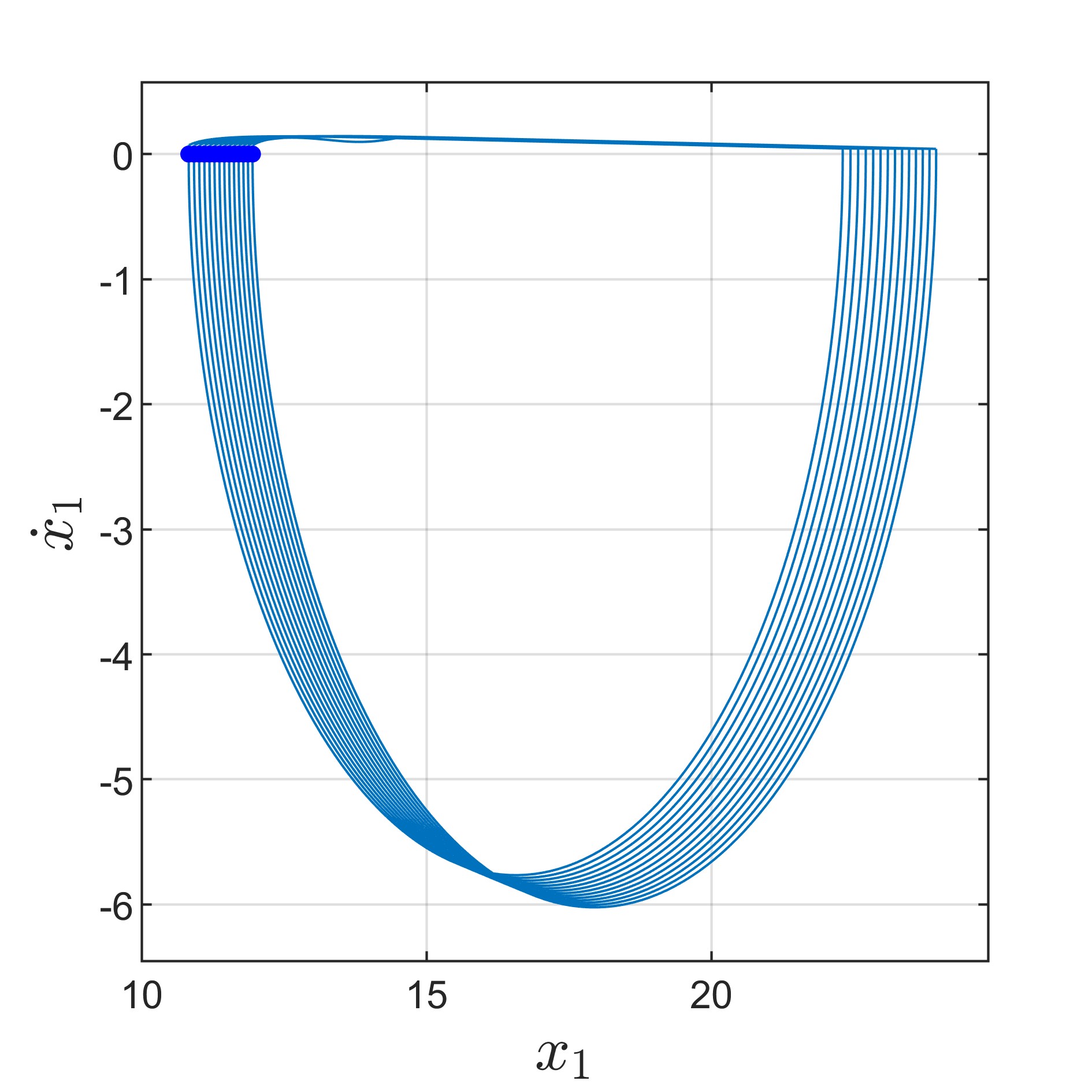}
        \caption{ }
        \label{fig10c}
    \end{subfigure}
           \begin{subfigure}{0.48\textwidth}
        \includegraphics[width=\linewidth]{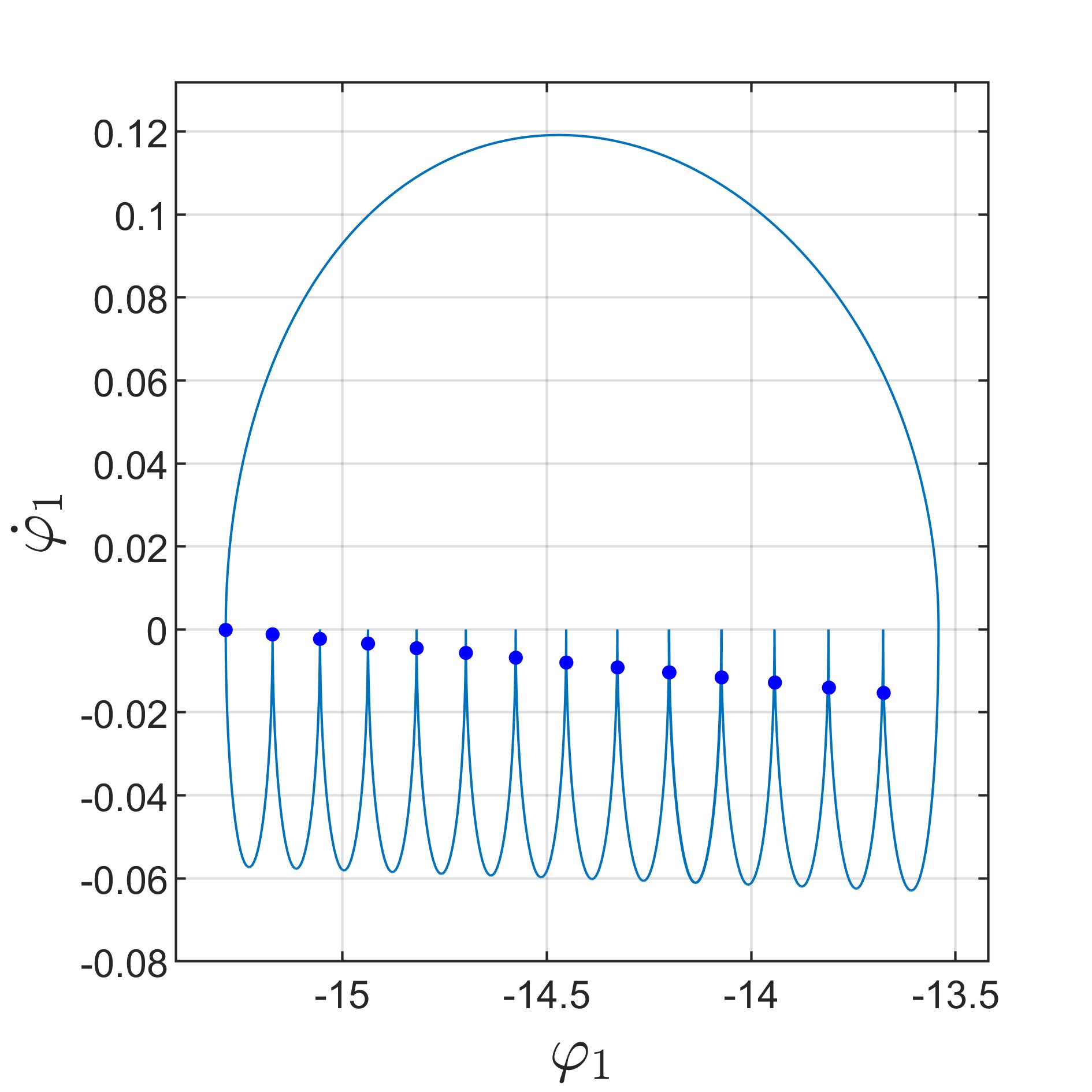}
        \caption{ }
        \label{fig10d}
    \end{subfigure}
\caption{
Time history (a), phase plots (b--d), and corresponding Poincaré points at $\eta = 2.7$;  
Integration starts from $(x_{10}, \dot{x}_{10}, \varphi_{10}, \dot{\varphi}_{10}) = (0,0,0,0)$; Other parameters follow Eq.~\ref{eq:nondimensional parameter values}.
}
    \label{fig10}
\end{figure}

Fig.~\ref{fig10} shows the time history and phase portrait of the system. These results are topologically similar to those observed under constant belt velocity (see Fig.~\ref{fig3}). The time response and phase plot have almost the same shapes and patterns as in the ideal excitation case, which means the main structure of the system's motion stays about the same after adding DC motor drive. The only difference is the period of oscillation.

\begin{figure}[H]
    \centering
    \begin{subfigure}{0.49\textwidth}
        \includegraphics[width=\linewidth]{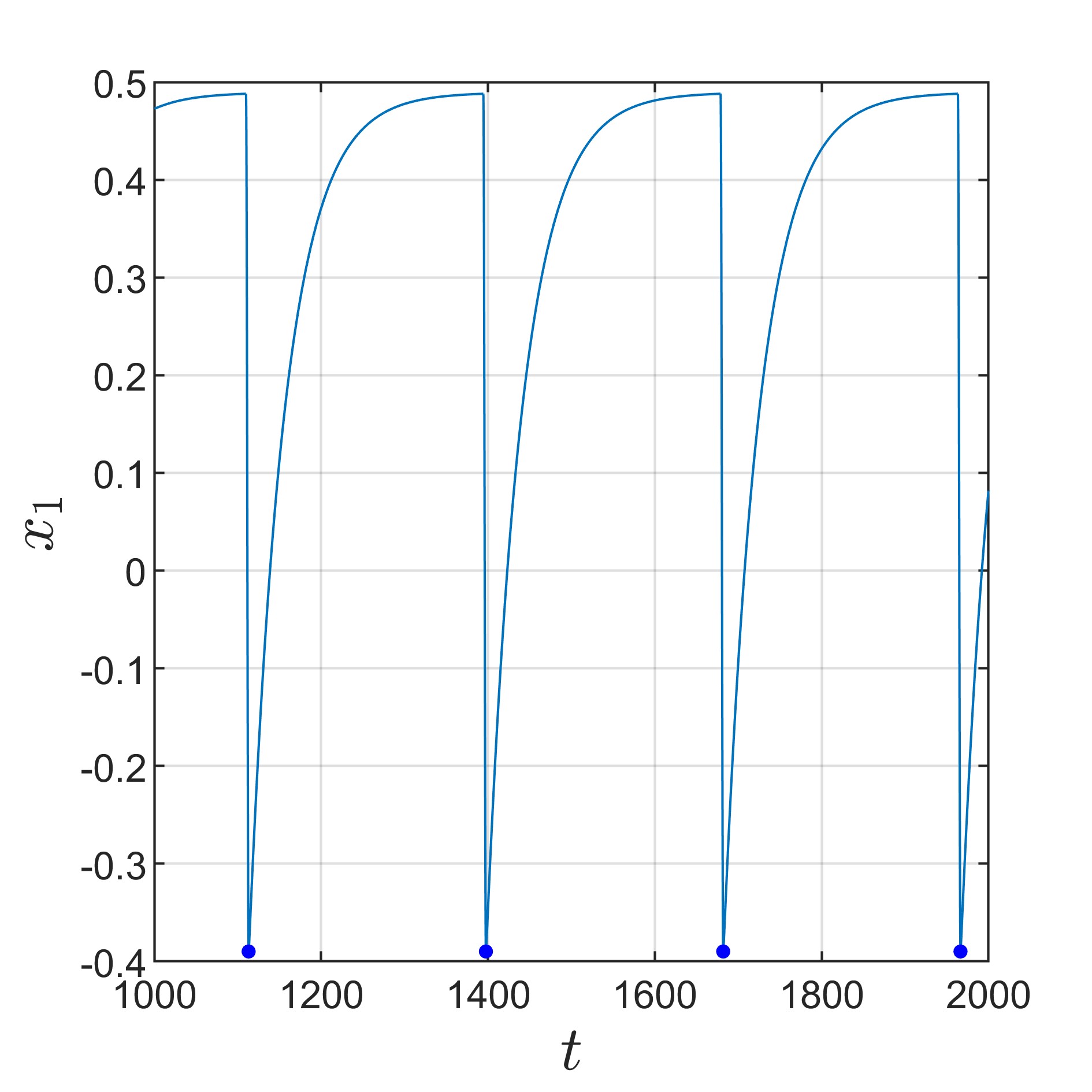}
        \caption{}
        \label{fig11a}
    \end{subfigure}
    \begin{subfigure}{0.49\textwidth}
        \includegraphics[width=\linewidth]{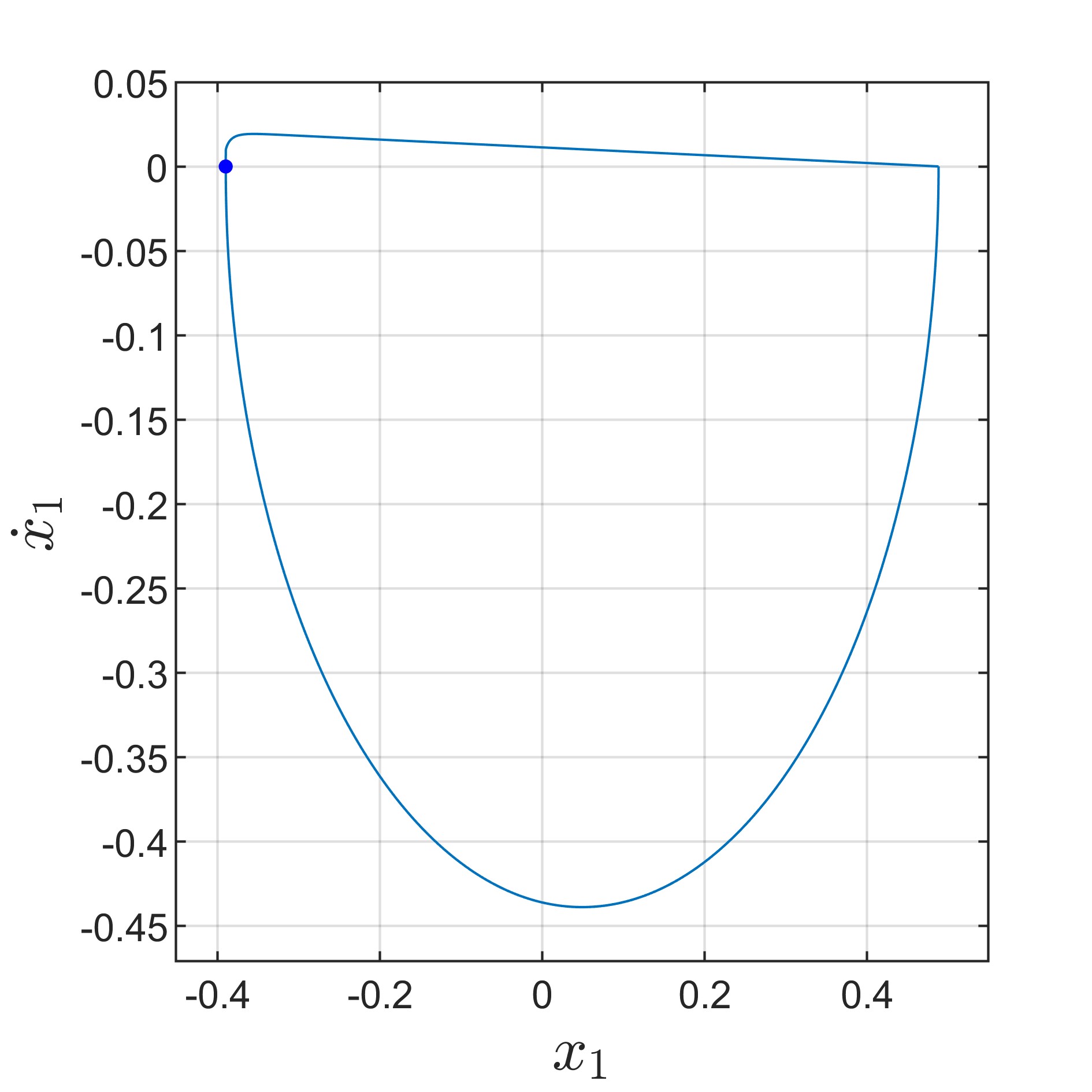}
        \caption{}
        \label{fig11b}
    \end{subfigure}
\caption{
Time history (a) and phase plot (b) with the corresponding Poincaré point at $\eta = 9.95$, for the symmetric spring configuration ($\hat{k}_1 = \hat{k}_2 = 5000$ N/m, so $k_{12} = 0$); Integration is started from $(x_{0},\,\dot{x}_{0},\,\varphi_{0},\,\dot{\varphi}_{0}) = (0,\,0,\,0,\,0)$; Other parameters follow Eq.~\ref{eq:nondimensional parameter values}.
}
    \label{fig11}
\end{figure}

For the symmetric spring case with equal stiffness ($k_{12} = 0$), the bifurcation diagram under DC motor drive is identical to that shown in Fig.~\ref{fig4}, and is therefore not repeated here. Fig.~\ref{fig11} presents the time history and phase plot for the system with equal spring stiffness, similar to the symmetric case examined in Fig.~\ref{fig5}. The system shows single-period motion in both cases. However, while the constant belt velocity case (Fig.~\ref{fig5}) produces sharp, nearly vertical transitions in the time history—characteristic of ideal stick-slip behavior—the DC motor drive (Fig.~\ref{fig11}) results in smooth, rounded transitions. This smoothing effect is due to the limited power supply from the motor, which cannot maintain perfectly constant belt velocity during the slip events, leading to a more gradual change in displacement compared to the ideal case.

\begin{figure}[H]
    \centering
    \begin{subfigure}{0.49\textwidth}
        \includegraphics[width=\linewidth]{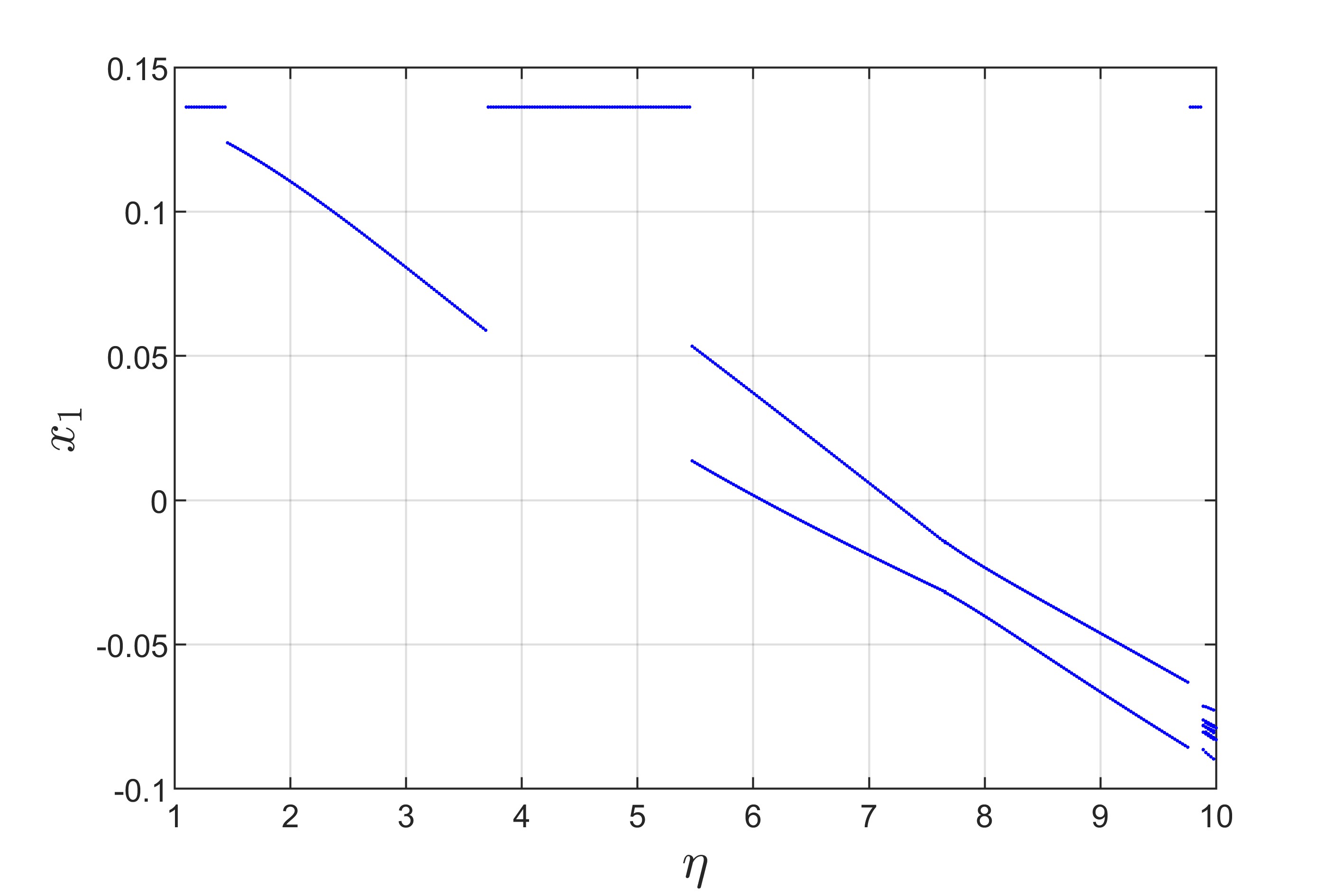}
        \caption{}
        \label{fig12a}
    \end{subfigure}
    \begin{subfigure}{0.49\textwidth}
        \includegraphics[width=\linewidth]{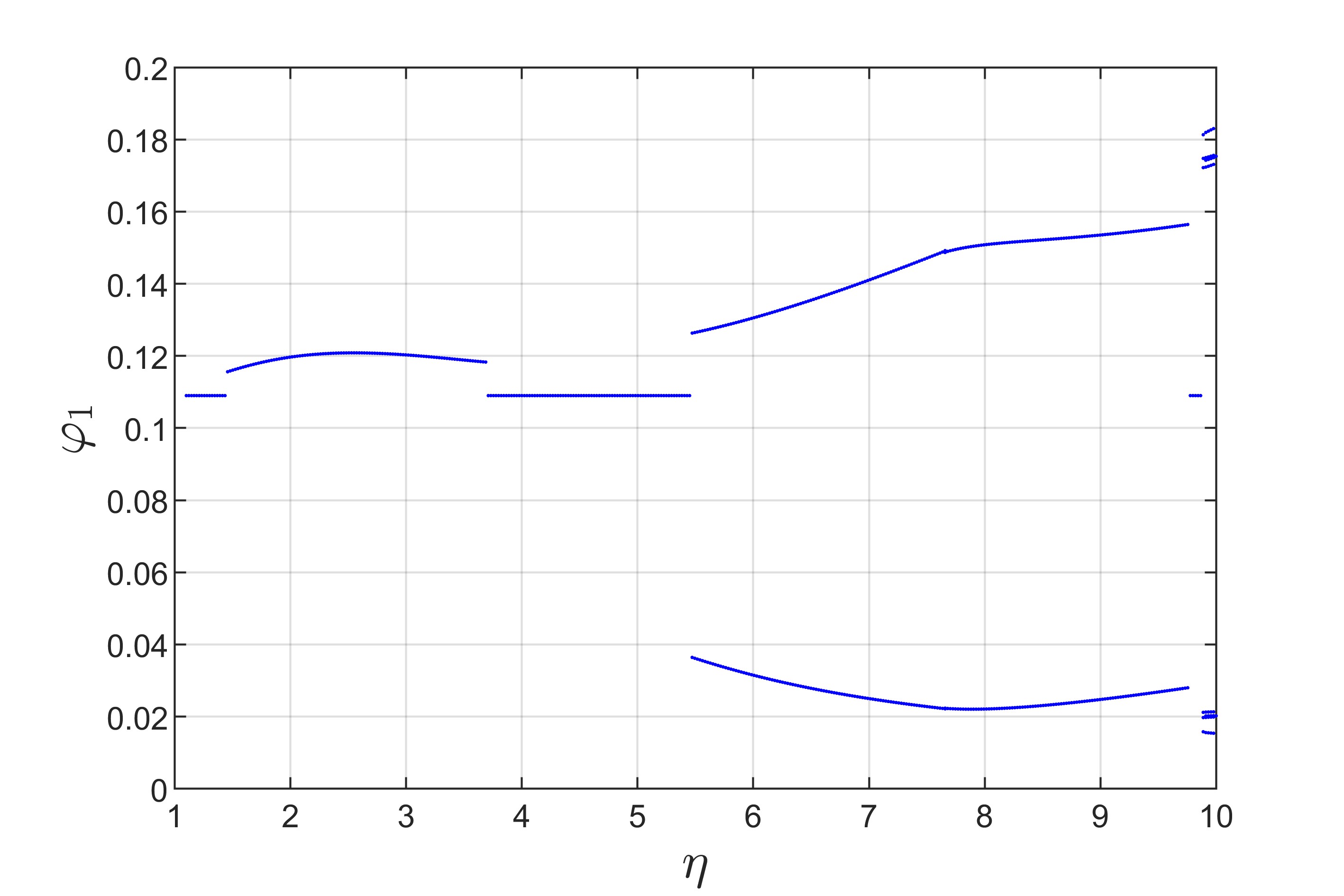}
        \caption{}
        \label{fig12b}
    \end{subfigure}
\caption{
Bifurcation diagrams for the translational ($x_1$, panel a) and rotational ($\varphi$, panel b) coordinates versus $\eta$ for the asymmetric spring configuration; For each $\eta$, integration is restarted from $(x_{0},\,\dot{x}_{0},\,\varphi_{0},\,\dot{\varphi}_{0}) = (0,\,0,\,0,\,0)$; Parameters follow Eq.~\ref{eq:nondimensional parameter values}.
}
    \label{fig12}
\end{figure}

Fig.~\ref{fig12} shows the bifurcation diagram for the system with asymmetric springs under DC motor drive. Comparing the $\varphi$ plot (Fig.~\ref{fig12}(b)), the overall pattern is similar to what is seen with constant belt velocity (see Fig.~\ref{fig6}). At the lower end of the sweeping range, both cases show an amplitude jump. However, the range where this first jump happens is smaller in the constant velocity case (Fig.~\ref{fig6}) compared to the DC motor case (Fig.~\ref{fig12}). At the higher end of the range, the two cases are different. In the constant velocity case (Fig.~\ref{fig6}), the period-2 sequence changes to period-6 for a short interval, then goes back to period-2, and finally becomes period-1 motion. In contrast, with the DC motor (Fig.~\ref{fig12}), the period-2 behavior continues almost to the end of the range, followed by a short period-1 regime and then period-7 behavior.
In the bifurcation diagram for $x_1$ (Fig.~\ref{fig12}(a)), single-period motion at the start of the sweeping range appears as a horizontal line, showing that the amplitude stays constant as $\eta$ increases. After the first amplitude jump, the single-period motion forms a tilted line. This means the amplitude now changes with $\eta$ instead of staying the same. There is another jump that returns the amplitude to the same value as at the beginning, again forming a flat segment. For double-period motion, the amplitude also changes with $\eta$, as shown by the sloped lines. In summary, constant-amplitude segments are seen at the beginning and after the second jump, while sloped lines show where amplitude varies with $\eta$.

\begin{figure}[H]
    \centering
    \begin{subfigure}{0.48\textwidth}
        \includegraphics[width=\linewidth]{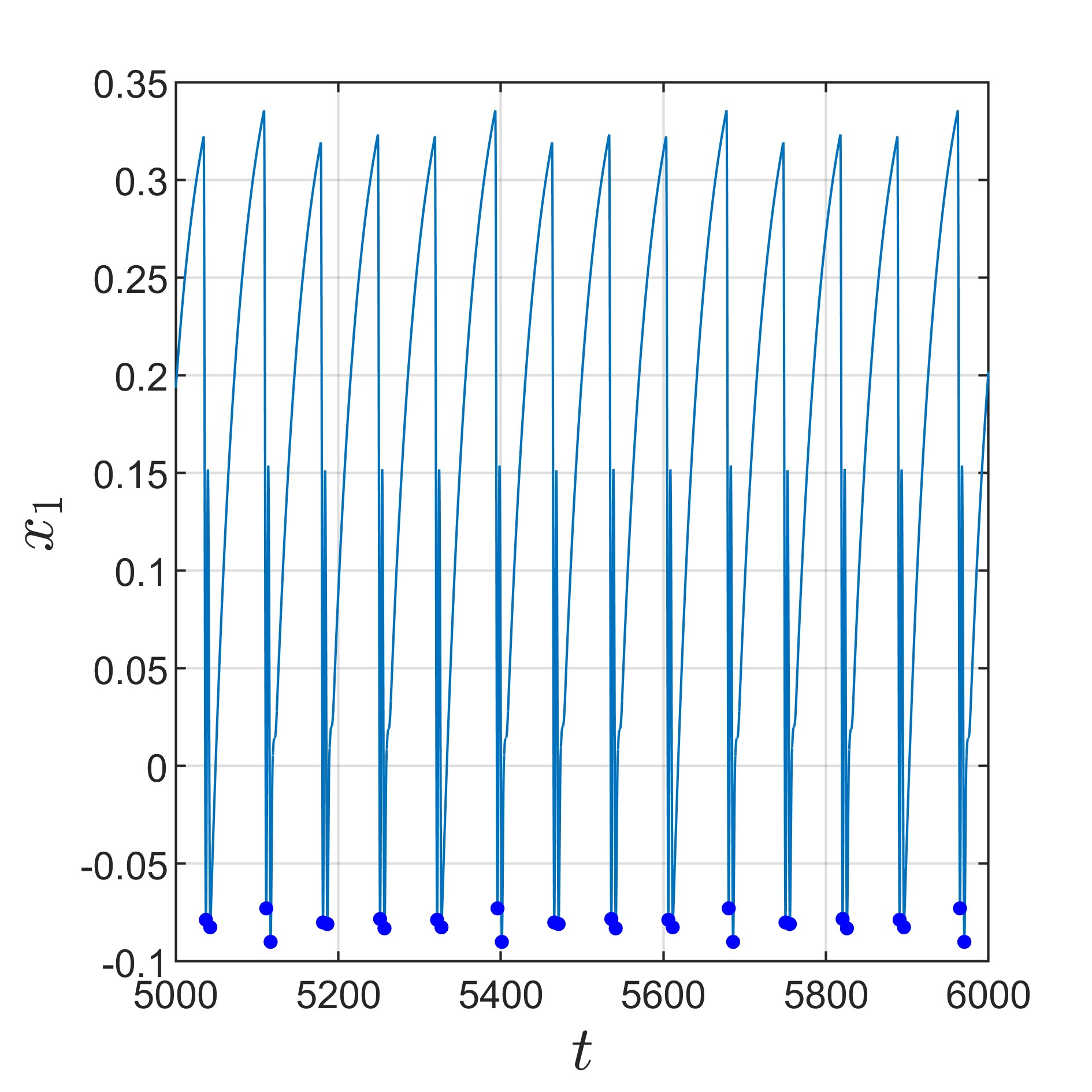}
        \caption{ }
        \label{fig13a}
    \end{subfigure}
    \begin{subfigure}{0.48\textwidth}
        \includegraphics[width=\linewidth]{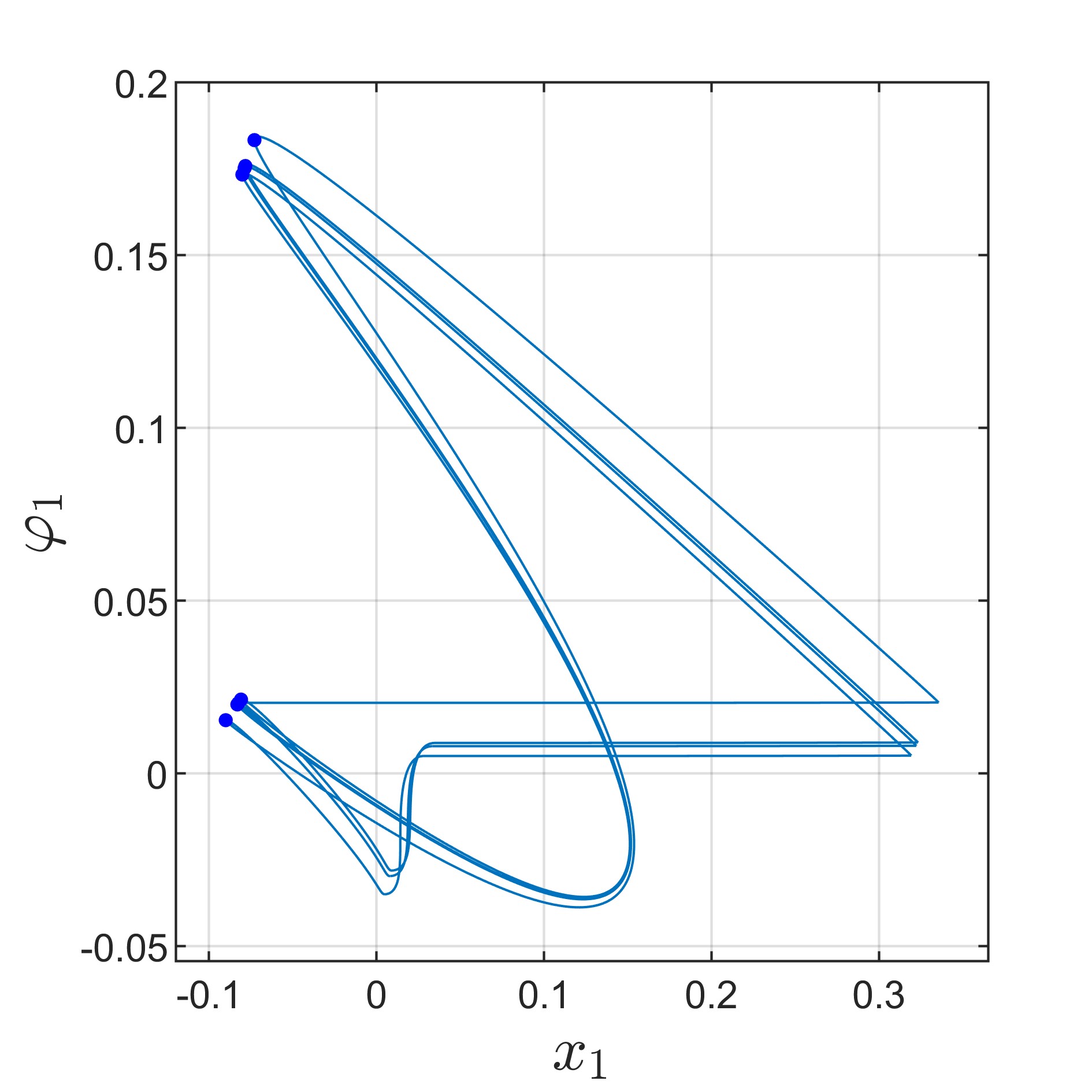}
        \caption{ }
        \label{fig13b}
    \end{subfigure}
        \begin{subfigure}{0.48\textwidth}
        \includegraphics[width=\linewidth]{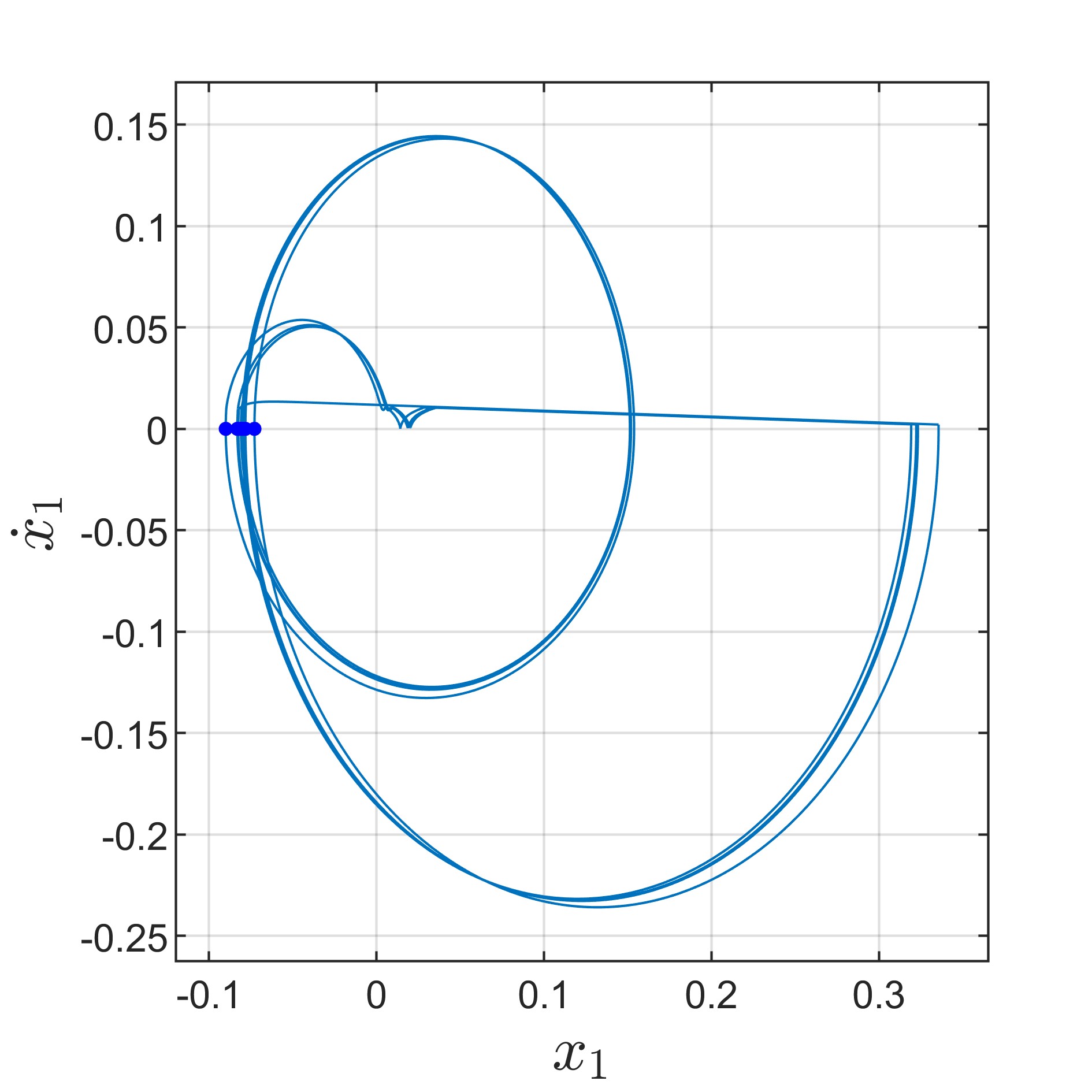}
        \caption{ }
        \label{fig13c}
    \end{subfigure}
           \begin{subfigure}{0.48\textwidth}
        \includegraphics[width=\linewidth]{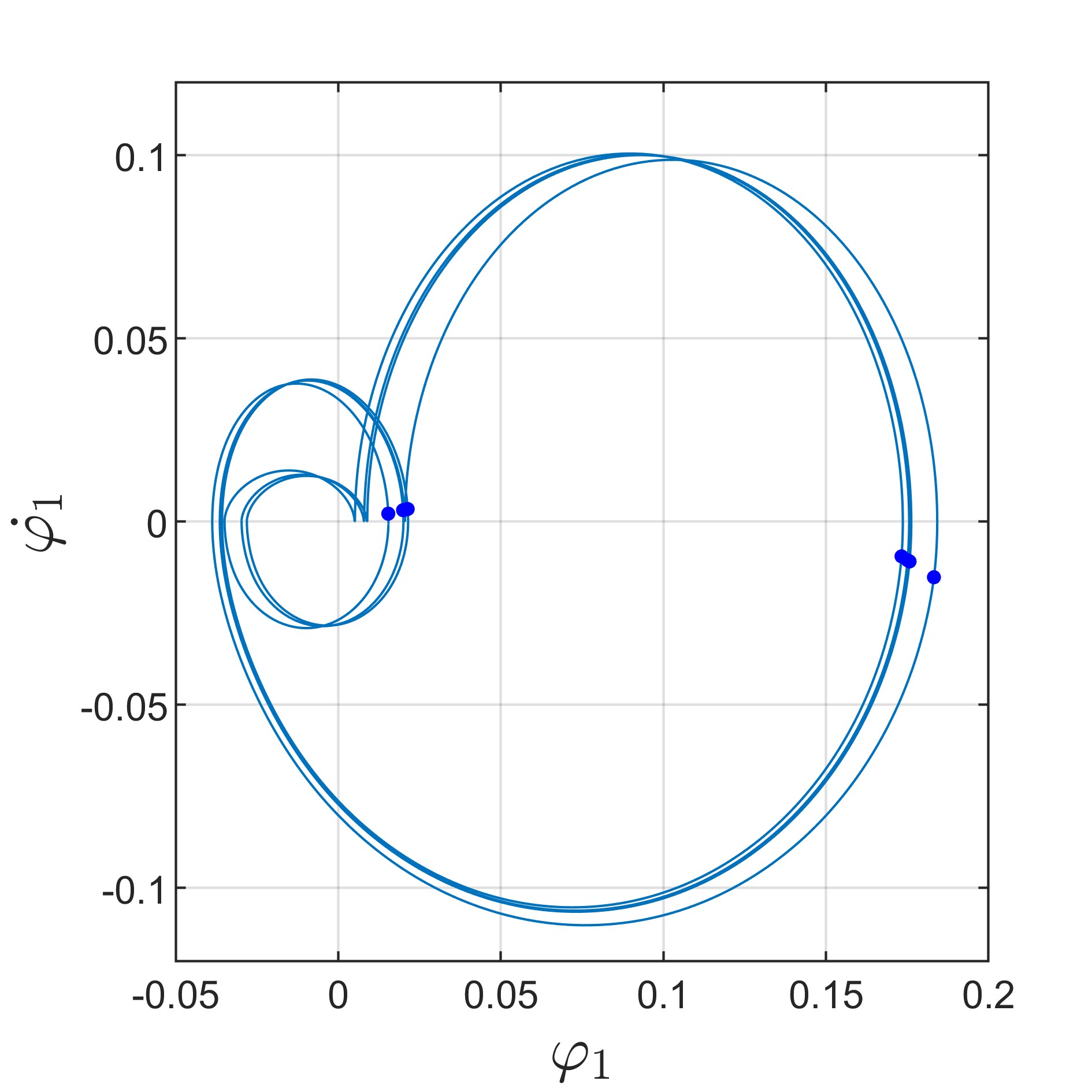}
        \caption{ }
        \label{fig13d}
    \end{subfigure}
\caption{
Time history (a) and phase plots (b--d), along with the corresponding Poincaré points (defined as local minima of $x$), at $\eta = 9.99$ for the asymmetric spring configuration; Integration is performed from the initial condition $(x_{0},\,\dot{x}_{0},\,\varphi_{0},\,\dot{\varphi}_{0}) = (0,\,0,\,0,\,0)$; Other parameters follow Eqs.~\ref{eq:nondimensional parameter values}.
}
    \label{fig13}
\end{figure}

Fig.~\ref{fig13} shows the time histories and phase plots for the system in the period-7 regime, observed at high values of $\eta$ as identified in the bifurcation diagram (see Fig.~\ref{fig12}). The time history reveals seven repeated cycles before the motion repeats, demonstrating the period-7 response. In the phase plots, this corresponds to seven distinct loops or branches, which clearly reflect the sevenfold periodicity. The presence of these multiple cycles and loops confirms that the system settles into a stable period-7 state in this high-$\eta$ region.

\begin{figure}[H]
  \centering
  \includegraphics[scale=0.15]{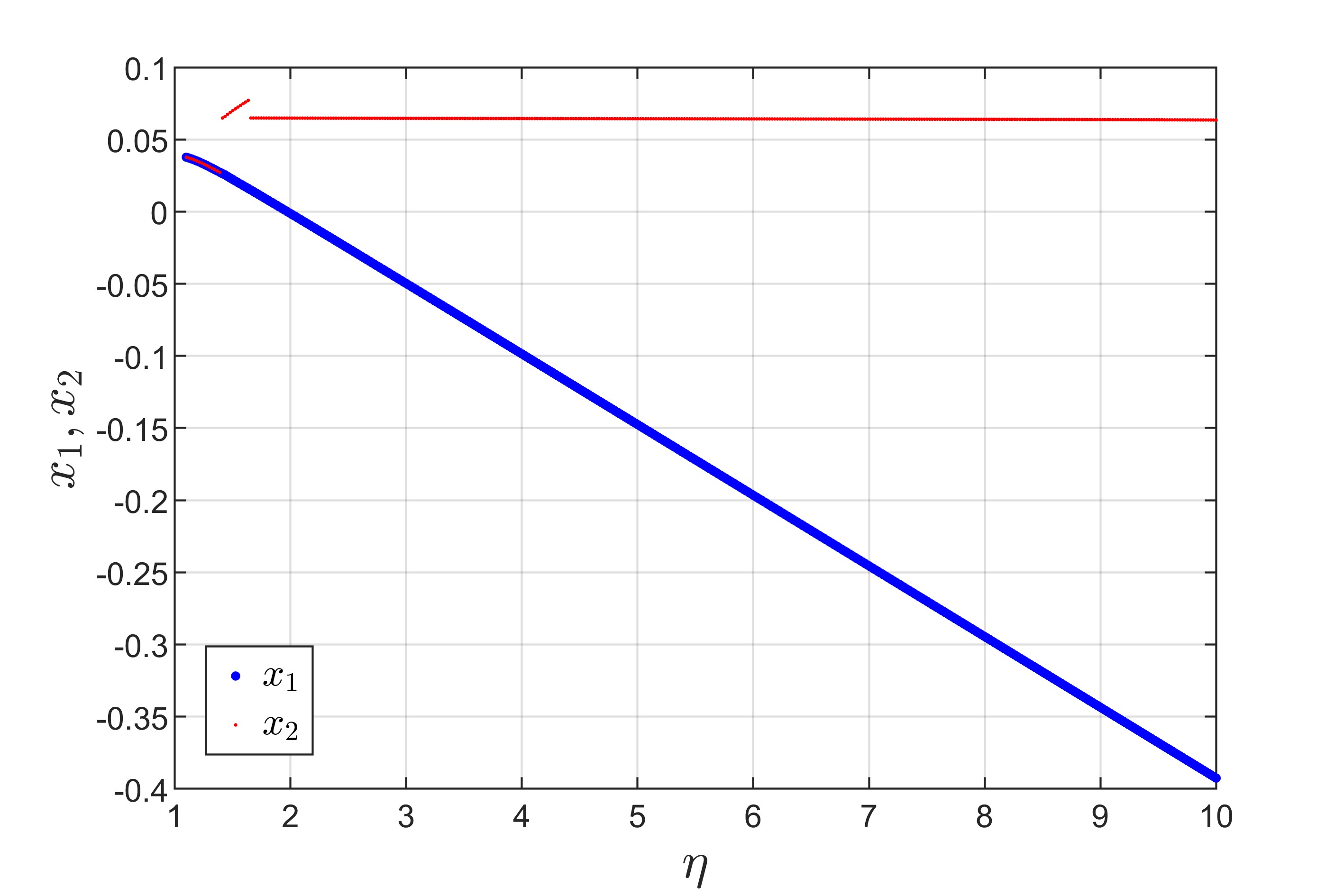} 
  \caption{
Bifurcation diagram of the Poincaré map for two oscillators defined by the local minima of $x_1$, as the friction coefficient ratio ($\eta$) is swept as the control parameter for the symmetric spring configuration ($\hat{k}_1 = \hat{k}_2 = 5000$ and thus $k_{12} = 0$); For each $\eta$, integration is restarted from the initial condition $(x_{1_0}, x_{2_0}, \dot{x}_{1_0}, \dot{x}_{2_0}, \varphi_{1_0}, \varphi_{2_0}, \dot{\varphi}_{1_0}, \dot{\varphi}_{2_0}) = (0, 0.05, 0, 0.05, 0, 0, 0, 0)$; Other parameters follow Eq.~\ref{eq:nondimensional parameter values}.
}
  \label{fig14}
\end{figure}

\begin{figure}
    \centering
    \begin{subfigure}{0.48\textwidth}
        \includegraphics[width=\linewidth]{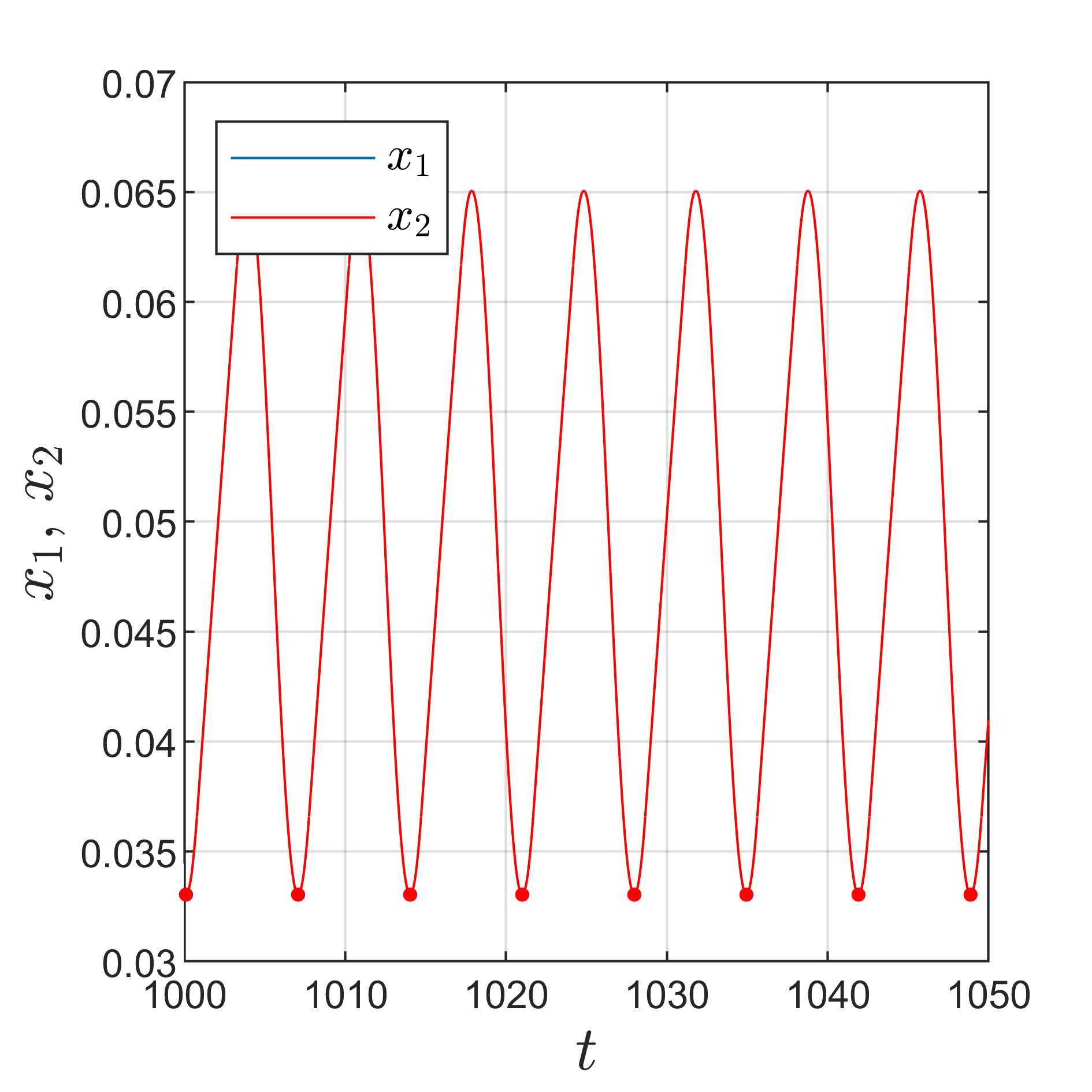}
        \caption{ }
        \label{fig15a}
    \end{subfigure}
    \begin{subfigure}{0.48\textwidth}
        \includegraphics[width=\linewidth]{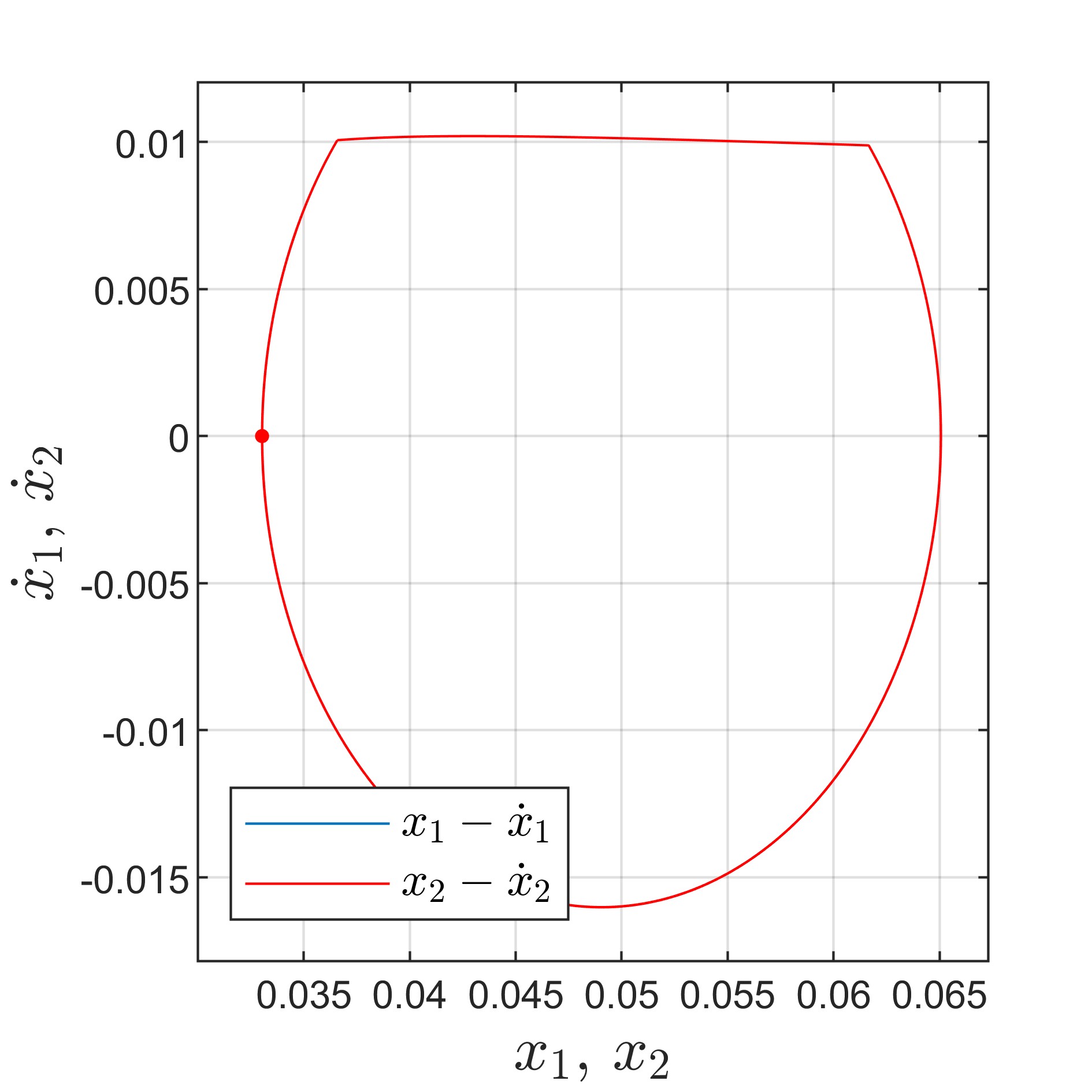}
        \caption{ }
        \label{fig15b}
    \end{subfigure}
        \begin{subfigure}{0.48\textwidth}
        \includegraphics[width=\linewidth]{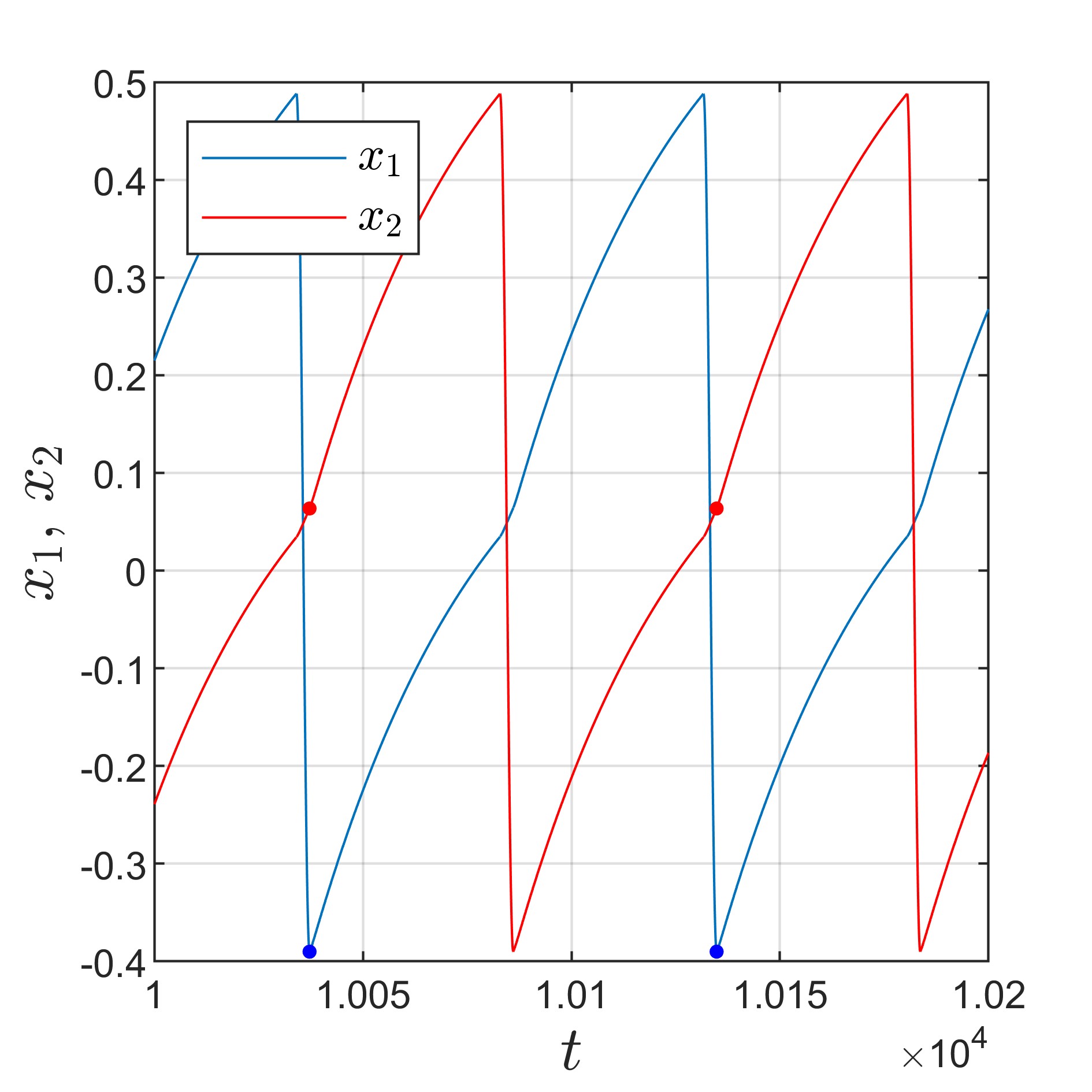}
        \caption{ }
        \label{fig15c}
    \end{subfigure}
           \begin{subfigure}{0.48\textwidth}
        \includegraphics[width=\linewidth]{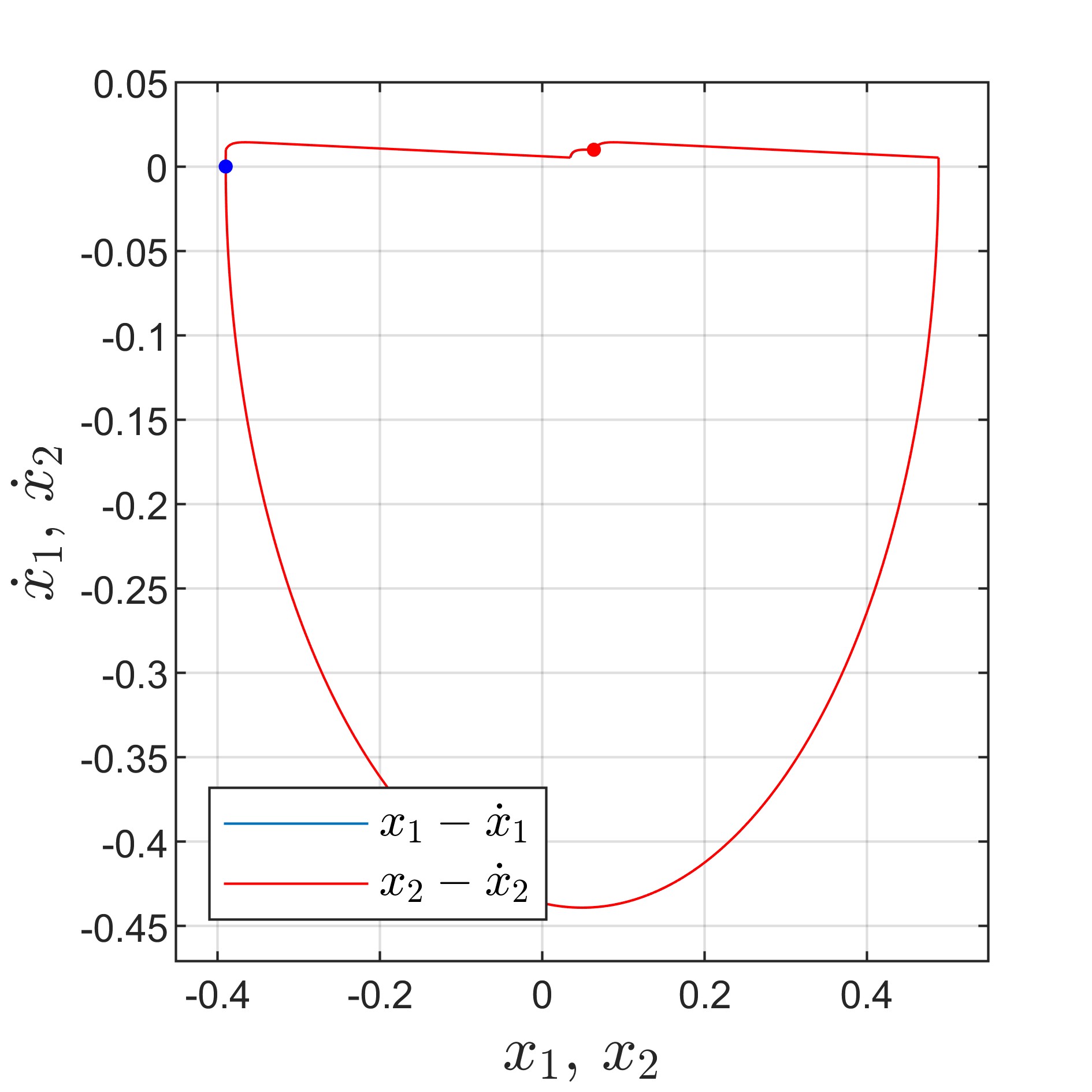}
        \caption{ }
        \label{fig15d}
    \end{subfigure}
\caption{
Time history (a) and phase plots (b--d), along with the corresponding Poincaré points (defined as local minima of $x_1$), at $\eta = 9.99$ for the symmetric spring configuration with two oscillators; Integration is performed from the initial condition $(x_{1_0}, x_{2_0}, \dot{x}_{1_0}, \dot{x}_{2_0}, \varphi_{1_0}, \varphi_{2_0}, \dot{\varphi}_{1_0}, \dot{\varphi}_{2_0}) = (0, 0.05, 0, 0.05, 0, 0, 0, 0)$; Other parameters follow Eqs.~\ref{eq:nondimensional parameter values}.
}
    \label{fig15}
\end{figure}

Fig.~\ref{fig14} shows the bifurcation diagram for the system with two oscillators ($N=2$) and symmetric springs ($k_{12}=0$). The diagram shows that both oscillators remain in single-period motion as $\eta$ is varied, just like in the single oscillator case. The variable $\varphi$ does not appear because of the system’s symmetry. Although the plot shows different amplitudes for $x_1$ and $x_2$, this is not because the second oscillator has a larger amplitude, but because of a phase difference between the two oscillators. The points in the diagram are recorded when $x_1$ reaches its minimum, so if $x_2$ is not at its minimum at the same time, a different value is recorded for $x_2$, and this produces the appearance of two separate branches. At very low values of $\eta$, both oscillators move in phase with each other, as shown in Fig.~\ref{fig15}a,b, which present the time history and phase portrait for this case. However, as $\eta$ increases, the second oscillator goes out of phase with the first one, as illustrated in Fig.~\ref{fig15}c,d. This out-of-phase behavior remains throughout the rest of the sweeping range of $\eta$.

\begin{figure}
    \centering
    \begin{subfigure}{0.49\textwidth}
        \includegraphics[width=\linewidth]{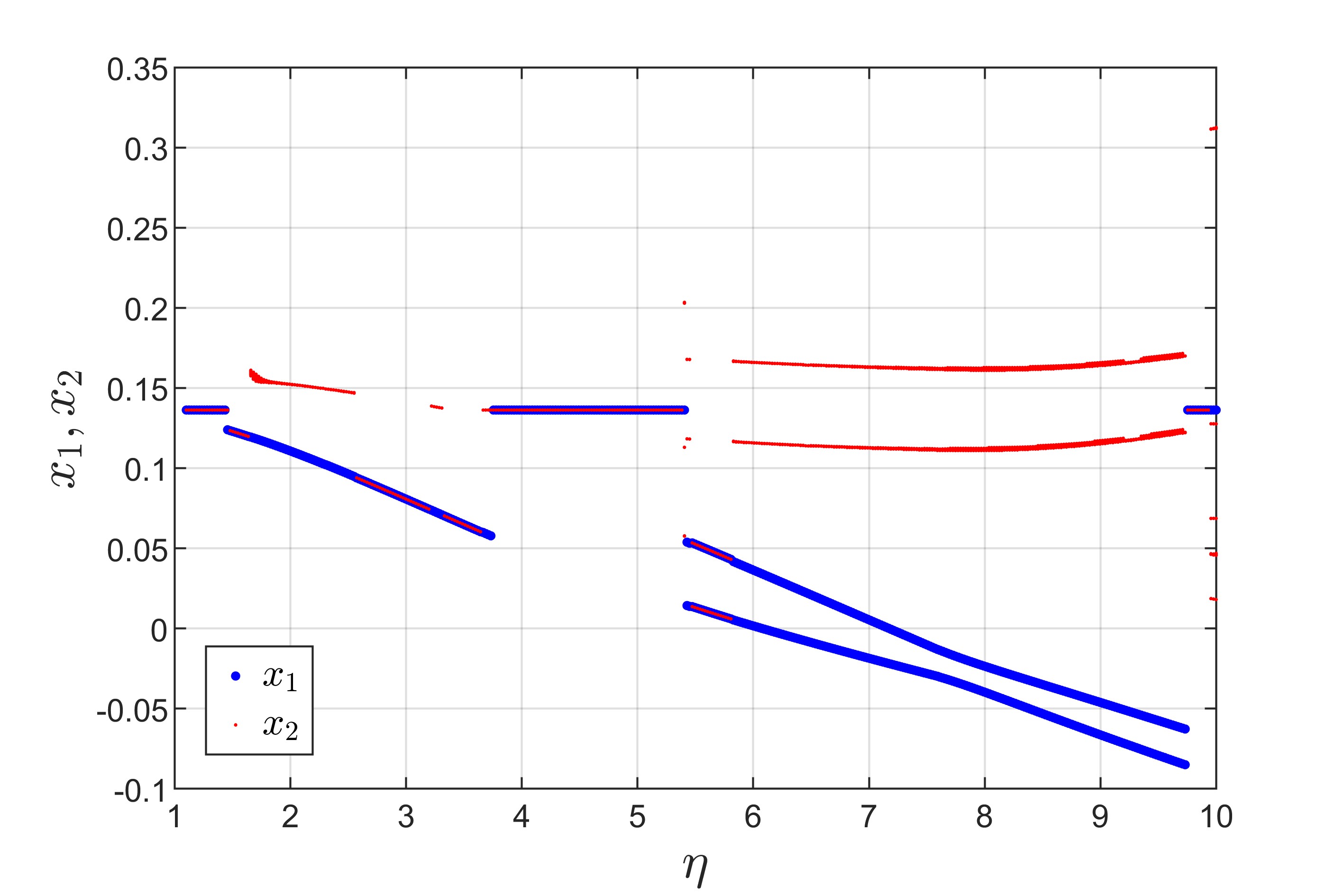}
        \caption{}
        \label{fig16a}
    \end{subfigure}
    \begin{subfigure}{0.49\textwidth}
        \includegraphics[width=\linewidth]{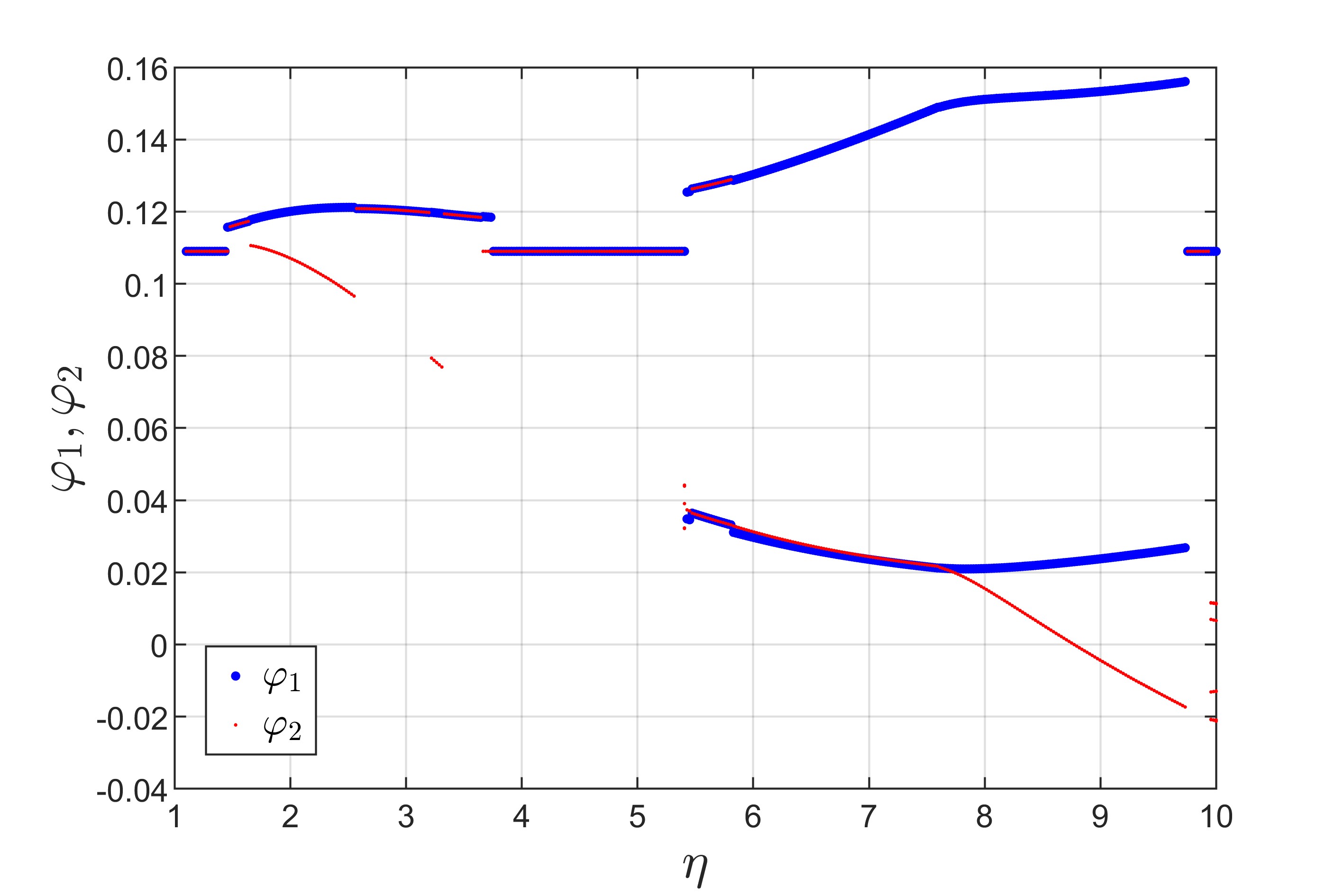}
        \caption{}
        \label{fig16b}
    \end{subfigure}
\caption{
Bifurcation diagrams for two oscillators defined by the translational ($x_1$, panel a) and rotational ($\varphi$, panel b) coordinates versus $\eta$ for the asymmetric spring configuration; For each $\eta$, integration is restarted from the initial condition $(x_{1_0}, x_{2_0}, \dot{x}_{1_0}, \dot{x}_{2_0}, \varphi_{1_0}, \varphi_{2_0}, \dot{\varphi}_{1_0}, \dot{\varphi}_{2_0}) = (0, 0.01, 0, 0.01, 0, 0.01, 0, 0.01)$; Parameters follow Eq.~\ref{eq:nondimensional parameter values}.
}

    \label{fig16}
\end{figure}

\begin{figure}
    \centering
    \begin{subfigure}{0.48\textwidth}
        \includegraphics[width=\linewidth]{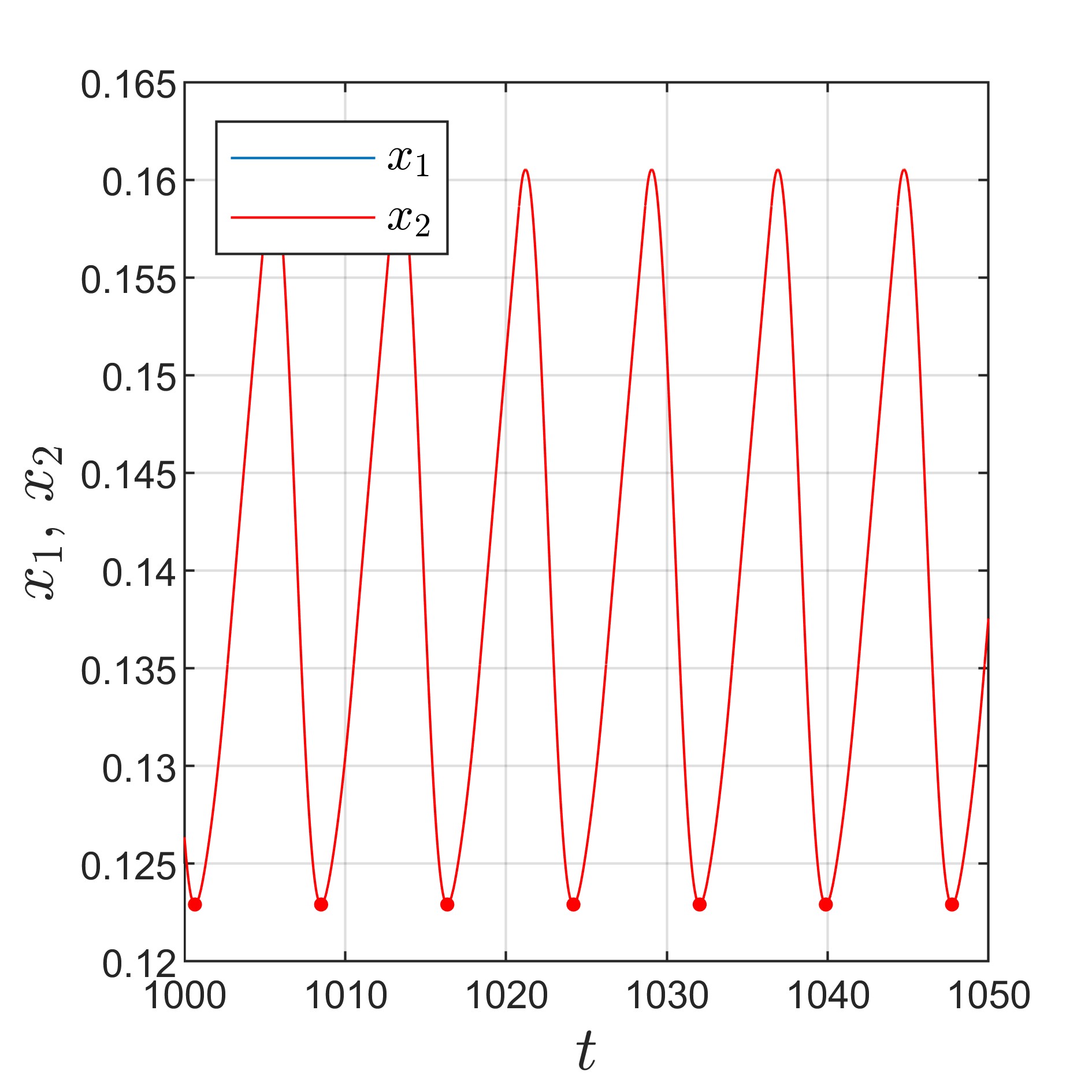}
        \caption{ }
        \label{fig17a}
    \end{subfigure}
    \begin{subfigure}{0.48\textwidth}
        \includegraphics[width=\linewidth]{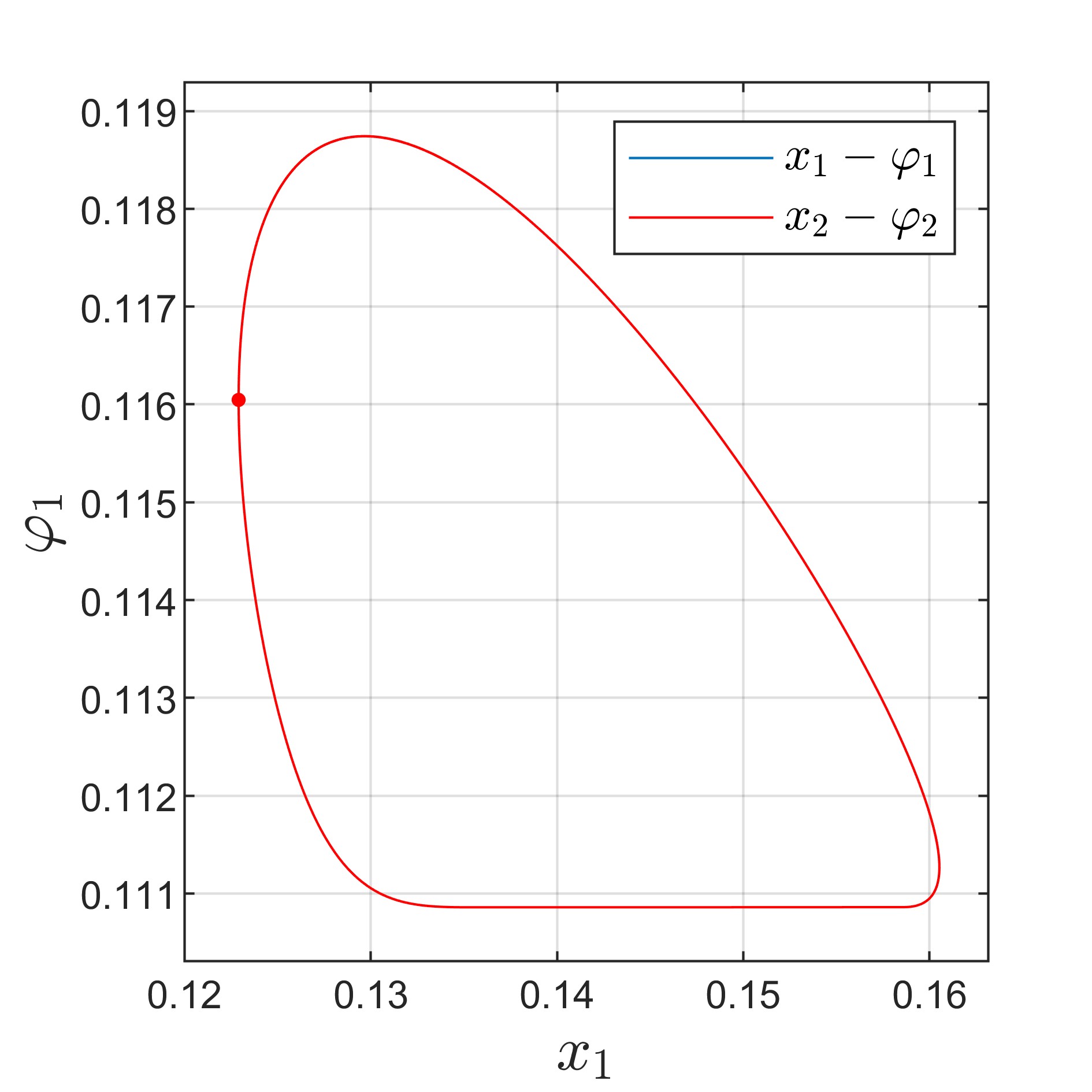}
        \caption{ }
        \label{fig17b}
    \end{subfigure}
        \begin{subfigure}{0.48\textwidth}
        \includegraphics[width=\linewidth]{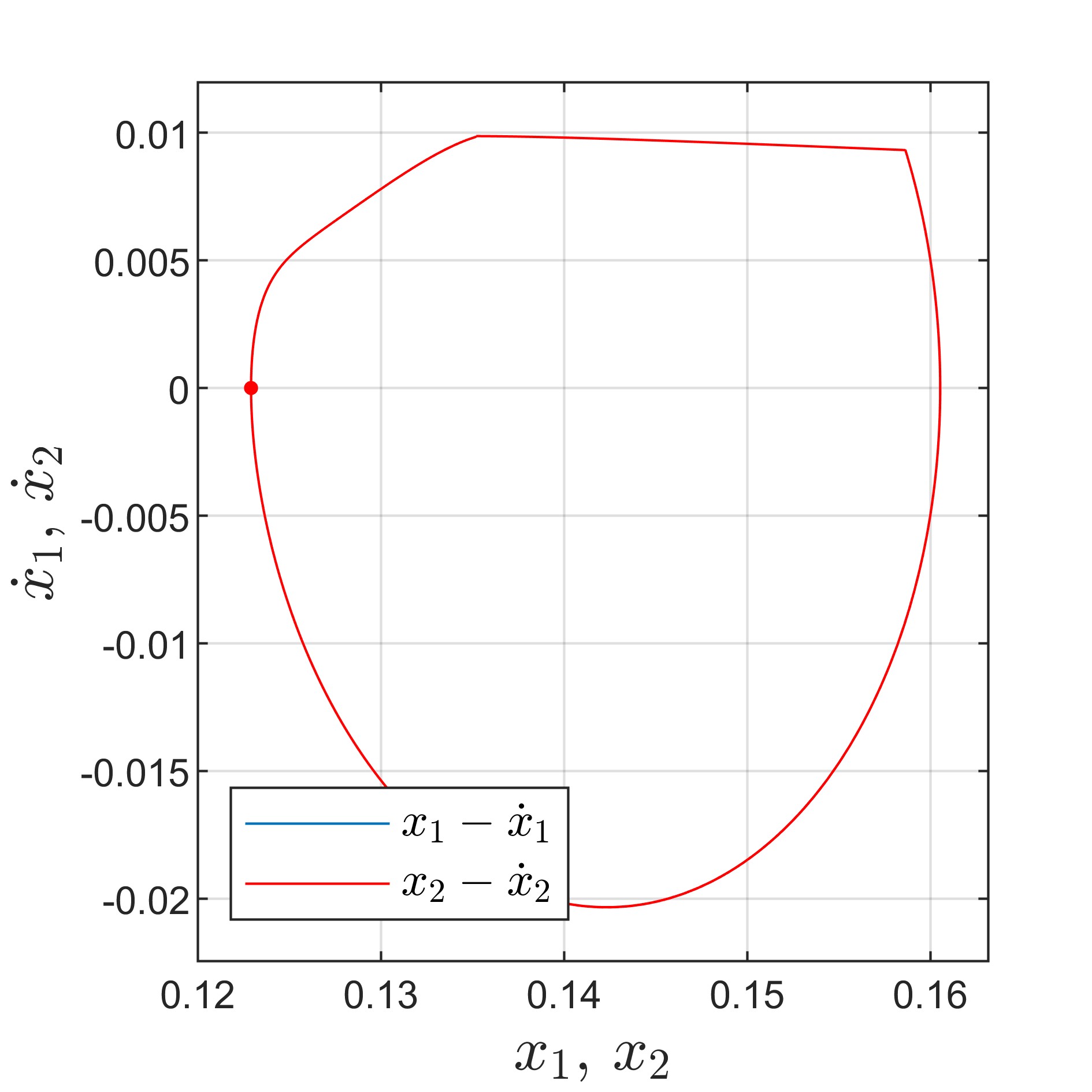}
        \caption{ }
        \label{fig17c}
    \end{subfigure}
           \begin{subfigure}{0.48\textwidth}
        \includegraphics[width=\linewidth]{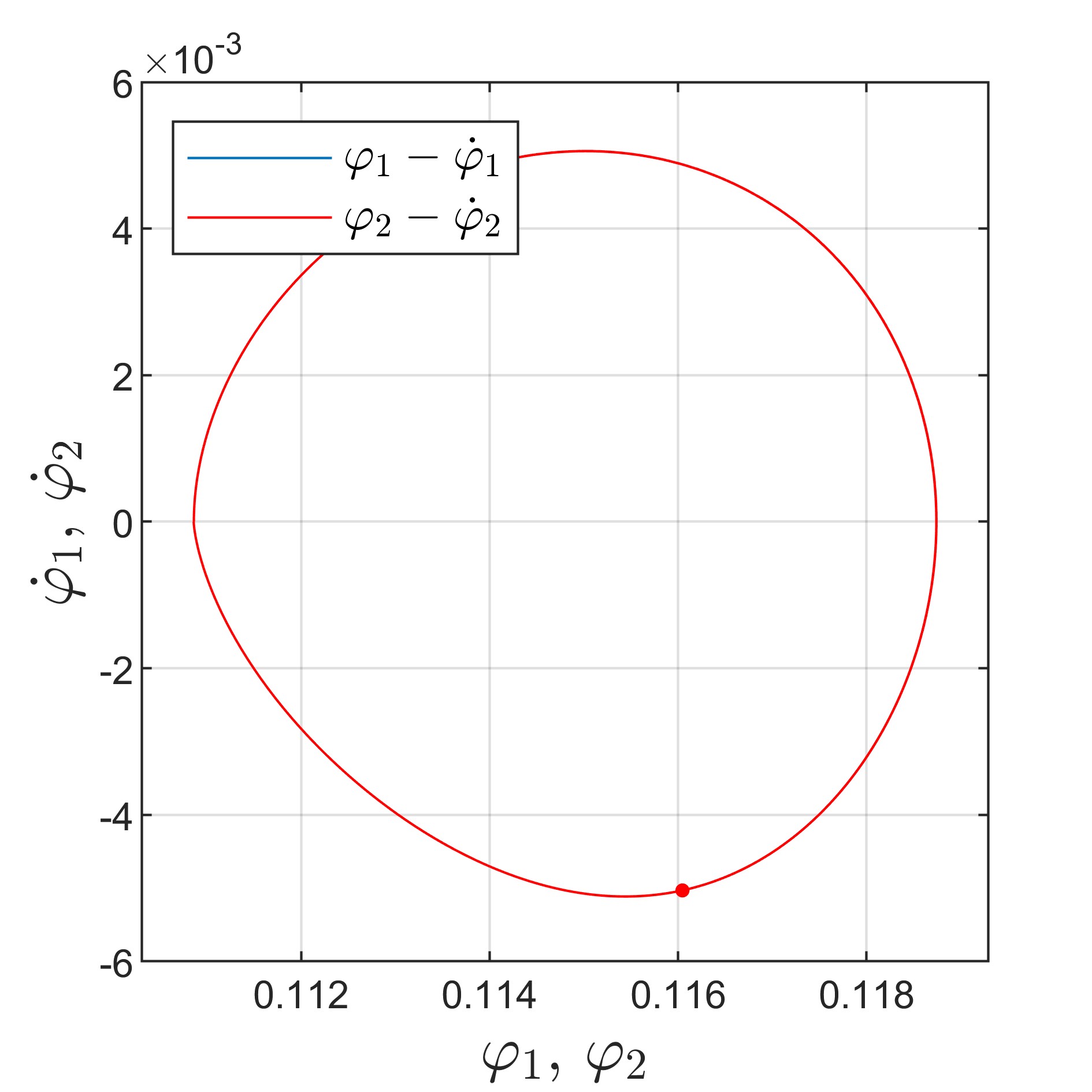}
        \caption{ }
        \label{fig17d}
    \end{subfigure}
\caption{
Time history (a) and phase plots (b--d), along with the corresponding Poincaré points (defined as local minima of $x_1$), at $\eta = 1.5$ for the symmetric spring configuration with two oscillators; Integration is performed from the initial condition $(x_{1_0}, x_{2_0}, \dot{x}_{1_0}, \dot{x}_{2_0}, \varphi_{1_0}, \varphi_{2_0}, \dot{\varphi}_{1_0}, \dot{\varphi}_{2_0}) = (0, 0.01, 0, 0.01, 0, 0.01, 0, 0.01)$; Other parameters follow Eqs.~\ref{eq:nondimensional parameter values}.
}
    \label{fig17}
\end{figure}

\begin{figure}
    \centering
    \begin{subfigure}{0.48\textwidth}
        \includegraphics[width=\linewidth]{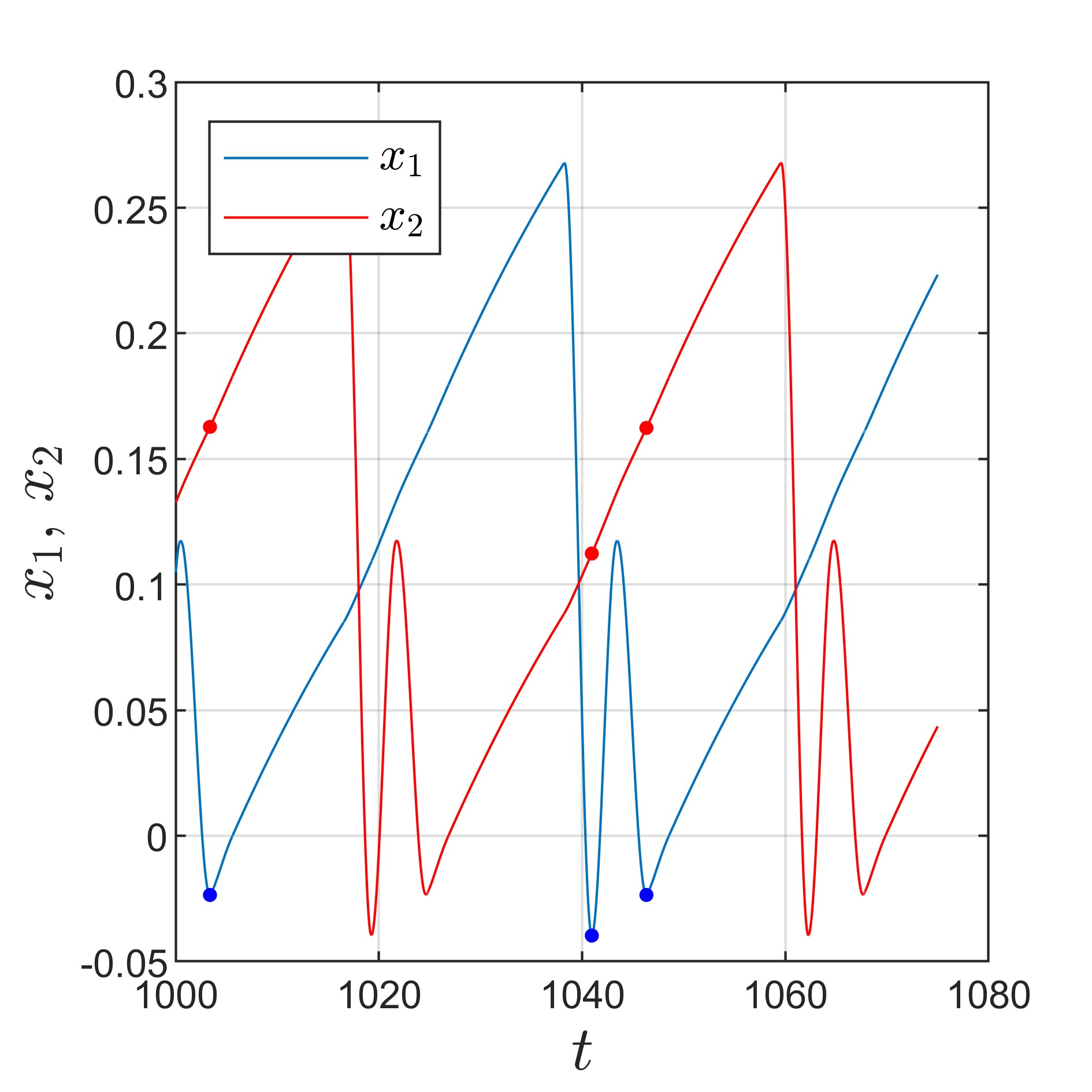}
        \caption{ }
        \label{fig18a}
    \end{subfigure}
    \begin{subfigure}{0.48\textwidth}
        \includegraphics[width=\linewidth]{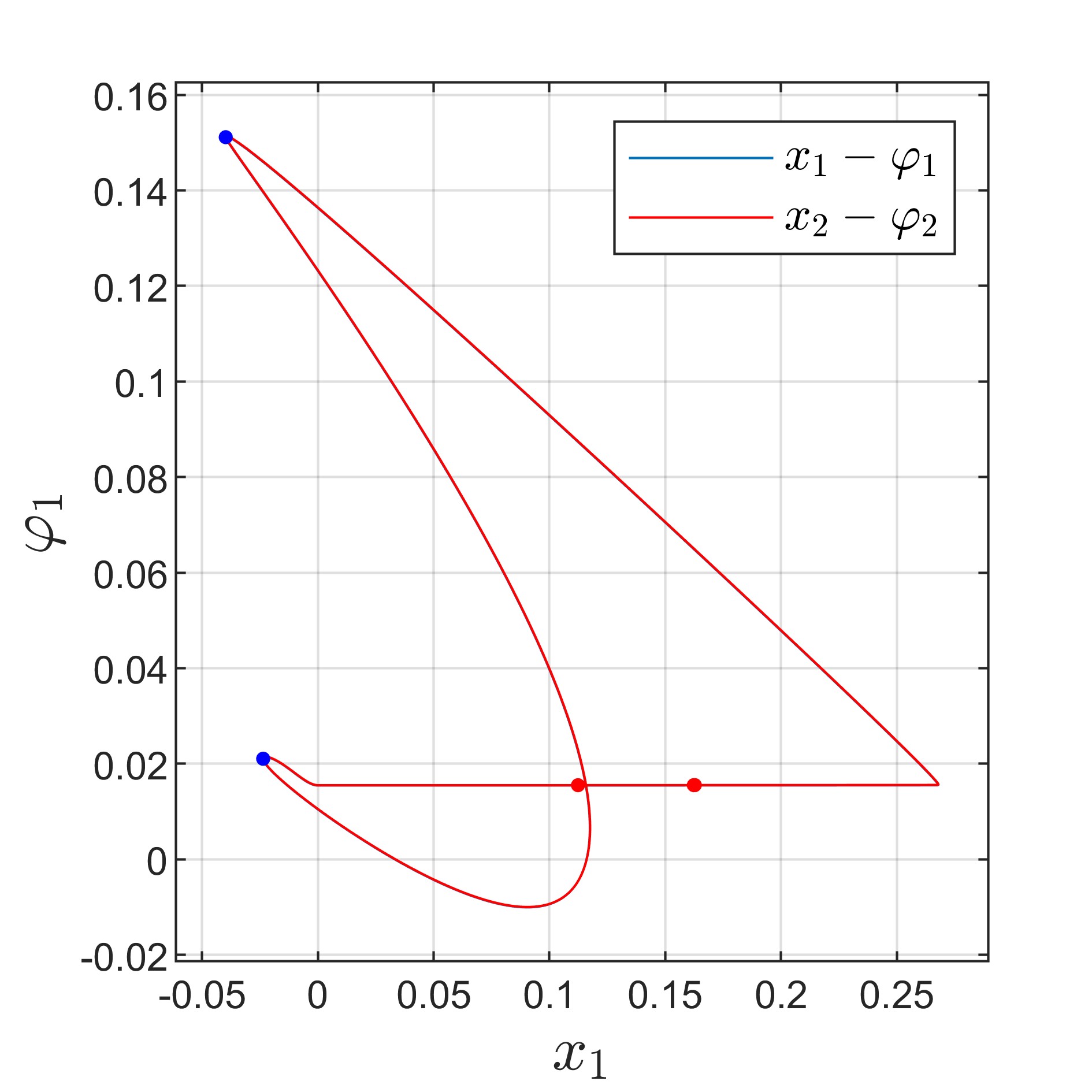}
        \caption{ }
        \label{fig18b}
    \end{subfigure}
        \begin{subfigure}{0.48\textwidth}
        \includegraphics[width=\linewidth]{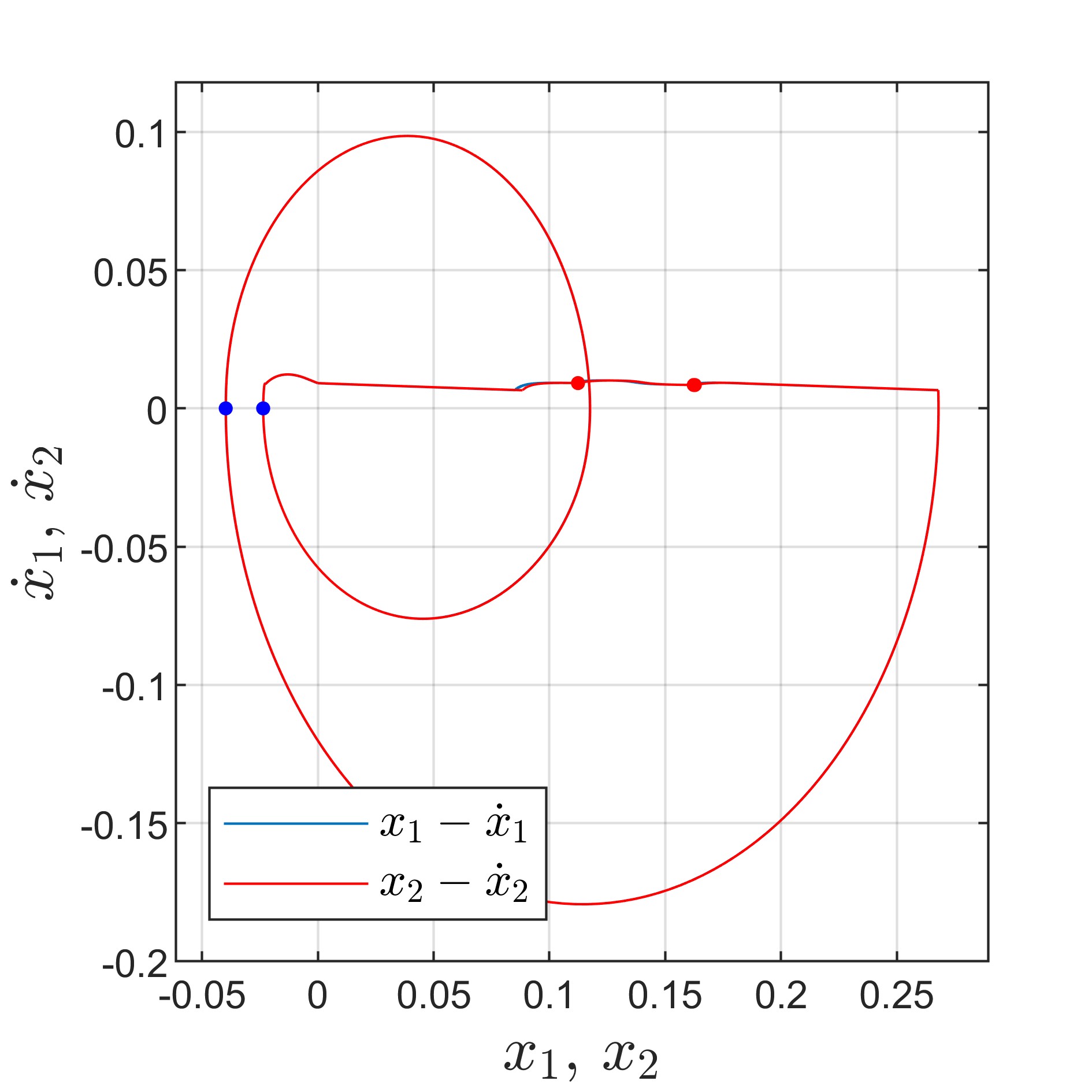}
        \caption{ }
        \label{fig18c}
    \end{subfigure}
           \begin{subfigure}{0.48\textwidth}
        \includegraphics[width=\linewidth]{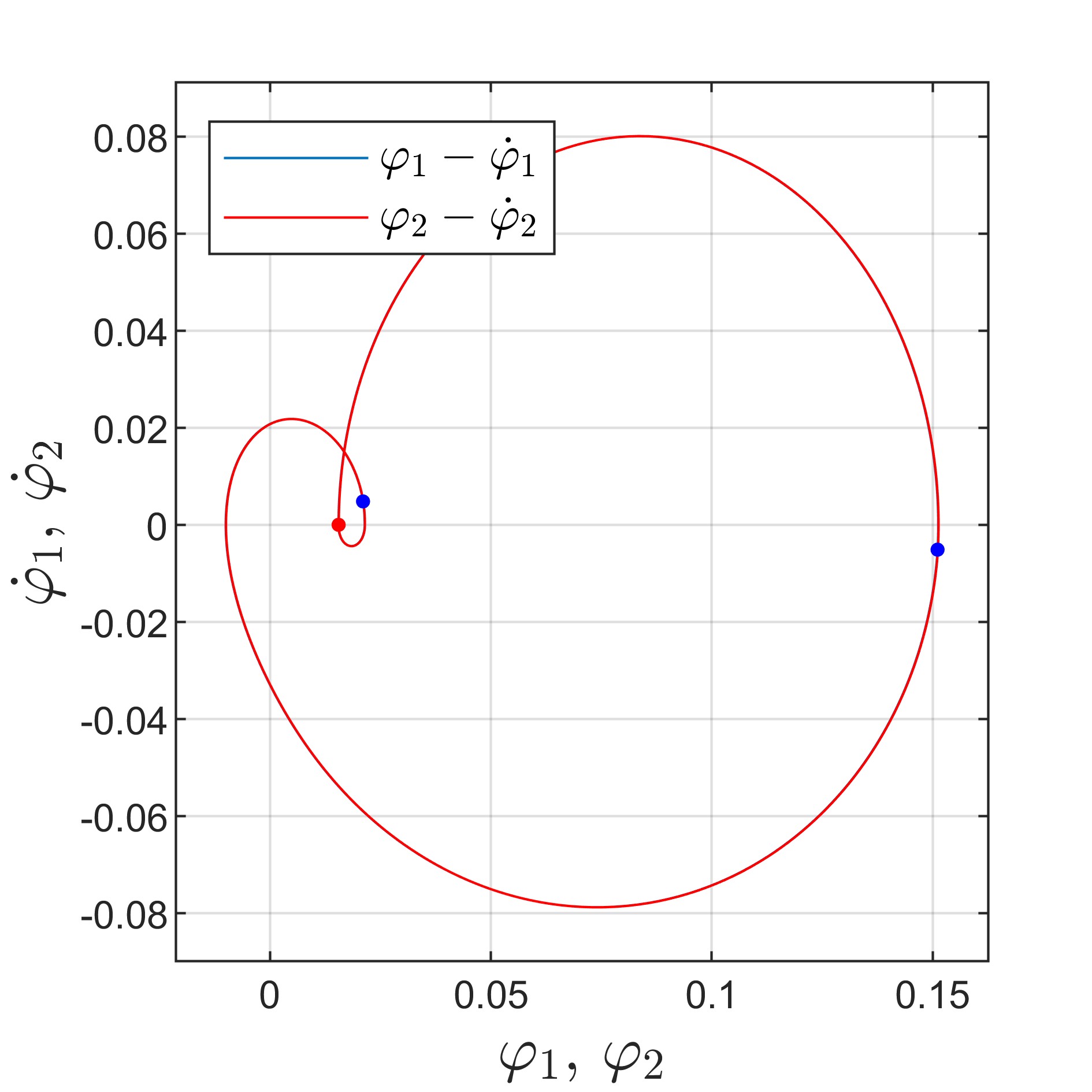}
        \caption{ }
        \label{fig18d}
    \end{subfigure}
\caption{Time history (a) and phase plots (b--d), along with the corresponding Poincaré points (defined as local minima of $x_1$), at $\eta = 8$ for the symmetric spring configuration with two oscillators; Integration is performed from the initial condition $(x_{1_0}, x_{2_0}, \dot{x}_{1_0}, \dot{x}_{2_0}, \varphi_{1_0}, \varphi_{2_0}, \dot{\varphi}_{1_0}, \dot{\varphi}_{2_0}) = (0, 0.01, 0, 0.01, 0, 0.01, 0, 0.01)$; Other parameters follow Eqs.~\ref{eq:nondimensional parameter values}.
}
    \label{fig18}
\end{figure}

Fig.~\ref{fig16} shows the bifurcation diagram for two coupled oscillators with asymmetric spring configuration. As in the previous case, differences in amplitude between $x_1$ and $x_2$ reflect underlying phase differences, not true amplitude changes. At the lowest values of $\eta$, both oscillators are synchronized. With a small increase in $\eta$, their amplitudes jump but remain synchronized (see Fig.~\ref{fig17}. As $\eta$ increases further, two intervals appear where the oscillators desynchronize—their phase shifts lead to different recorded values when $x_1$ is at its minimum. Between these intervals, and after each major transition, the oscillators return to synchronized motion. As $\eta$ increases more, a bifurcation occurs, producing period-2 motions which are also desynchronized and persist for much of the range. The details of this desynchronized period-2 motion can be seen in Fig.~\ref{fig18} (for $\eta = 8$). Near the end of the sweeping range, a final amplitude jump returns both oscillators to synchronized, single-period motion.

for two oscillators defined

\section{Concluding remarks}\label{sec:Concluding remarks}

This study presents a comprehensive framework for analyzing coupled translational-rotational stick-slip oscillators driven by finite-area friction and non-ideal energy sources. The closed-loop belt-motor formulation successfully extends from single-oscillator systems to chains of $N$ oscillators powered by a common DC motor.

For single oscillators, symmetric spring configurations exhibit only simple periodic motion with no bifurcations, reducing to classical one-dimensional stick-slip behavior. Introducing stiffness asymmetry activates rotational-translational coupling, producing amplitude jumps, period-2 cascades, and multi-periodic regimes up to period-28. The friction coefficient ratio emerges as the key control parameter governing these transitions.

Replacing constant belt velocity with DC motor drive shifts bifurcation thresholds to lower friction ratios, alters the multi-periodic sequence structure, and produces amplitude-dependent responses absent under ideal excitation. Limited motor power smooths sharp stick-slip transitions, resulting in more gradual displacement changes during slip events.

For coupled systems with symmetric springs, both oscillators maintain periodic motion but transition from in-phase to out-of-phase configurations as friction increases. In asymmetric configurations, the system alternates between synchronized and desynchronized states, exhibiting period-adding bifurcations and amplitude jumps. At higher friction ratios, extended intervals of desynchronized period-2 motion appear before returning to synchronized single-periodic states.

The smooth friction formulations based on the Contensou-Zhuravlev framework enable long-term simulations and detailed bifurcation analysis while avoiding discontinuity-related singularities. These findings provide theoretical insight into friction-driven systems and practical guidance for engineering applications including clutches, brakes, precision positioning mechanisms, and vibration-driven actuators

\section*{Author contribution statement}
{\textbf{Grzegorz Kudra:} Conceptualization, Methodology, Validation, Software, Visualization, Supervision, Project Administration; \textbf{Ali Fasihi:} Software, Investigation, Validation, Visualization, Result interpretation,  Writing—Reviewing and Editing; \textbf{Mohammad Parsa Rezaei:} Software, Investigation, Validation,  Literature Review, Visualization, , Data Curation (publishing code to online platforms) \textbf{Muhammad Junaid-U-Rehman:} Software, Visualization, Writing (Original draft); \textbf{Jan Awrejcewicz:} Supervision, Formal Analysis.}
\section*{Acknowledgements}
This work has been supported by the Polish National Science Centre under the grant OPUS 18 No. 2019/35/B/ST8/00980. This article was completed while Ali Fasihi, Mohammad Parsa Rezaei and Muhammad Junaid-U-Rehman were PhD students at the Interdisciplinary Doctoral School of Lodz University of Technology, Poland.

 For the purpose of Open Access, the authors have applied a CC-BY public copyright license to any Author Accepted Manuscript (AAM) version arising from this submission.
 
\section*{Declaration of competing interest}
The authors affirm that no personal relationships or recognized competing financial interests could have influenced the research provided in this study.
\section*{Data availability}
\sloppy
The data supporting these findings are available at:\\ \href{https://repod.icm.edu.pl/dataset.xhtml?persistentId=doi:10.18150/RQ5P7F}{https://repod.icm.edu.pl/dataset.xhtml?persistentId=doi:10.18150/RQ5P7F}

\end{document}